\documentclass[12pt]{article}
\usepackage[T1]{fontenc}
\usepackage[utf8]{inputenc}
\usepackage{textcomp,lmodern}
\usepackage{microtype}
\UseMicrotypeSet[protrusion]{basicmath}
\usepackage{parskip}
\usepackage{amsmath}
\usepackage{graphicx}
\usepackage{enumerate}
\usepackage[authoryear]{natbib}
\usepackage{url} 
\usepackage{amssymb,bbm}
\usepackage{amsthm}
\usepackage{algorithm}
\usepackage{algpseudocode}
\usepackage{multirow}
\RequirePackage{amsfonts}
\usepackage{verbatim}
\usepackage{subfigure}
\usepackage{bm}
\usepackage{enumitem}
\usepackage{xr}
\usepackage{tablefootnote}    
\usepackage{setspace}
\usepackage{authblk}

\newtheorem{theorem}{Theorem}
\newtheorem{lemma}{Lemma}
\newtheorem{corollary}{Corollary}
\newtheorem{assumption}{Assumption}

\newtheorem{example}{Example}

\newtheorem{remark}{Remark}
\newtheorem{proposition}{Proposition}

\usepackage[dvipsnames]{xcolor}
\usepackage{hyperref}
\hypersetup{hypertexnames=false,hidelinks}

\newcommand{\XL}{a}
\newcommand{\XR}{b}
\newcommand{\tb}{\textbf}
\newcommand{\bs}{\boldsymbol}

\title{Weighted Score Matching on Bounded Domains:
Nonparametric Density Estimation and Interaction
Graph Recovery}
\author[1]{Xinyuan Fan}
\author[2]{Rui Feng}
\author[3]{Chenlei Leng\thanks{Co-corresponding author: \texttt{chenlei.leng@polyu.edu.hk}}}
\author[1]{Weichi Wu\thanks{Co-corresponding author: \texttt{wuweichi@tsinghua.edu.cn}}}

\affil[1]{Department of Statistics and Data Science, Tsinghua University\\
	\texttt{fxy22@mails.tsinghua.edu.cn}, \texttt{wuweichi@tsinghua.edu.cn}}

\affil[2]{Department of Statistics, University of Warwick\\
	\texttt{rui.feng.1@warwick.ac.uk}}

\affil[3]{Department of Applied Mathematics, Hong Kong Polytechnic University\\
	\texttt{chenlei.leng@polyu.edu.hk}}

\date{}

\begin{document}
	
	\maketitle
	
\begin{abstract}
We study nonparametric density estimation and interaction-graph
recovery on bounded product domains using boundary-weighted
score matching. In a canonically centered pairwise log-density
model, graph edges are determined by bivariate interactions,
but penalizing the univariate main effects can induce false
edges. We derive the resulting shift in interaction scores
and study partially penalized weighted score matching
(P-WSM), which leaves main effects unpenalized.
At the population level, P-WSM excludes edges between
independent subsystems even when the marginal log densities
are not represented exactly by the sieve. A nonparametric
construction separates partial and joint penalization:
throughout a common deterministic tuning interval, P-WSM
recovers the graph while joint penalization selects a false
edge. We also give growing-dimensional recovery conditions
that do not require full-Gram invertibility.
For density estimation at fixed dimension, we establish
weighted Fisher-divergence rates and, under stronger
basis-specific conditions, Kullback--Leibler and Hellinger
rates for the normalized estimator. Simulations examine
the benefits and limitations of partial penalization.
An analysis of bone-marrow signaling data illustrates
nonlinear conditional-association estimation.
\end{abstract}
	\setstretch{1.25}
	\clearpage
\section{Introduction}

Bounded multivariate measurements arise in applications involving proportions,
physical constraints, and normalized scientific data, including representations
of single-cell cytometry measurements \citep{bendall2011single}. Two related
objectives are flexible joint density estimation and identification of
conditional associations. In an undirected graphical model, a missing edge
represents conditional independence given the remaining variables
\citep{lauritzen1996graphical,jordan2004graphical}.

Gaussian graphical models encode conditional independence through precision
matrix zeros \citep{yuan2007model}. Gaussian densities and their rectangular
truncations, however, have quadratic log densities with bilinear interactions
and cannot represent arbitrary nonlinear interaction functions. Fully
nonparametric multivariate density estimation offers greater flexibility but
faces dimensionality and boundary challenges. We therefore study density
estimation and graph recovery within a nonparametric pairwise log-density
model on a bounded product domain.

Let \(X=(X_1,\ldots,X_p)\) take values in
\(\Omega=\mathcal X^p\), where \(\mathcal X=[a,b]\), and write
\[
\log p_0(x)
=
C+\sum_{i=1}^p f_i(x_i)
+\sum_{1\le i<j\le p} f_{ij}(x_i,x_j).
\]
Canonical centering uniquely identifies the main effects \(f_i\) and
interactions \(f_{ij}\). For positive densities in this family,
\(f_{ij}\equiv0\) is equivalent to
\(X_i\perp X_j\mid X_{-\{i,j\}}\). Thus graph recovery amounts to selecting
the nonzero bivariate components, whereas the univariate components are
nuisance parameters for edge selection.

Likelihood-based estimation generally requires evaluating an intractable
normalizing constant. Score matching avoids this calculation by targeting
the log-density score \citep{hyvarinen2005estimation}. On bounded domains,
however, the integration-by-parts boundary terms underlying classical
score matching need not vanish. Coordinatewise boundary-vanishing weights
remove these terms under suitable regularity conditions
\citep{liu2022estimating}. Combining this construction with centered sieve
expansions yields a convex quadratic loss independent of the normalizing
constant.

Main effects do not determine the graph, but their estimation
can affect edge selection. The score-matching loss couples
main-effect and interaction coefficients through its
cross-Hessian. Penalizing the main effects therefore shifts
the interaction scores and can cause a missing edge to be
selected. This can occur even when both penalties tend to zero.

We study partially penalized weighted score matching
(P-WSM), which penalizes interaction blocks and leaves main
effects unpenalized. The main effects can be profiled out
when their Hessian block is nonsingular. We compare P-WSM
with Joint-WSM, which penalizes both types of component
using the same basis, weights, loss, and interaction penalty.
The comparison isolates the effect of nuisance penalization;
it does not imply that leaving main effects unpenalized
always improves recovery.

Our framework builds on several established developments. Log-density
ANOVA connects nonparametric interaction modeling with conditional
independence
\citep{jeon2006effective,gu2013nonparametric,tan2019layer}.
Score matching has been developed for non-Gaussian graphical models and
infinite-dimensional exponential families
\citep{lin2016estimation,sun2015learning,sriperumbudur2017density}.
In particular, \citet{janofsky2018learning} studies regularized score
matching for nonparametric exponential-series graphical models, including
model-selection theory with growing truncation. Nuisance profiling through
a Schur complement also appears in generalized score matching
\citep{yu2019generalized,yu2022generalized}. Here we study the effect of nuisance penalization on edge
selection, rather than profiling as a computational device.

\noindent\textbf{Contributions.}
Our contributions are as follows.

First, we derive the contribution of main-effect shrinkage
to the interaction scores. We then construct a three-variable
model with infinitely many nonzero interaction coefficients
in which P-WSM recovers the graph while Joint-WSM selects
a false edge throughout a common deterministic tuning interval.
Both objectives attain unique minimizers throughout the
interval, so the separation concerns graph selection
rather than existence or uniqueness. 

Second, we prove that population P-WSM excludes edges
between independent subsystems, even when the main-effect
sieve does not represent the marginal log densities exactly.
We also establish growing-dimensional edge recovery under
restricted conditioning and inactive-margin assumptions,
without requiring full-Gram invertibility. Companion
conditions for Joint-WSM describe when joint penalization
also recovers the graph.

Third, for density estimation at fixed dimension, we obtain
weighted Fisher-divergence rates for the unrestricted
quadratic estimator. Weighted Poincar\'e inequalities and
oscillation bounds give Kullback-Leibler (KL) and Hellinger rates for the
exactly normalized density under stronger basis-specific
conditions.

Section~\ref{sec:methodology} introduces the model, estimators, and
computational procedures. Section~\ref{sec:theoretical-results} develops
density-risk bounds and graph-recovery theory.
Sections~\ref{sec:numerical-validation} and~\ref{sec:realdata} present
simulations and an analysis of bone-marrow signaling data.
Section~\ref{sec:conclusion} concludes. Technical proofs and additional
empirical results appear in the Supplementary Material.

\section{Weighted score matching on bounded domains: modeling, estimation, and computation}
\label{sec:methodology}
\label{sec:problem_setup}
Density estimation and graph recovery use the same pairwise
log-density model and weighted score-matching loss.
They differ in regularization: density estimation uses an
unpenalized sieve fit, whereas graph recovery penalizes
interaction blocks while leaving main effects unpenalized.

\subsection{Pairwise interaction model and conditional independence}
\label{sec:pairwise_interaction_model}

Let \(\tb X=(X_1,\ldots,X_p)\) be a continuous random vector on
\(\Omega=\mathcal X^p\), where \(\mathcal X=[\XL,\XR]\) with
\(\XL<\XR\). Let \(p_0\) denote its density, assumed positive almost
everywhere, and suppose \(h_0=\log p_0\in L_2(\Omega)\).
Throughout, function equalities hold almost everywhere with respect
to Lebesgue measure.

Assume that \(h_0\) belongs to the canonically centered pairwise
ANOVA family
\begin{equation}
    \label{eq:pairwise_log_density}
    h_0(\tb x)
    =
    C+\sum_{i=1}^p f_i(x_i)
    +\sum_{1\le i<j\le p}f_{ij}(x_i,x_j),
    \qquad \tb x\in\Omega,
\end{equation}
where \(C\) is determined by normalization,
\(f_i\in L_2(\mathcal X)\), and \(f_{ij}\in L_2(\mathcal X^2)\).
The canonical centering constraints are
\(\int_{\mathcal X}f_i(u)\,du=0\),
\(\int_{\mathcal X}f_{ij}(u,x_j)\,du=0\), and
\(\int_{\mathcal X}f_{ij}(x_i,u)\,du=0\) for every \(i<j\).

These constraints separate marginal structure from genuine pairwise
interaction. To make this distinction precise, let \(A_i h\) denote
averaging \(h\) over coordinate \(i\) under the uniform distribution
on \(\mathcal X\). The commuting averaging operators generate the
orthogonal projections
\(\Pi_S=\prod_{i\in S}(I-A_i)\prod_{i\notin S}A_i\),
\(S\subseteq[p]\), and the functional ANOVA decomposition
\(h=\sum_{S\subseteq[p]}\Pi_S h\) for \(h\in L_2(\Omega)\).
Consequently, \eqref{eq:pairwise_log_density} is precisely the
structural restriction \(\Pi_S h_0=0\) for \(|S|\ge3\), with unique
components \(C=\Pi_\emptyset h_0\),
\(f_i=\Pi_{\{i\}}h_0\), and \(f_{ij}=\Pi_{\{i,j\}}h_0\).
In particular, the interaction components cannot absorb arbitrary
main effects.

Define the undirected graph \(G_0=([p],E_0)\), where
\(E_0=\{\{i,j\}:i<j,\ \|f_{ij}\|_{L_2(\mathcal X^2)}>0\}\).

\begin{proposition}[Pairwise interactions and conditional independence]
    \label{prop:pairwise_ci}
    Suppose \(p_0>0\) almost everywhere on \(\Omega\) and
    satisfies \eqref{eq:pairwise_log_density} with canonical
    centering. For every \(i<j\),
    \(X_i\perp X_j\mid\tb X_{-\{i,j\}}\) if and only if
    \(f_{ij}\equiv0\). Thus \(G_0\) is its
    conditional-independence graph.
\end{proposition}

\begin{proof}
    For almost every fixed \(\tb x_{-\{i,j\}}\), the conditional
    log density of \((X_i,X_j)\) is
    \(C(\tb x_{-\{i,j\}})
    +u_i(x_i,\tb x_{-\{i,j\}})
    +u_j(x_j,\tb x_{-\{i,j\}})
    +f_{ij}(x_i,x_j)\).
    If \(f_{ij}=0\), exponentiation gives conditional factorization.
    Conversely, conditional independence and positivity make this
    conditional log density additive in \(x_i,x_j\).
    Applying the two-coordinate centering operator
    \((I-A_i)(I-A_j)\) to its \(L_2(\mathcal X^2)\) slice annihilates
    the constant and additive terms, leaving \(f_{ij}=0\).
\end{proof}

This is the log-density ANOVA formulation of conditional
density factorization
\citep{dawid1979conditional,jeon2006effective}. 
Our graph-recovery theory assumes correct pairwise specification.
With higher-order log-density interactions, minimizing the population
criterion instead targets a weighted-score approximation within the
fitted pairwise family, when a minimizer exists; its interaction
support need not equal the true conditional-independence graph.
The tensor-product basis and coordinatewise boundary construction
also exploit the product domain \(\Omega\); general domains require
a different boundary treatment \citep{yu2022generalized}.

\subsection{Weighted Fisher divergence}
\label{sec:weighted_fisher_divergence}

For a positive candidate density \(q\), write
\(s_q=\nabla\log q\) and \(s_{q,i}=\partial_i\log q\).
Let \(g=(g_1,\ldots,g_p)^\top\) be a deterministic nonnegative
weight function such that \(g_i\) vanishes on the faces
\(x_i=\XL,\XR\), with regularity as in
Assumption~\ref{ass:1}. Define
\begin{equation}
    \label{def:weighted_fisher_divergence}
    J_g(p_0\|q)
    =
    \mathbb E_0\sum_{i=1}^p
    g_i(\tb X)
    \{s_{p_0,i}(\tb X)-s_{q,i}(\tb X)\}^2,
\end{equation}
where \(\mathbb E_0\) denotes expectation under \(p_0\).
The weights serve two distinct purposes: positivity in the interior
preserves identification, whereas boundary behavior permits
integration by parts. The first property does not require the second.

\begin{lemma}[Identification through weak derivatives]
    \label{lem:identification}
    Let \(U=(\XL,\XR)^p\), and let \(p_0,q\) be normalized densities
    positive almost everywhere on \(U\). Suppose
    \(\log p_0,\log q\in W^{1,1}_{\mathrm{loc}}(U)\), and each \(g_i\)
    is measurable, finite, and strictly positive almost everywhere.
    Interpreting \eqref{def:weighted_fisher_divergence} through weak
    derivatives, \(J_g(p_0\|q)=0\) implies \(q=p_0\) almost everywhere.
\end{lemma}

\begin{proof}
    Positivity of \(p_0g_i\) gives
    \(\nabla(\log q-\log p_0)=0\) almost everywhere.
    A locally Sobolev function with zero weak gradient on connected
    \(U\) is constant almost everywhere. Normalization makes this
    constant zero.
\end{proof}

Thus, weights may vanish at the boundary without compromising
identification. This is a qualitative statement: it does not by
itself bound density error by weighted score error or guarantee
well-conditioned estimation.

For candidate potentials satisfying the conditions of
Supplementary Lemma~\ref{lem:1}, integration by parts eliminates
the unknown score of \(p_0\) \citep{liu2022estimating}:
\begin{equation}
    \label{eq:population_weighted_score_matching_loss}
    J_g(p_0\|q)=C_{0,g}+\mathcal L_g(q),
    \qquad
    \mathcal L_g(q)
    =
    \mathbb E_0\sum_{i=1}^p
    \left[g_i s_{q,i}^2+2\,\partial_i(g_i s_{q,i})\right](\tb X),
\end{equation}
with the relevant terms finite and \(C_{0,g}\) independent of \(q\).
For \(h=\log q\),
\(\partial_i(g_i s_{q,i})
=(\partial_i g_i)(\partial_i h)+g_i\partial_i^2h\).
Consequently, evaluating the loss requires derivatives of the
candidate log density, but neither its normalizing constant nor
the unknown score. This yields a tractable fitting criterion even
when normalizing the pairwise model is costly.

\subsection{Sieve expansions and unrestricted density estimation}
\label{sec:score_matching_density_estimation}

We approximate the ANOVA components by functional sieves.
Density estimation and graph recovery share the same centered basis
and empirical loss; graph recovery additionally uses componentwise
group penalties. Here, ``unrestricted'' means that the finite
coefficient vector is unpenalized and unconstrained, not that the
pairwise structural restriction is removed. 
Given a univariate basis \(\{\ell_r:r\ge1\}\) on \(\mathcal X\), set
\(L_r(x)=\ell_r(x)-|\mathcal X|^{-1}
\int_{\mathcal X}\ell_r(u)\,du\).
The functions \(L_r(x_i)\) and \(L_r(x_i)L_s(x_j)\) then satisfy
the main-effect and interaction centering constraints.

\begin{assumption}[Sieve expandability]
    \label{ass:expandability}
    There exist constants \(B,\nu_1,\nu_2>1\) such that
    \(f_i(x)=\sum_{r\ge1}\beta_{ir}^*L_r(x)\), with
    \(|\beta_{ir}^*|\le Br^{-\nu_1}\), and
    \(f_{ij}(x,y)=\sum_{r,s\ge1}
    \gamma_{ijrs}^*L_r(x)L_s(y)\), with
    \(|\gamma_{ijrs}^*|\le B(rs)^{-\nu_2}\).
    These expansions hold in the corresponding \(L_2\) spaces.
    The centered basis functions are linearly independent and
    sufficiently smooth for the score-matching derivatives.
\end{assumption}

Let \(\theta^*=\{\beta_{ir}^*,\gamma_{ijrs}^*\}\) denote the full
coefficient sequence, so that \(p_{\theta^*}=p_0\) and
\(s_{\theta^*}=\nabla\log p_0\).
For truncation level \(d_n\to\infty\), let
\(D_n=pd_n+\binom p2d_n^2\), collect the retained coefficients in
\(\theta_n\in\mathbb R^{D_n}\), and let \(\psi_n(\tb x)\) contain
the corresponding centered basis functions. Define
\(\eta(\tb x,\theta_n)=\psi_n(\tb x)^\top\theta_n\),
\(p_\theta(\tb x)=
\exp\{\eta(\tb x,\theta_n)\}/
\int_\Omega\exp\{\eta(\tb u,\theta_n)\}\,d\tb u\),
and \(s_\theta(\tb x)=\nabla_{\tb x}\eta(\tb x,\theta_n)\).
The normalizing constant affects the density but not its score.

Given i.i.d.\ observations \(\tb Z_1,\ldots,\tb Z_n\sim p_0\), write
\(\dot\psi_{n,j}=\partial_j\psi_n\) and
\(\ddot\psi_{n,j}=\partial_j^2\psi_n\).
For \(g\) satisfying Assumption~\ref{ass:1}, the half-scaled
empirical loss is
\begin{equation}
    \label{eq:empirical_wsm_loss}
    \begin{aligned}
        M_n(\theta)
        &:=
        \frac12\widehat{\mathcal L}_g(\theta)
        =\frac1n\sum_{k=1}^n\ell(\theta,\tb Z_k),
        \\
        \ell(\theta,\tb x)
        &:=
        \sum_{j=1}^p
        \left[
        \{(\partial_jg_j)\dot\psi_{n,j}
          +g_j\ddot\psi_{n,j}\}^\top\theta
        +\frac12g_j(\dot\psi_{n,j}^\top\theta)^2
        \right](\tb x).
    \end{aligned}
\end{equation}
Equivalently,
\begin{equation}
    \label{def:loss_function}
    M_n(\theta)
    =
    \frac12\theta^\top\widehat\Gamma_n\theta
    +\widehat\rho_n^\top\theta,
\end{equation}
where
\(\widehat\Gamma_n=n^{-1}\sum_{k=1}^n\sum_{j=1}^p
[g_j\dot\psi_{n,j}\dot\psi_{n,j}^\top](\tb Z_k)\) and
\(\widehat\rho_n=n^{-1}\sum_{k=1}^n\sum_{j=1}^p
[(\partial_jg_j)\dot\psi_{n,j}
+g_j\ddot\psi_{n,j}](\tb Z_k)\).
Thus \(M_n\) is a convex quadratic objective.

The population counterpart clarifies the approximation target.
Let \(\Gamma_n=\mathbb E_0\widehat\Gamma_n\) and
\(\rho_n=\mathbb E_0\widehat\rho_n\).
When these quantities are finite and \(\Gamma_n\succ0\), the
population minimizer is
\(\theta_n^\dagger=-\Gamma_n^{-1}\rho_n\).
Under the integration-by-parts conditions above, it satisfies
\begin{equation}
    \label{eq:sieve_score_projection}
    J_g(p_0\|p_\theta)
    =
    J_g(p_0\|p_{\theta_n^\dagger})
    +(\theta-\theta_n^\dagger)^\top
      \Gamma_n(\theta-\theta_n^\dagger).
\end{equation}
The population minimizer need not equal the coefficientwise
truncation of \(\theta^*\): the score projection uses the
\(p_0g\)-weighted derivative Gram matrix, whereas canonical
centering uses the uniform measure.

For density estimation, define
\begin{equation}
    \label{eq:density-unrestricted}
    u_n=
    \begin{cases}
        -\widehat\Gamma_n^{-1}\widehat\rho_n,
        & \widehat\Gamma_n\succ0,\\
        0, & \text{otherwise}.
    \end{cases}
\end{equation}
On the positive-definite event, \(u_n\) is the unique unrestricted
minimizer; the fallback need not minimize the empirical loss.
Population identification does not ensure empirical invertibility:
\(\widehat\Gamma_n\) is a sum of at most \(np\) rank-one matrices,
so it is necessarily singular when \(D_n>np\).
The unrestricted estimator therefore serves as the full-rank
density-estimation procedure, whereas the regularized graph
estimator below addresses sparse interaction selection.

The decay conditions in Assumption~\ref{ass:expandability} constrain
the truth, not the fitted coefficients.
Theorem~\ref{th:density-unrestricted} studies \(u_n\), which is also
used in the full-rank, ridge-free density experiments.
Its exactly normalized density is
\begin{equation}
    \label{eq:estimated_density}
    \hat p(\tb x)=p_{u_n}(\tb x)=
    \frac{\exp\{\eta(\tb x,u_n)\}}
    {\int_\Omega\exp\{\eta(\tb u,u_n)\}\,d\tb u}.
\end{equation}
The denominator is unnecessary for fitting or identifying interaction
blocks, but must be evaluated or approximated separately for density
evaluation.

\subsection{Graph recovery: P-WSM and the Joint-WSM benchmark}
\label{sec:penalized_sparse_structure_recovery}

By Proposition~\ref{prop:pairwise_ci}, graph recovery requires selecting
whole interaction components rather than individual coefficients.
Let
\(\mathcal B=\{1,\ldots,p\}\cup\{(j,k):1\le j<k\le p\}\)
index the main-effect and interaction blocks.
For \(G=j\), set \(\theta_G=\beta_j\in\mathbb R^{d_n}\) and
\(q_G=d_n\); for \(G=(j,k)\), set
\(\theta_G=\gamma_{jk}\in\mathbb R^{d_n^2}\) and \(q_G=d_n^2\).
Writing \(f_{G,\theta_G}=\psi_G^\top\theta_G\), define
\begin{equation}
    \label{eq:block_score_gram}
    \Phi_G=
    \frac1n\sum_{i=1}^n\sum_{\ell=1}^p
    g_\ell(\tb Z_i)
    \partial_\ell\psi_G(\tb Z_i)
    \partial_\ell\psi_G(\tb Z_i)^\top,
    \qquad
    \|\theta_G\|_{\Phi_G}
    =(\theta_G^\top\Phi_G\theta_G)^{1/2}.
\end{equation}
Only coordinates on which \(\psi_G\) depends contribute.
The squared penalty equals the component's empirical weighted
derivative energy,
\(n^{-1}\sum_{i,\ell}g_\ell(\tb Z_i)
\{\partial_\ell f_{G,\theta_G}(\tb Z_i)\}^2\).
It is a seminorm in general and a norm when \(\Phi_G\succ0\).
Under the stated smoothness, positivity, and centering assumptions,
its population counterpart vanishes exactly when the component
is zero.

Write \(\theta=(\beta^\top,\gamma^\top)^\top\), with \(\beta\)
stacking all main effects and \(\gamma\) stacking the interaction
blocks indexed by \(\mathcal E=\{(j,k):j<k\}\).
Partition
\(\widehat\Gamma_n=
\left(\begin{smallmatrix}
\widehat A&\widehat B\\
\widehat B^\top&\widehat C
\end{smallmatrix}\right)\)
and
\(\widehat\rho_n=
\left(\begin{smallmatrix}\widehat a\\\widehat b\end{smallmatrix}\right)\).
Then
\(M_n(\beta,\gamma)
=\frac12\beta^\top\widehat A\beta
+\beta^\top\widehat B\gamma
+\frac12\gamma^\top\widehat C\gamma
+\widehat a^\top\beta+\widehat b^\top\gamma\).
Distinct main effects have disjoint derivative coordinates, so
\(\widehat A=\operatorname{diag}(\widehat A_1,\ldots,\widehat A_p)\).
However, canonical ANOVA centering does not imply
\(\widehat B=0\): main effects and interactions remain coupled
in the weighted score geometry.

Because graph edges depend only on interactions, partially
penalized weighted score matching (P-WSM) leaves the nuisance
main effects unpenalized:
\begin{equation}
    \label{def:pwsm}
    (\widehat\beta^{\rm P},\widehat\gamma^{\rm P})
    \in\operatorname*{arg\,min}_{\beta,\gamma}
    \left\{M_n(\beta,\gamma)
    +\lambda_n\sum_{e\in\mathcal E}\|\gamma_e\|_{\Phi_e}\right\},
    \qquad \Phi_e=\widehat C_{ee}.
\end{equation}
The estimated edge set is
\(\widehat E^{\rm P}=\{e:\widehat\gamma_e^{\rm P}\ne0\}\).
When \(\widehat A\succ0\), the nuisance minimizer is
\(\widehat\beta(\gamma)
=-\widehat A^{-1}(\widehat a+\widehat B\gamma)\).
Substitution gives, up to an additive constant,
\begin{equation}
    \label{eq:pwsm-profile}
    \frac12\gamma^\top\widehat S\gamma
    +\widehat q_{\!P}^\top\gamma
    +\lambda_n\sum_{e\in\mathcal E}\|\gamma_e\|_{\Phi_e},
\end{equation}
where
\(\widehat S=\widehat C-\widehat B^\top\widehat A^{-1}\widehat B\)
and
\(\widehat q_{\!P}
=\widehat b-\widehat B^\top\widehat A^{-1}\widehat a\).
Thus profiling changes the quadratic curvature but retains the
raw penalty blocks \(\Phi_e=\widehat C_{ee}\).
Recover
\(\widehat\beta^{\rm P}
=\widehat\beta(\widehat\gamma^{\rm P})\).

The benchmark \emph{Joint-WSM} instead penalizes all ANOVA blocks:
\begin{equation}
    \label{def:loss_function_with_penalization}
    \hat\theta_n^{\rm J}
    \in\operatorname*{arg\,min}_{\theta\in\mathbb R^{D_n}}
    \left\{M_n(\theta)
    +\lambda_n\sum_{G\in\mathcal B}\|\theta_G\|_{\Phi_G}\right\}.
\end{equation}
Its edge estimate is
\(\widehat E^{\rm J}=\{e:\widehat\gamma_e^{\rm J}\ne0\}\).
The distinction is explicit in the optimality conditions.
Let \(z_\beta\) and \(z_\gamma\) denote the main-effect and
interaction penalty subgradients at a Joint-WSM minimizer.
For \(\widehat A\succ0\), eliminating the main effects yields
\begin{equation}
    \label{eq:joint_wsm_leakage}
    \widehat S\widehat\gamma^{\rm J}
    +\widehat q_{\!P}
    +\lambda_n z_\gamma
    -\lambda_n\widehat B^\top\widehat A^{-1}z_\beta=0.
\end{equation}
The last term is absent from the P-WSM equations.
It is the contribution of nuisance shrinkage to interaction
selection and can remain consequential relative to the
interaction penalty as \(\lambda_n\to0\).
Schur-complement profiling has been used in generalized
score matching \citep{yu2019generalized,yu2022generalized};
Section~\ref{sec:pwsm-theory} studies its implications
for edge selection here.

For both estimators, we use the minimum-Euclidean-norm minimizer
when all penalty blocks are positive definite and the minimum
is attained; otherwise we record a failure, not an empty graph.
This selection convention is included in the recovery events;
see Lemma~\ref{iface-lem:selector} in the Supplementary Material.

\subsection{Computation}
\label{sec:alg}

We fit both estimators along decreasing penalty paths with warm
starts. For Joint-WSM, set
\(H_G=\Phi_G+\varepsilon I_{q_G}\), \(\varepsilon\ge0\), and minimize
\begin{equation}
    \label{eqn:stabilized-objective}
    F_{n,\varepsilon}(\theta)
    =
    M_n(\theta)+\frac{\varepsilon}{2}\|\theta\|_2^2
    +\lambda\sum_{G\in\mathcal B}\|\theta_G\|_{H_G}.
\end{equation}
We take \(\varepsilon=0\) when the block Gram matrices are positive
definite, subject to the existence convention above.
Positive \(\varepsilon\) gives a unique stabilized fit but changes
both the loss and penalty; it is distinguished from the
unstabilized estimator covered by the recovery theory.

Let
\(Q_{n,\varepsilon}=M_n+\varepsilon\|\theta\|_2^2/2\).
Its \(G\)th diagonal Hessian block is \(H_G\), so block minimization
reduces to a linear solve followed by group thresholding, as in
Algorithm~\ref{alg:1}.

\begin{algorithm}[!htbp]
    \caption{Cyclic block-coordinate descent for \(F_{n,\varepsilon}\)}
    \label{alg:1}
    \begin{algorithmic}[1]
        \Require \(\lambda\), \(\varepsilon\ge0\), tolerance \(\tau>0\)
        \State Initialize \(\hat\theta\); factorize
        \(H_G=\Phi_G+\varepsilon I_{q_G}\) for each \(G\)
        \Repeat
            \State \(F^{\rm old}\gets F_{n,\varepsilon}(\hat\theta)\)
            \For{\(G\in\mathcal B\)}
                \State Set
                \(r_G=\nabla_G Q_{n,\varepsilon}(\hat\theta)
                -H_G\hat\theta_G\); solve \(H_Gu_G=r_G\);
                set \(a_G=(r_G^\top u_G)^{1/2}\)
                \State Set \(\hat\theta_G=0\) if \(a_G\le\lambda\);
                otherwise set
                \(\hat\theta_G=-(1-\lambda/a_G)u_G\)
            \EndFor
        \Until{\(\left|F_{n,\varepsilon}(\hat\theta)-F^{\rm old}\right|
        /\{1+|F^{\rm old}|\}\le\tau\)}
        \Ensure Final iterate \(\hat\theta_{n,\varepsilon}\)
    \end{algorithmic}
\end{algorithm}

For unstabilized P-WSM with \(\widehat A\succ0\), each sweep begins
with
\(\beta\leftarrow-\widehat A^{-1}
(\widehat a+\widehat B\gamma)\),
followed by interaction updates with \(\beta\) fixed and
\(H_e=\Phi_e=\widehat C_{ee}\).
This is block-coordinate descent on \eqref{def:pwsm}, whose
minimizers correspond to those of \eqref{eq:pwsm-profile};
it is not direct coordinate descent on the profiled quadratic,
whose diagonal blocks are \(\widehat S_{ee}\).

For Joint-WSM, the all-zero vector minimizes
\(F_{n,\varepsilon}\) whenever
\(\lambda\ge\lambda_{\max}
=\max_G\|\nabla_G M_n(0)\|_{H_G^{-1}}\).
For P-WSM, assuming \(\widehat A\succ0\) and every \(\Phi_e\succ0\),
the corresponding zero-edge endpoint is
\(\lambda_{\max}^{\rm P}
=\max_e\|\widehat q_{\!P,e}\|_{\Phi_e^{-1}}\).
At equality, a minimizer has \(\gamma=0\) and
\(\beta=-\widehat A^{-1}\widehat a\); above it, every minimizer
has zero interactions.

Numerical supports use
\(\widehat S_1(\lambda;\tau_n)
=\{j:\|\widehat\beta_j(\lambda)\|_{H_j}>\tau_n\}\) and
\(\widehat S_2(\lambda;\tau_n)
=\{e:\|\widehat\gamma_e(\lambda)\|_{H_e}>\tau_n\}\),
with \(\tau_n=10^{-10}\) and \(H_G=\Phi_G\) for unstabilized
P-WSM. The theorems concern exact nonzero blocks; thresholded
recovery additionally requires separation of estimation error,
threshold, and minimum active signal.

We select \(\lambda\) by five-fold cross-validation after an
unpenalized refit on each training fold's selected support.
Joint-WSM refits its selected main effects and interactions;
P-WSM retains all main effects and refits selected interactions.
The primary rule minimizes the average raw validation score,
\(\operatorname{CV}_{\mathrm{raw},k}(\lambda)
=M_{\mathrm{val},k}
\{\widehat\theta_{-k}^{\mathrm{refit}}(\lambda)\}\).
This targets weighted score fit, not graph recovery directly.

As sensitivity analyses, we use
\(\operatorname{CV}_{\mathrm{adj},k}
=\operatorname{CV}_{\mathrm{raw},k}
+\{\log n_{\mathrm{tr},k}+\log D_n\}
\widehat q_k(\lambda)/(2n_{\mathrm{tr},k})\)
and apply the one-standard-error rule to both criteria,
choosing the largest penalty within one estimated standard error
of the minimum.
Here
\(\widehat q_{{\rm J},k}
=d_n|\widehat S_{1,-k}|+d_n^2|\widehat S_{2,-k}|\), whereas
\(\widehat q_{{\rm P},k}
=pd_n+d_n^2|\widehat E^{\rm P}_{-k}|\).
The adjusted score is a dimension-based sensitivity criterion,
not a likelihood information criterion.
For \(\widehat A\succ0\), a unique unstabilized P-WSM refit is
obtained when
\(\widehat S_{\widehat E^{\rm P},\widehat E^{\rm P}}\succ0\);
failed fits or refits are not interpreted as empty graphs.

A dense implementation with incrementally updated gradients costs
\(O(D_n^2)\) operations per sweep and stores \(O(D_n^2)\) Hessian
entries. One-time block factorizations cost
\(O\{pd_n^3+\binom p2d_n^6\}\).

\section{Theoretical results}
\label{sec:theoretical-results}

We first study density estimation with fixed \(p\) and
increasing \(d_n\), then graph recovery with both dimensions
allowed to grow. For fixed \(p\ge2\),
\(D_n=pd_n+\binom p2d_n^2\asymp d_n^2\);
for \(p=1\), \(D_n\asymp d_n\).
The following conditions control the basis derivatives
and boundary terms.

\begin{assumption}[Score-matching regularity]
    \label{ass:1}
    Assume the following conditions.
    \begin{enumerate}
        \item The centered basis functions satisfy
        \(\sup_{k\ge1}\|L_k\|_\infty<\infty\), and, for some
        \(\nu_4,\nu_5\ge0\),
        \(\|L_k'\|_\infty\lesssim k^{\nu_4}\) and
        \(\|L_k''\|_\infty\lesssim k^{\nu_5}\).

        \item The coefficient-decay exponents in
        Assumption~\ref{ass:expandability} satisfy
        \(\min(\nu_1,\nu_2)>\max(\nu_4,\nu_5)+1\).

        \item The weights are coordinatewise separable,
        \(g(\tb x)=(g_1(x_1),\ldots,g_p(x_p))\).
        Each \(g_j\) is absolutely continuous on
        \(\mathcal X=[\XL,\XR]\), vanishes at both endpoints, and
        satisfies \(|g_j'|^2\le g_j\le1\) almost everywhere.
    \end{enumerate}
\end{assumption}

Condition (2) makes coefficient decay dominate derivative growth,
ensuring summability of differentiated expansions and truncation
tails. Condition (3) controls the boundary and derivative terms
introduced by integration by parts.

\subsection{Density estimation}
\label{sec:density-estimation-theory}

We first bound the weighted Fisher risk of the unrestricted
estimator \eqref{eq:density-unrestricted}, and then transfer the
bound to KL and Hellinger loss for the exactly normalized density
\eqref{eq:estimated_density}.

\begin{theorem}[Unrestricted density risk for trigonometric and monomial bases]
    \label{th:density-unrestricted}
    Fix \(p\), take \(\mathcal X=[-\pi,\pi]\) and
    \(g_j(x)=(1-|x|/\pi)^2\), and use either
    \(L_{2r-1}(x)=\sin(rx)\), \(L_{2r}(x)=\cos(rx)\), or
    \(L_r(x)=(x/\pi)^r-\mu_r\), where \(\mu_r=0\) for odd \(r\)
    and \(\mu_r=1/(r+1)\) for even \(r\).
    Suppose the observations are i.i.d.\ from the normalized true
    pairwise expansion in Assumption~\ref{ass:expandability}, with
    \(\nu_1,\nu_2>3\). If \(d_n^4\log(eD_n)=o(n)\), then
    \begin{equation}
        \label{eq:density-unrestricted-rate}
        J_g(p_0\|p_{u_n})
        =
        O_p\!\left(
        \frac{d_n^2D_n}{n}
        +d_n^{-2(\nu_1-2)}
        +d_n^{-2(\nu_2-2)}
        \right).
    \end{equation}
    In particular, \(D_n=O(d_n^2)\) at fixed \(p\), and
    \(d_n\asymp n^q\), \(0<q<1/4\), is sufficient for either basis.
\end{theorem}

The bound separates stochastic error from main-effect and
interaction approximation errors. The proof combines relative
Gram concentration with covariance domination; for monomials,
a weighted polynomial derivative inequality gives the same
\(d_n^2D_n/n\) stochastic order. Thus the functional-risk bound
does not require a uniform lower bound on the smallest eigenvalue
of the unscaled coefficient Gram matrix.
Section~\ref{supp:density-inverse-closure} of the Supplementary
Material gives the proof and a normal-equation residual bound
for approximate solutions.

\begin{corollary}[Exactly normalized density loss]
    \label{cor:density-normalization}
    Under Theorem~\ref{th:density-unrestricted}, let
    \(d_n\asymp n^q\). For the trigonometric basis, suppose
    \(0<q\le1/6\). For the centered monomial basis, instead suppose
    \(\min(\nu_1,\nu_2)\ge4\) and \(0<q<1/6\).
    With \(H^2(p_0,p)=\int(\sqrt{p_0}-\sqrt p)^2\),
    \begin{equation}
        \label{eq:density-normalized-risk}
        \mathrm{KL}(p_0\|p_{u_n})+H^2(p_0,p_{u_n})
        =
        O_p\!\left(
        \frac{d_n^4}{n}
        +d_n^{-2(\nu_1-2)}
        +d_n^{-2(\nu_2-2)}
        \right).
    \end{equation}
\end{corollary}

The transfer to density loss is not a consequence of score
identification alone. It combines a weighted Poincar\'e bound
for the log-density error with oscillation control.
For monomials, the latter additionally uses a deterministic
norming grid, concentration of the leading random log-sieve
term, and an eventwise sample-inverse bound.

For \(p\ge2\), write \(\nu=\min(\nu_1,\nu_2)\).
Balancing \(d_n^4/n\) against \(d_n^{-2(\nu-2)}\) gives
\(d_n\asymp n^{1/(2\nu)}\) and the rate
\(O_p\{n^{-(\nu-2)/\nu}\}\).
This choice lies within the normalized-loss regime for
trigonometric bases when \(\nu>3\), and for monomials under
the additional condition \(\nu\ge4\).
This is a specialization of the upper bounds, not a claim of
minimax optimality. The density experiments instead use
\(q=1/6\) for trigonometric functions and \(q=1/7\) for monomials,
both within the stated regimes.

\subsection{Population preservation and edge-only recovery for P-WSM}
\label{sec:pwsm-theory}

We now allow \(p=p_n\) and \(d_n\) to increase.
Our analysis separates three questions: how nuisance shrinkage
changes population edge scores, whether profiling preserves
independent subsystems under approximation, and when empirical
P-WSM recovers the exact edge set.

\begin{proposition}[Main-effect shrinkage propagates to edge scores]
    \label{prop:pwsm-propagation-main}
    Suppose the finite-dimensional population model is correctly
    specified and the integration-by-parts identity holds.
    Partition its quadratic Hessian as
    \(\Gamma=
    \left(\begin{smallmatrix}A&B\\B^\top&C\end{smallmatrix}\right)\),
    with \(A\succ0\), and set \(S=C-B^\top A^{-1}B\).
    At a point satisfying main-effect stationarity with penalty
    level \(\lambda_\beta\) and subgradient \(z_\beta\),
    \begin{equation}
        \label{eq:population-main-effect-leakage}
        \nabla_\gamma M(\beta,\gamma)
        =
        S(\gamma-\gamma^*)
        -\lambda_\beta B^\top A^{-1}z_\beta.
    \end{equation}
    P-WSM removes the direct nuisance-shrinkage term by taking
    \(\lambda_\beta=0\).
\end{proposition}

The nuisance-shrinkage term
\(-\lambda_\beta B^\top A^{-1}z_\beta\)
is present at the population level. It vanishes for P-WSM,
or when \(B^\top A^{-1}z_\beta=0\). 
Partial penalization still leaves interaction shrinkage,
approximation error, and sampling error.

The next result goes beyond correct finite-dimensional
specification. It shows that nuisance profiling preserves
independence even when the chosen main-effect sieve cannot
represent the marginal log densities.

\begin{proposition}[Product and subsystem preservation]
    \label{prop:pwsm-preservation-main}
    Let \(L(x)=(L_1(x),\ldots,L_{d_n}(x))^\top\), with main-effect
    and interaction features \(L(x_i)\) and
    \(L(x_i)\otimes L(x_j)\).
    Suppose \(g_i=g_i(x_i)\ge0\), all population loss terms and
    moments below are finite, and
    \(A_i=\mathbb E\{g_i(X_i)L'(X_i)L'(X_i)^\top\}\succ0\)
    for every \(i\).
    Main-effect coefficients are unrestricted and unpenalized.
    Fix \(\lambda>0\), and suppose \(\Phi_e\succ0\) for every edge.

    If \(P=\bigotimes_{i=1}^pP_i\), define
    \(\mu_i=\mathbb E\{L(X_i)\}\) and
    \(V_i=\operatorname{Cov}\{L(X_i)\}\).
    The profiled population loss satisfies
    \begin{equation}
        \label{eq:pwsm-product-preservation}
        q_{\!P}=0,\qquad
        S_{ij,ij}=A_i\otimes V_j+V_i\otimes A_j,\qquad
        S_{e,f}=0\quad(e\ne f).
    \end{equation}
    Consequently, population P-WSM has the unique minimizer
    \(\gamma=0\), \(\beta=-A^{-1}a\), without requiring the marginal
    log densities to belong to the main-effect sieve.

    More generally, suppose the coordinates partition into
    mutually independent subvectors, and the population problem
    with all cross-subsystem interactions fixed at zero has a
    minimizer. Every such minimizer also minimizes the unrestricted
    problem, and every unrestricted minimizer has zero
    cross-subsystem interactions. If \(\Gamma\succ0\), the
    minimizer is unique.
\end{proposition}

The covariance factors in
\eqref{eq:pwsm-product-preservation} describe the role of nuisance
profiling: although the basis is centered under Lebesgue measure,
profiling removes the marginal feature means
\(\mu_i\), which need not be zero under \(P_i\).
Thus limited marginal resolution does not itself force
population P-WSM to introduce edges between independent
subsystems. This is a preservation result, not a claim that
every edge within a dependent subsystem is recovered under
misspecification.

Finite-sample recovery additionally requires stable profiling,
sufficient active signal, and a strict inactive optimality
margin. The following conditions quantify these requirements
without assuming invertibility of the full Gram matrix.

\begin{assumption}[Sufficient conditions for P-WSM edge recovery]
    \label{ass:pwsm-edge}
    For each \(n\), let \(p\ge2\), \(d\ge1\),
    \(\mathcal E=\{(i,j):1\le i<j\le p\}\), and
    \(D=pd+|\mathcal E|d^2\).
    All bounds below may depend on \(n\).

    \textnormal{(i) Model and approximation.}
    The observations are independent with common positive density
    \(p_0\) on \(\mathcal X^p=[\XL,\XR]^p\) and differentiable
    log density. Its canonically centered pairwise log density
    has edge set \(E\subseteq\mathcal E\).
    The feature vector \(\psi\in\mathbb R^D\) contains \(d\)
    main-effect features per node and their \(d^2\) tensor
    products per edge.
    Fix a deterministic sieve coefficient
    \(\theta_d=(\beta_d,\gamma_d)\) with
    \(\{e:\gamma_{d,e}\ne0\}=E\), and write
    \(s_d=\nabla(\psi^\top\theta_d)\),
    \(s_0=\nabla\log p_0\).
    The weights are coordinatewise, nonnegative, and
    boundary-vanishing, and integration by parts holds for every
    feature. Finite bounds satisfy
    \(0\le g_i\le G\), \(|g_i'|\le G_1\),
    \(|\partial_i\psi_j|\le U\),
    \(|\partial_i^2\psi_j|\le V\),
    \(\max_i\|s_{d,i}\|_\infty\le T\), and
    \(\max_i\|s_{d,i}-s_{0,i}\|_\infty\le H\).

    \textnormal{(ii) Population matrices and perturbation bounds.}
    Partition
    \(\Gamma=\mathbb E\sum_i
    g_i\partial_i\psi\,\partial_i\psi^\top
    =\left(\begin{smallmatrix}A&B\\B^\top&C\end{smallmatrix}\right)\),
    with \(A=\operatorname{diag}(A_1,\ldots,A_p)\), and set
    \(S=C-B^\top A^{-1}B\).
    Define
    \(a_0=\min_i\lambda_{\min}(A_i)>0\),
    \(B_0=\max_e\max_{i\in e}\|B_{i,e}\|_2\),
    \(\phi_0=\min_e\lambda_{\min}(C_{ee})>0\), and
    \(M_0=\max_e\lambda_{\max}(C_{ee})\), where \(B_{i,e}\)
    is the main-block-\(i\)/edge-block-\(e\) submatrix.
    Choose \(\delta_n\in(0,1)\) with \(\delta_n\to0\), and set
    \begin{align*}
        t_\Gamma
        &=2GU^2\sqrt{\frac{2\log(4D^2/\delta_n)}{n}},\\
        t_w
        &=2(GUT+G_1U+GV)
          \sqrt{\frac{2\log(4D/\delta_n)}{n}}+2GUH.
    \end{align*}
    Let
    \(e_A=dt_\Gamma\), \(e_B=d^{3/2}t_\Gamma\),
    \(e_\Phi=d^2t_\Gamma\),
    \(\underline a=a_0-e_A>0\),
    \(\underline\phi=\phi_0-e_\Phi>0\), and
    \(\overline M=M_0+e_\Phi\).
    Define
    \[
        e_S=d^2t_\Gamma+
        2\left\{
        \frac{e_B(2B_0+e_B)}{\underline a}
        +\frac{B_0^2e_A}{a_0\underline a}
        \right\},
        \qquad
        R_n=\left\{
        d+\frac{2(B_0+e_B)\sqrt d}{\underline a}
        \right\}t_w.
    \]

    \textnormal{(iii) Signal, restricted conditioning, and inactive margin.}
    The penalty \(\lambda_n>0\) is deterministic.
    For \(s=|E|>0\), let
    \(\rho_d=\min_{e\in E}\|\gamma_{d,e}\|_2\),
    \(\kappa_0=\lambda_{\min}(S_{EE})\), and require
    \(\underline\kappa=\kappa_0-se_S>0\) and
    \[
        \rho_d>
        \frac{\sqrt{s}}{\underline\kappa}
        \left(R_n+\lambda_n\sqrt{\overline M}\right).
    \]
    If \(0<s<|\mathcal E|\), also define
    \(\mu_0=\max_{e\notin E}\|S_{eE}\|_2\),
    \(\overline\mu=\mu_0+\sqrt{s}e_S\), and require
    \[
        \frac{\overline\mu\sqrt{s\overline M}}
        {\underline\kappa\sqrt{\underline\phi}}
        +
        \frac{R_n}{\lambda_n\sqrt{\underline\phi}}
        \left(1+
        \frac{\overline\mu\sqrt{s}}{\underline\kappa}\right)
        <1.
    \]
    For a full graph, omit the inactive condition.
    For an empty graph, replace all active quantities and
    conditions in (iii) by
    \(\lambda_n>R_n/\sqrt{\underline\phi}\).
\end{assumption}

\begin{theorem}[P-WSM edge recovery with growing dimension and sieve]
    \label{th:pwsm-main}
    Under Assumption~\ref{ass:pwsm-edge}, the unstabilized P-WSM
    estimator exists uniquely and satisfies
    \(\widehat E^{\rm P}=E\) with probability at least
    \(1-\delta_n\to1\).
\end{theorem}

For a nonempty, nonfull graph, the conditions reveal an explicit
penalty window. Define
\(\eta_n=
\overline\mu\sqrt{s\overline M}/
(\underline\kappa\sqrt{\underline\phi})\).
When \(\eta_n<1\), the inactive-margin and signal requirements
are equivalent to
\begin{equation}
    \label{eq:pwsm-penalty-window}
    \underbrace{
    \frac{R_n}
    {\sqrt{\underline\phi}(1-\eta_n)}
    \left(1+
    \frac{\overline\mu\sqrt{s}}{\underline\kappa}\right)
    }_{\lambda_{-,n}}
    <\lambda_n<
    \underbrace{
    \frac{\underline\kappa\rho_d/\sqrt{s}-R_n}
    {\sqrt{\overline M}}
    }_{\lambda_{+,n}}.
\end{equation}
The lower endpoint suppresses inactive edges; the upper endpoint
protects active signals from shrinkage. Both sampling error
and sieve approximation enter through \(R_n\), so increasing
\(n\) alone does not remove a nonvanishing approximation error.
The window closes as the inactive coupling approaches its
margin or the minimum signal becomes too small.

Crucially, the theorem requires positive nuisance blocks,
positive penalty blocks, and restricted curvature on \(E\),
not invertibility of the full empirical Gram matrix.
It therefore separates the requirements for sparse edge recovery
from those for unrestricted density fitting.
The empty-graph case follows from the strict zero-edge dual
feasibility condition in Theorem~\ref{th:pwsm-finite}
of the Supplementary Material.

Proposition~\ref{prop:pwsm-nonvacuity} in the Supplementary
Material verifies a concrete regime in which \(p_n\), \(d_n\),
and \(s_n\) all diverge, with nonvanishing main effects and
shrinking interaction signals. Thus the conditions accommodate
simultaneous growth of ambient dimension, sieve resolution,
and graph size, rather than only a fixed-dimensional limit.

Restricted conditioning should not, however, be confused with
blockwise positivity alone. The following counterexample
identifies the missing existence guarantee.

\begin{proposition}[Positive block Grams do not ensure existence]
    \label{prop:pwsm-existence}
    The conditions \(\widehat A\succ0\) and
    \(\widehat\Phi_e\succ0\) alone do not guarantee existence of
    an empirical P-WSM minimizer at every \(\lambda>0\), even
    with smooth positive interior weights and compatible
    centered polynomial bases.
\end{proposition}

The proof of Proposition~\ref{prop:pwsm-existence} is given in
Section~\ref{supp:pwsm-existence-proof} of the Supplementary Material.
The obstruction concerns the profiled loss, not the separate
diagonal blocks: if
\(\widehat S v=0\) and
\(|\widehat q_{\!P}^\top v|
>\lambda\sum_e\|v_e\|_{\Phi_e}\), the profiled objective is
unbounded below along one direction of \(v\).
Singularity alone does not imply nonexistence; the issue is
whether the linear drift exceeds the penalty along a
zero-curvature direction.

\subsection{Structure recovery for Joint-WSM}
\label{sec:structure-recovery-theory}

We next analyze Joint-WSM
\eqref{def:loss_function_with_penalization}, which penalizes
both main effects and interactions. We establish joint support
recovery under explicit conditions, then show that nuisance
shrinkage can force a false edge throughout a tuning interval
on which P-WSM recovers the graph.

For the block collection \(\mathcal B\), define
\(S_1=\{j:\bs\beta_j^\ast\neq0\}\),
\(S_2=\{(j,k):\bs\gamma_{jk}^\ast\neq0\}\), and
\(S=S_1\cup S_2\neq\varnothing\).
Let \(\tilde\theta_G^\ast\) collect the first \(d_n\)
main-effect coefficients when \(G=j\), or the first
\(d_n\times d_n\) interaction coefficients when \(G=(j,k)\).
Write
\(S_n=\{G:\tilde\theta_G^\ast\neq0\}\) and
\(\rho_n^\ast=\min_{G\in S}\|\tilde\theta_G^\ast\|_\infty\).
Let \(\Sigma_{SS}\) and \(\Sigma^c_{G,S}\) denote the active
block and inactive--active block row of
\(\widehat\Gamma_n=\nabla^2M_n(\theta)\), respectively.
Set
\(\Lambda_{\max}=\lambda_{\max}(\mathbb E\Sigma_{SS})\),
\(\Lambda_{\min}=\lambda_{\min}(\mathbb E\Sigma_{SS})\), and
\(\mu_n=\max_{G\in S^c}\|\mathbb E\Sigma^c_{G,S}\|_2\).
If \(S^c=\varnothing\), set \(\mu_n=0\) and omit all
inactive-group conditions.

To summarize the concentration and approximation bounds, define
\(\tau_n=d_n^{\nu_4}
\{d_n^{-\nu_1+\nu_4+1}+d_n^{-\nu_2+\nu_4+1}\}\) and
\begin{align*}
R_{H,n}
&=|S|d_n^{2+2\nu_4}
  \sqrt{\log(e p d_n)/n},\\
R_{A,n}
&=|S|d_n^{\nu_4\vee\nu_5}
  \sqrt{\log(e|S|d_n)/n}+p\tau_n,\\
R_{I,n}
&=d_n^{\nu_4\vee\nu_5}
  \sqrt{\log(e p d_n)/n}+\tau_n.
\end{align*}
These control active-Hessian concentration, active score error,
and uniform inactive score error, respectively.

\begin{assumption}[Support-recovery regularity]
\label{ass:2}
The following conditions hold.
\begin{enumerate}
\item
\(S_n=S\), \(\rho_n^\ast>0\), and \(\Lambda_{\min}>0\).
For some \(c_\Phi>0\),
\(\mathbb P\{\min_{G\in S^c}\lambda_{\min}(\Phi_G)
\ge c_\Phi\Lambda_{\min}\}\to1\).

\item
\(R_{H,n}\Lambda_{\min}^{-3/2}=O(1)\),
\(R_{H,n}(\Lambda_{\max}/\Lambda_{\min}^3)^{1/2}\to0\), and
\(R_{H,n}\|(\mathbb E\Sigma_{SS})^{-1}\|_\infty\to0\).

\item
\((\lambda_n\sqrt{\Lambda_{\max}}+R_{A,n})
\|(\mathbb E\Sigma_{SS})^{-1}\|_\infty
=o(\rho_n^\ast)\).

\item
\(\displaystyle
\frac{p d_nR_{I,n}}{\lambda_n\sqrt{\Lambda_{\min}}}
\left[1\vee\frac{\sqrt{|S|}\mu_n}{\Lambda_{\min}}\right]
\to0.
\)

\item
For some \(\delta_0\in(0,1)\) and all sufficiently large \(n\),
\(\mu_n(\Lambda_{\max}/\Lambda_{\min}^3)^{1/2}
<(1-\delta_0)\sqrt{c_\Phi/|S|}\).
\end{enumerate}
\end{assumption}

For growing \(p=p_n\), Assumptions~\ref{ass:expandability},
\ref{ass:1}, and \ref{ass:2} hold uniformly in \(n\), with
fixed constants independent of \(p\).
Conditions (1)--(3) retain active blocks; (4)--(5) ensure
strict inactive KKT inequalities. Condition (5) is a sufficient
irrepresentability condition controlling inactive--active
coupling.

\begin{theorem}[Selection consistency]
\label{th:2}
Under Assumptions~\ref{ass:expandability}, \ref{ass:1},
and \ref{ass:2}, with deterministic \(\lambda_n>0\), let
\(\widehat S^{\rm J}
=\{G\in\mathcal B:\hat\theta_G^{\rm J}\neq0\}\).
Then \(\mathbb P(\widehat S^{\rm J}=S)\to1\).
Consequently, canonical identifiability implies
\(\mathbb P(\widehat E^{\rm J}=E)\to1\), where
\(\widehat E^{\rm J}
=\{(j,k):\hat{\bs\gamma}_{jk}^{\rm J}\neq0\}\)
and \(E=S_2\).
\end{theorem}

\begin{corollary}[A transparent scaling regime]
\label{cor:support-transparent}
Let \(\ell_n=\log(e p d_n)\) and
\(\nu_\star=\min(\nu_1,\nu_2)>4\).
Suppose \(S\neq\varnothing\), \(S_n=S\), and
Assumption~\ref{ass:1} holds with \(\nu_4=1,\nu_5=2\).
Assume, for fixed positive constants,
\(c_\Lambda\le\Lambda_{\min}\le\Lambda_{\max}\le C_\Lambda\)
and \(\|(\mathbb E\Sigma_{SS})^{-1}\|_\infty\le C_\Sigma\),
together with Assumption~\ref{ass:2}(1)'s inactive-Gram
condition and Assumption~\ref{ass:2}(5).
Define
\[
r_{H,n}=|S|d_n^4\sqrt{\ell_n/n},\quad
r_{A,n}=|S|d_n^2\sqrt{\ell_n/n}
       +p d_n^{-(\nu_\star-3)},\quad
r_{\lambda,n}=p d_n^3\sqrt{\ell_n/n}
       +p d_n^{-(\nu_\star-4)}.
\]
If \(r_{H,n}\to0\), \(r_{A,n}=o(\rho_n^\ast)\),
\(r_{\lambda,n}=o(\lambda_n)\), and
\(\lambda_n=o(\rho_n^\ast)\), then
\(\mathbb P(\widehat S^{\rm J}=S)\to1\) and
\(\mathbb P(\widehat E^{\rm J}=E)\to1\).

In particular, let
\(d_n\asymp n^\alpha\), \(p\lesssim n^\zeta\),
\(|S|\lesssim n^\sigma\), and
\(\rho_n^\ast\gtrsim n^{-b}\), with
\(\alpha>0\) and \(\zeta,\sigma,b\ge0\).
The preceding conditions follow, up to logarithmic factors, if
\(\sigma+4\alpha<1/2\),
\(b+\sigma+2\alpha<1/2\),
\(b+\zeta+3\alpha<1/2\), and
\(b+\zeta<\alpha(\nu_\star-4)\), with
\(\lambda_n=n^{-c}\) for any
\[
b<c<\min\{1/2-\zeta-3\alpha,\,
                 \alpha(\nu_\star-4)-\zeta\}.
\]
\end{corollary}

The tuning window requires the penalty to dominate inactive
errors while remaining below the minimum signal.
Increasing \(d_n\) reduces approximation bias but raises
stochastic error. The bounded-spectrum specialization above
does not apply to growing, unscaled trigonometric bases;
see Remark~\ref{iface-rem:trig-spectrum} in the
Supplementary Material.

The sufficient incoherence condition is not a necessary
characterization of Joint-WSM recovery. Nevertheless, the
following construction realizes the nuisance-shrinkage
mechanism of Proposition~\ref{prop:pwsm-propagation-main}
as a false edge over a common tuning interval.

\begin{proposition}[Common-interval support separation]
\label{prop:separation-main}
Consider the three-variable non-Gaussian log-density
construction in Section~\ref{supp:comparison} of the
Supplementary Material, with true graph
\(E=\{\{1,2\},\{2,3\}\}\).
Let \(d_n=\lceil\log n\rceil\) and
\(I_n=[n^{-1/4},n^{-1/8}]\).
For all sufficiently large \(n\), on an event of probability
at least \(1-n^{-2}\), the full and raw-block empirical Gram
matrices are strictly positive definite, and simultaneously
for every \(\lambda\in I_n\), the attained unique minimizers
satisfy
\[
\widehat E^{\rm P}(\lambda)=E,
\qquad
\{1,3\}\in\widehat E^{\rm J}(\lambda).
\]
\end{proposition}

Throughout \(I_n\), the Joint-WSM minimizer constrained by
\(\gamma_{13}=0\) violates the missing-edge dual KKT condition,
forcing the false edge. Thus the separation is not caused by
nonexistence, nonuniqueness, or full-Gram singularity.
It holds over an expanding multiplicative tuning range,
\(\sup I_n/\inf I_n=n^{1/8}\), but does not assert failure
of Joint-WSM for every tuning sequence.

	\section{Numerical studies}
\label{sec:numerical-validation}

We examine density estimation, boundary weighting, and graph
recovery. The graph experiments compare P-WSM with Joint-WSM
and four existing methods, then vary main-effect strength,
graph topology, nuisance dimension, and sieve size.

Each graph setting uses 100 independent paired replicates, common data and
folds, and a 31-point relative penalty grid. Tuning minimizes the mean raw
held-out score after unpenalized refitting over five folds. The Supplementary Material
reports alternative tuning rules and the best graph
along each fitted path. Besides Joint-WSM, we compare P-WSM with graphical lasso
\citep{yuan2007model}, \texttt{highscore}
\citep{lin2016estimation}, BEAM \citep{leday2019fast}, and BGGM
\citep{williams2021bayesian} in the benchmark, preservation, and
graph-topology experiments.

On \(( -\pi,\pi )\), the primary weighted score-matching experiments and
the primary Bendall analysis use
\(g_j(x)=\frac{(\pi-|x|)^2}{\pi^2}\), \(j=1,\ldots,p\).
The density-estimation and boundary-weight experiments use 50 repetitions.
	
	\subsection{Bounded-domain density estimation}
	\label{sec:numerics-density}
	\label{5.1}
	
	We evaluate the normalized unrestricted estimator \eqref{eq:density-unrestricted}
	on the multivariate von Mises-type density \citep{mardia2008multivariate}
	\(p(\tb x) = \widetilde C \exp \left\{ \sum_{i=1}^p \sin(x_i) + \frac12 \sum_{i=1}^{p-1} \sin(x_i)\sin(x_{i+1}) \right\} \mathbbm 1\{\tb x\in[-\pi,\pi]^p\}\).
	We consider
	\(
	(p,n)\in\{(4,500),(4,1000),(6,500),(6,1000)\}.
	\)
	Following Corollary~\ref{cor:density-normalization}, we use
	\(
	d=\lceil n^{1/6}\rceil\) 
	{for the trigonometric basis} and 
	\(
	d=\lceil n^{1/7}\rceil\) 
	{for the centered-monomial basis}.
	Importance sampling from an equal mixture of the uniform and product von Mises
	main-effect densities approximates the normalizing constant. We compare the two fits
	(WSM-Tri and WSM-Poly) with bounded kernel density estimation (KDE) and
	truncated-Gaussian maximum likelihood (TGM), using mean negative log-likelihood
	on \(10^4\) independent test observations.
	
	\begin{table}[htbp]
		\centering
		\caption{Density-estimation performance measured by average negative test
			log-likelihood over 50 repetitions: median [first quartile, third
			quartile]. Smaller values are better.}
		\label{tab:0}
		\resizebox{\textwidth}{!}{%
			\begin{tabular}{cccccc}
				\hline
				\(p\) & \(n\) & WSM-Tri & WSM-Poly & KDE & TGM\\
				\hline
				4 & 500  & 6.071 [6.039, 6.110] & 6.138 [6.108, 6.192]
				& 6.344 [6.325, 6.359] & 6.429 [6.415, 6.450]\\
				4 & 1000 & 6.030 [6.012, 6.059] & 6.015 [5.995, 6.043]
				& 6.256 [6.232, 6.264] & 6.446 [6.436, 6.456]\\
				6 & 500  & 9.175 [9.128, 9.220] & 9.400 [9.285, 9.715]
				& 9.615 [9.596, 9.636] & 9.567 [9.550, 9.585]\\
				6 & 1000 & 9.072 [9.045, 9.113] & 8.990 [8.956, 9.033]
				& 9.484 [9.469, 9.496] & 9.635 [9.578, 9.662]\\
				\hline
		\end{tabular}}
	\end{table}
	
	Both WSM variants improve on KDE and TGM in all four settings, particularly
	for \(p=6\).
	
	\subsection{Effect of boundary weighting}
	\label{sec:numerics-boundary}
	\label{5.2}
	
	To isolate boundary weighting, consider
	\(p(x) = \widetilde C \exp\{2x+0.5\sin(2x)\} \mathbbm 1\{x\in[-\pi,\pi]\}\),
	which places substantial mass near the right endpoint of the interval. We
	compare the boundary-vanishing weight
	\(
	g(x)=\frac{(\pi-|x|)^2}{\pi^2}
	\)
	with ordinary unweighted score matching, corresponding to \(g\equiv1\). Both
	{fits use the same centered-monomial basis with \(d=6\) and a
		standardized ridge coefficient \(\eta=10^{-4}\). Specifically, the empirical
		Hessian \(\widehat\Gamma_n\) is replaced by
		\(
		\widehat\Gamma_n+\eta D_{\Gamma,n},
		\) with 
		\(D_{\Gamma,n}
		=
		\operatorname{diag}
		\left\{
		(\widehat\Gamma_n)_{11},\ldots,(\widehat\Gamma_n)_{D_nD_n}
		\right\}.
		\)
		}
	Table~\ref{tab:boundary-ablation} reports average negative test
	log-likelihood and empirical weighted Fisher divergence on an independent test
	sample of size \(20000\).
	
	\begin{table}[htbp]
		\centering
		\caption{Boundary-weight ablation over 50 repetitions: median [first
			quartile, third quartile]. Smaller values are better.}
		\label{tab:boundary-ablation}
		\resizebox{0.92\textwidth}{!}{%
			\begin{tabular}{ccll}
				\hline
				\(n\) & Method & Negative test log-likelihood
				& Weighted Fisher divergence\\
				\hline
				500 & Weighted SM
				& 0.322 [0.316, 0.328] & 0.040 [0.015, 0.119]\\
				500 & Unweighted SM
				& 0.525 [0.504, 0.550] & 0.552 [0.419, 0.842]\\
				1000 & Weighted SM
				& 0.320 [0.312, 0.325] & 0.015 [0.006, 0.057]\\
				1000 & Unweighted SM
				& 0.524 [0.504, 0.536] & 0.545 [0.458, 0.727]\\
				\hline
		\end{tabular}}
	\end{table}
	
	Boundary weighting improves both criteria. Its median weighted Fisher error
	decreases with \(n\), whereas the unweighted fit retains a large error under
	the same weighted target.
	
	{
		\subsection{Graph recovery benchmarks}
		\label{sec:numerics-graph-recovery}
		\label{5.3}
		
		We first evaluate P-WSM on Gaussian and nonlinear models whose interaction graph is a chain,
		\(E=\{(1,2),(2,3),\ldots,(p-1,p)\}\).
		For the Gaussian setting, we draw
		\(\tb X\sim N_p(0,K^{-1})\mid \tb X\in[-\pi,\pi]^p\),
		with \(p=6\), \(n=1500\), unit diagonal entries in \(K\), and \(K_{j,j+1}=K_{j+1,j}=-0.35\). We use the centered-monomial basis with \(d=2\). For the nonlinear setting, we use \(p=8\), \(n=2000\), a trigonometric basis with \(d=2\), and
		\(p(\tb x)\propto \exp\left\{ \sum_{j=1}^p\sin(x_j) +0.5\sum_{(j,k)\in E}\sin(x_j)\sin(x_k) \right\} \mathbbm 1\{\tb x\in[-\pi,\pi]^p\}\).
		
		We report structural Hamming distance (SHD), true positive rate (TPR), false
		discovery rate (FDR), and exact-recovery frequency in
		Table~\ref{tab:pwsm-benchmarks}, with F1 added in the topology comparison.
		
		\begin{table}[!htbp]
			\centering
			\caption{Graph recovery in the Gaussian and nonlinear chain settings. SHD is reported as mean (Monte Carlo standard error); TPR, FDR, and Exact are Monte Carlo means.}
			\label{tab:pwsm-benchmarks}
			\resizebox{0.96\textwidth}{!}{%
				\begin{tabular}{clcccc}
					\hline
					Setting & Method & SHD & TPR & FDR & Exact \\
					\hline
					Gaussian chain & P-WSM & 0.100 (0.041) & 1.000 & 0.015 & 0.930 \\
					& Joint-WSM & 8.170 (0.120) & 1.000 & 0.617 & 0.000 \\
					& Graphical lasso & 1.480 (0.118) & 1.000 & 0.204 & 0.240 \\
					& \texttt{highscore} & 0.700 (0.088) & 1.000 & 0.104 & 0.540 \\
					& BEAM & 0.430 (0.070) & 1.000 & 0.066 & 0.680 \\
					& BGGM & 0.460 (0.073) & 1.000 & 0.070 & 0.660 \\
					\hline
					Nonlinear chain & P-WSM & 1.720 (0.185) & 0.919 & 0.116 & 0.250 \\
					& Joint-WSM & 15.320 (0.207) & 1.000 & 0.683 & 0.000 \\
					& Graphical lasso & 4.550 (0.171) & 0.371 & 0.020 & 0.030 \\
					& \texttt{highscore} & 6.070 (0.175) & 0.141 & 0.010 & 0.010 \\
					& BEAM & 2.280 (0.149) & 0.847 & 0.153 & 0.090 \\
					& BGGM & 2.260 (0.154) & 0.840 & 0.146 & 0.090 \\
					\hline
			\end{tabular}}
		\end{table}
		
		P-WSM attains a mean SHD of 0.10 and an exact-recovery rate of 93\% in the Gaussian setting;
		in the nonlinear setting, its mean SHD, TPR, and FDR are 1.72, 0.919, and 0.116, respectively.
		Joint-WSM selects substantially more false edges in both settings.
		
		\subsection{Effect of penalizing the main effects}
		\label{sec:numerics-main-effects}
		\label{5.4}
		
		We vary only the main-effect amplitude in a nonlinear six-node graph of
		three disjoint interacting pairs, fixing \(n=1800\), \(d=2\), and
		interaction strength \(\rho=0.75\). The amplitudes are
		\(0,0.25,0.5,0.75,1,1.5,\) and \(2\); both methods use the paired design
		and primary tuning rule above.
		
		\begin{figure}[!htbp]
			\centering
			\subfigure[Graph-recovery error.]{\includegraphics[width=0.48\textwidth]{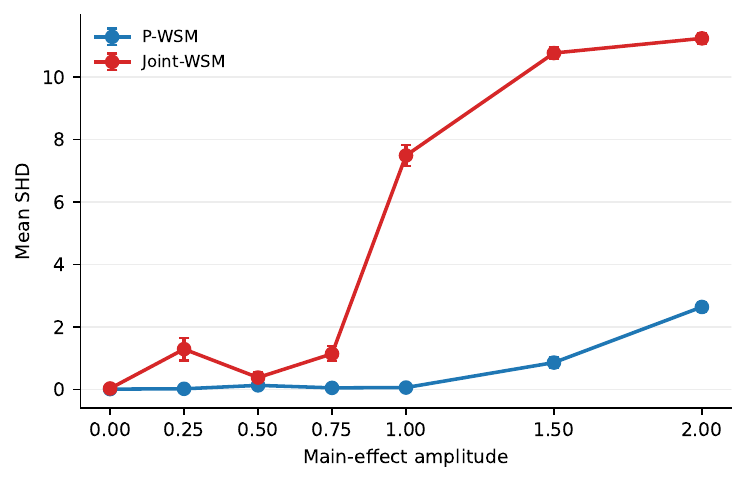}}
			\hfill
			\subfigure[Contribution of main-effect shrinkage to the interaction gradients.]{\includegraphics[width=0.48\textwidth]{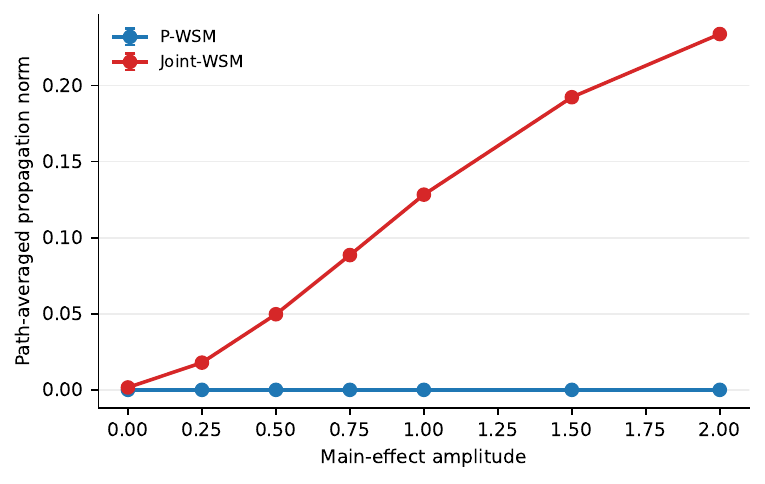}}
			\caption{Effect of increasing the main-effect amplitude. The interaction graph and interaction strength are held fixed across the seven settings.}
			\label{fig:pwsm-main-effect-amplitude}
		\end{figure}
		
		At amplitude \(0\), mean SHD is 0.01 for P-WSM and 0.03 for Joint-WSM.
		The respective values become 0.06 and 7.49 at amplitude \(1\), and 0.86
		and 10.77 at \(1.5\). At amplitude \(2\), P-WSM also deteriorates,
		mainly through missed interactions; Joint-WSM continues to select many false
		edges.
		
		The shrinkage term in Proposition~\ref{prop:pwsm-propagation-main}, averaged
		along the fitted Joint-WSM path, increases from 0.0016 at amplitude \(0\)
		to 0.2339 at \(2\); for P-WSM it is numerically zero. To isolate the
		main-effect update locally, we fix the interactions at the selected
		Joint-WSM fit and minimize over unpenalized main effects. At amplitude
		\(1.5\), the resulting loss-gradient norm is below the penalty level for
		94.2\% of fitted false-positive blocks, supporting the proposed nuisance-shrinkage mechanism. 
		Further diagnostics appear
		in the Supplementary Material.
		
		\subsection{Preservation of independent structure}
		\label{sec:numerics-preservation}
		\label{5.5}
		
		For six independent nonlinear variables, as in
		Proposition~\ref{prop:pwsm-preservation-main}, P-WSM has mean SHD 0.07 and
		empty-graph recovery 95\%, compared with 0.65 and 61\% for Joint-WSM.
		
		For two independent three-node chains, Figure~\ref{fig:pwsm-subsystems}
		separates within-chain errors from false cross-subsystem edges.
		
		\begin{figure}[!htbp]
			\centering
			\includegraphics[width=0.80\textwidth]{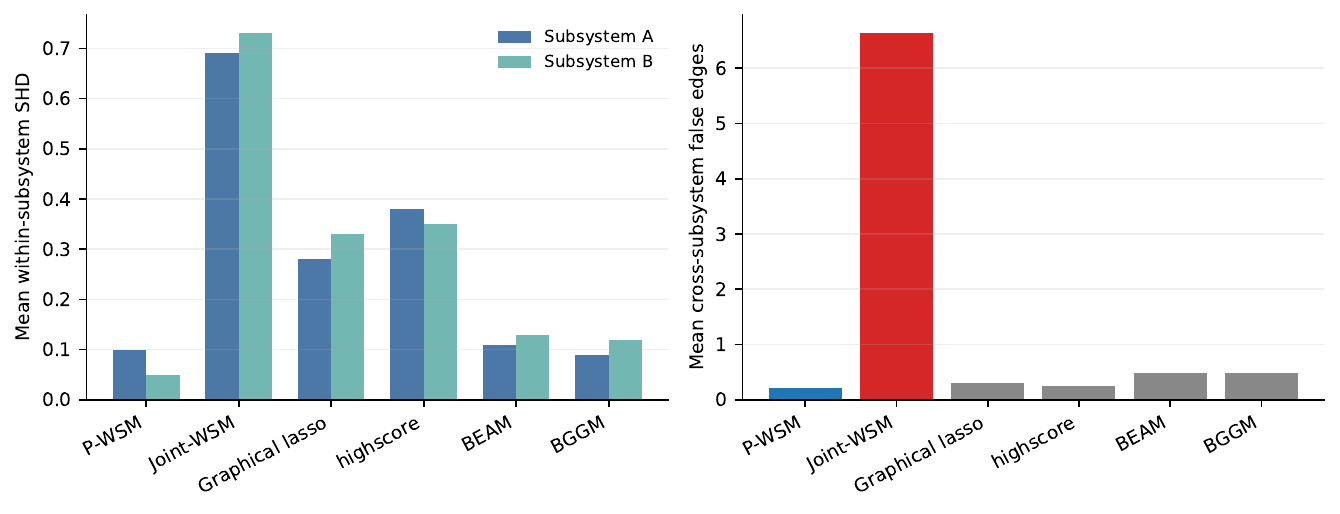}
			\caption{Recovery within two independent three-node subsystems and false edges between them.}
			\label{fig:pwsm-subsystems}
		\end{figure}
		
		P-WSM has within-chain mean SHDs 0.10 and 0.05, and an average of 0.22
		cross-subsystem edges, with at least one in 13\% of replicates. Joint-WSM
		selects 6.63 cross-subsystem edges on average, with at least one in every
		replicate.
		
		\subsection{Robustness and finite-sample behavior}
		\label{sec:numerics-robustness}
		\label{5.6}
		
		We also compare all six methods on an eight-node hub and a fixed eight-node
		sparse-random graph, each with seven edges, under truncated Gaussian and
		nonlinear sine models with \(n=2000\) and \(d=2\)
		(Figure~\ref{fig:pwsm-topologies}).
		
		\begin{figure}[!htbp]
			\centering
			\includegraphics[width=0.82\textwidth]{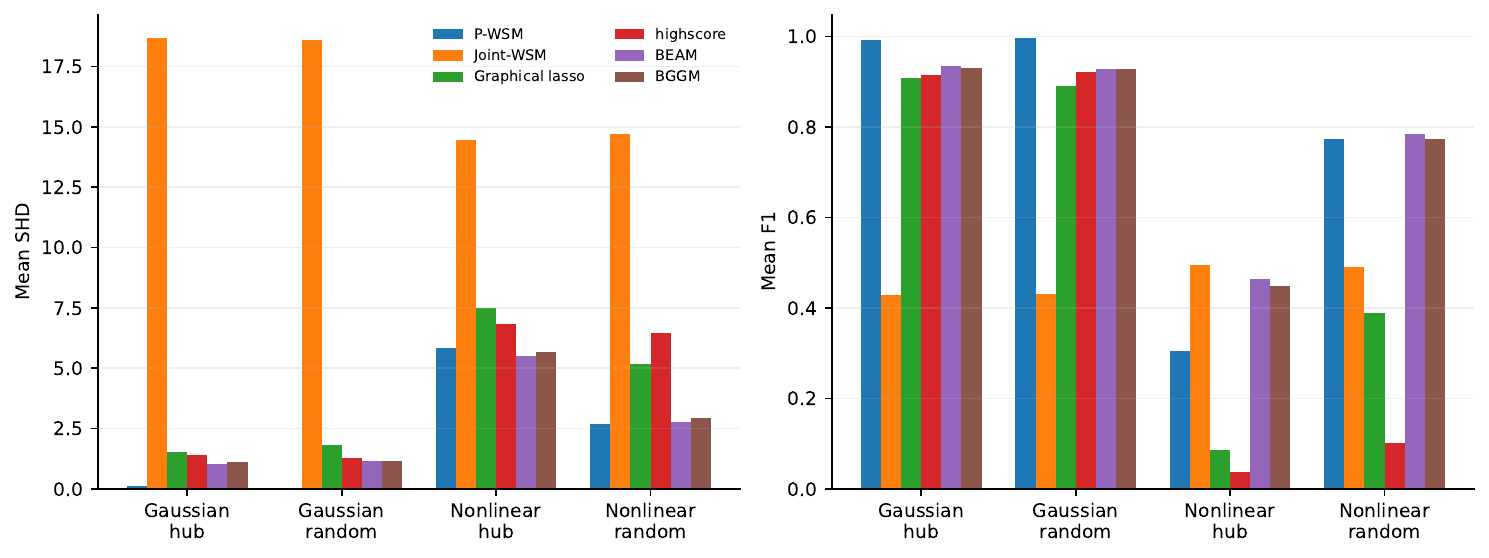}
			\caption{Graph recovery for Gaussian and nonlinear hub and sparse-random graphs. {The panels report mean SHD (left) and mean F1 (right); smaller SHD and larger F1 are better.}}
			\label{fig:pwsm-topologies}
		\end{figure}
		
		P-WSM has mean SHDs 0.12 and 0.04 for the Gaussian hub and sparse-random
		graphs. For the nonlinear hub, its SHD is 5.85, versus 5.52 for BEAM and
		5.66 for BGGM; for the nonlinear sparse-random graph, the values are 2.67,
		2.77, and 2.93, respectively. Its advantage therefore depends on topology.
		
		As \(pd/n\) increases from 0.01 to 0.10, P-WSM exact recovery falls from
		0.76 to 0, and zero-edge-endpoint selection rises from 0 to 0.80. Holding
		the generating model fixed and changing \(d=2,3,4\) gives mean SHDs 1.79,
		6.48, and 6.97. This degradation reflects the combined effects of main-effect estimation,
		interaction estimation, and tuning. In an overfitted-sieve setting, the selected SHD is 6.21 compared to an
		oracle-path minimum of 3.17, illustrating the interplay between model complexity and
		finite-sample tuning. Complete scaling, basis, tuning, and path results appear in the
		Supplementary Material.
	}
	
	\section{A real-data example}
	\label{sec:realdata} 
	
	We analyze the healthy bone-marrow mass-cytometry data of
	\citet{bendall2011single}: the first donor's lipopolysaccharide
	(LPS)-stimulated sample, restricted to CD11b$^{-}$, CD11b$^{\mathrm{hi}}$,
	and CD11b$^{\mathrm{mid}}$ monocytes. The graph summarizes conditional
	associations among intracellular signaling states.

	There are \(p=18\) markers and \(n=19987\) cells. We center the
	\(\operatorname{asinh}\)-transformed measurements by the sample median and
	divide by the median absolute deviation. Graphical lasso and \texttt{highscore} use these standardized
	data; WSM additionally applies a logistic transformation and linear rescaling
	to \((-\pi,\pi)\). Fixed coordinatewise one-to-one transformations preserve
	population conditional independence. Each validation fold uses training-fold
	transformations, and the fitted finite-sieve graph describes the pooled cells.
	
\begin{figure}[!ht]
		\centering
		\includegraphics[width=0.8\textwidth]
		{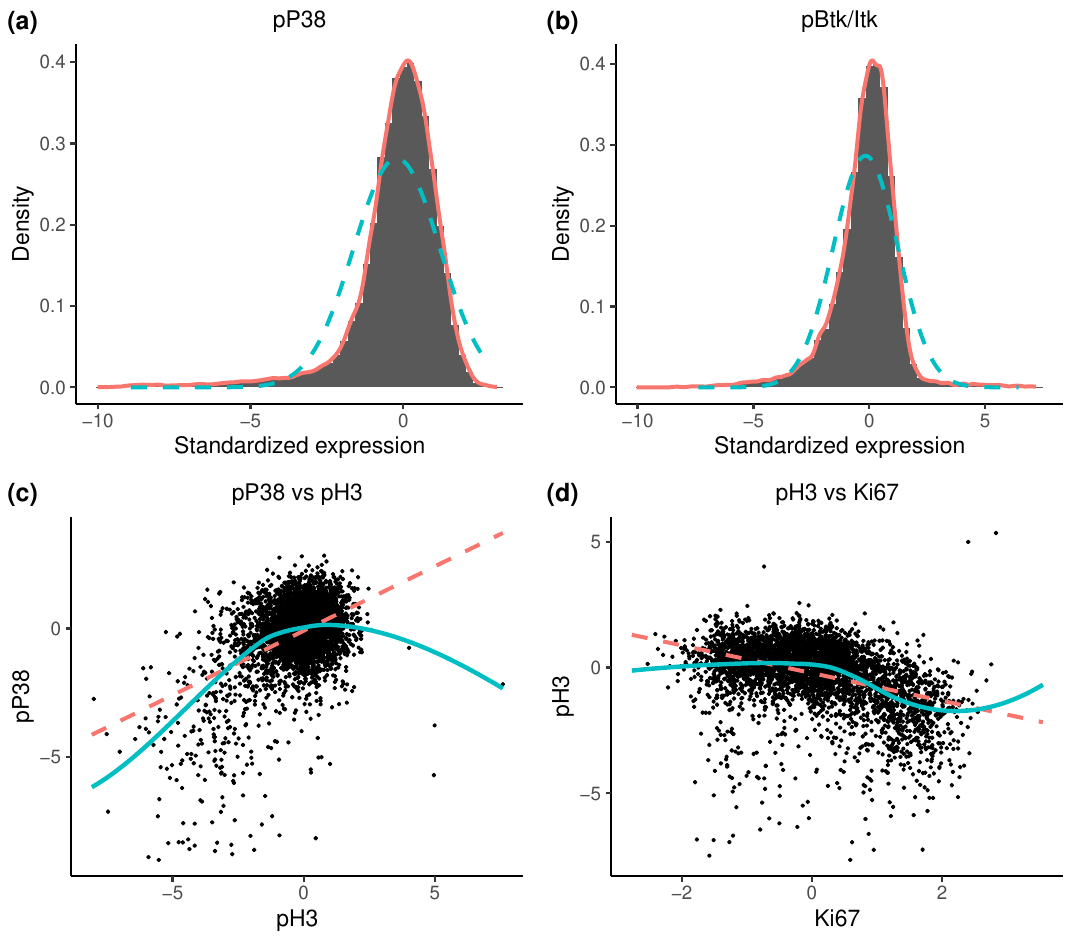}
		\caption{Representative distributional features of the Bendall bone-marrow data
			before the logistic transformation. The top panels show empirical marginal
			densities (solid) together with Gaussian densities having the same empirical
			mean and variance (dashed). The bottom panels show representative pairwise
			relationships, together with a LOESS smooth (solid) and a linear fit
			(dashed). 
		}
		\label{fig:bendall-nongaussian}
	\end{figure}

	Figure~\ref{fig:bendall-nongaussian} motivates a non-Gaussian interaction
	model: pP38 has skewness \(-2.19\), excess kurtosis \(8.31\), and a long
	left tail; pBtk/Itk has excess kurtosis \(4.98\). The pH3--pP38 and
	Ki67--pH3 relationships are visibly nonlinear.
	
	{
		With centered monomials, \(d=2\), and the squared boundary weight, P-WSM
		uses 31 relative fractions of each training sample's \(\lambda_{\max}\).
		Minimizing the raw five-fold score after unpenalized refitting selects fraction
		0.04967, corresponding to full-sample \(\lambda=0.02095\), and 131 of 153 interactions.
		
		Under the same basis, folds, grid, and tuning, Joint-WSM selects 141 edges:
		all 131 P-WSM edges plus 10, giving Jaccard similarity 0.929. Refits retain
		all main effects for P-WSM, only selected main effects for Joint-WSM, and
		selected interactions for both. Mean refitted held-out scores are
		\(-1.30199\) and \(-1.30271\), respectively. The paired P-WSM-minus-Joint-WSM
		difference is 0.00072, with a descriptive five-fold 95\% interval
		\([-0.00014,0.00158]\).
		
		\begin{table}[H]
			\centering
			\caption{Estimated interaction graphs for the Bendall bone-marrow data.}
			\label{tab:bendall-graph-summary}
			\begin{tabular}{lccc}
				\hline
				Method & Edges & Density & Common with P-WSM \\
				\hline
				P-WSM & 131 & 85.6\% & 131 \\
				Joint-WSM & 141 & 92.2\% & 131 \\
				Graphical lasso & 137 & 89.5\% & 123 \\
				\texttt{highscore} & 144 & 94.1\% & 127 \\
				\hline
			\end{tabular}
		\end{table}
		
		Graphical lasso and \texttt{highscore} have Jaccard overlaps 0.848 and 0.858
		with P-WSM. They fit Gaussian conditional-dependence graphs to pre-logistic
		data, whereas P-WSM fits interactions on the bounded representation; we
		therefore compare supports rather than their scores. 
		Figure~\ref{fig:bendall-support} highlights Joint-WSM's 10 additional edges. 
		These additional edges show that the fitted support is
                 sensitive to main-effect penalization. The data do not
                 establish which of the discordant edges are genuine.
		
		\begin{figure}[!ht]
			\centering
			\includegraphics[width=0.98\textwidth]{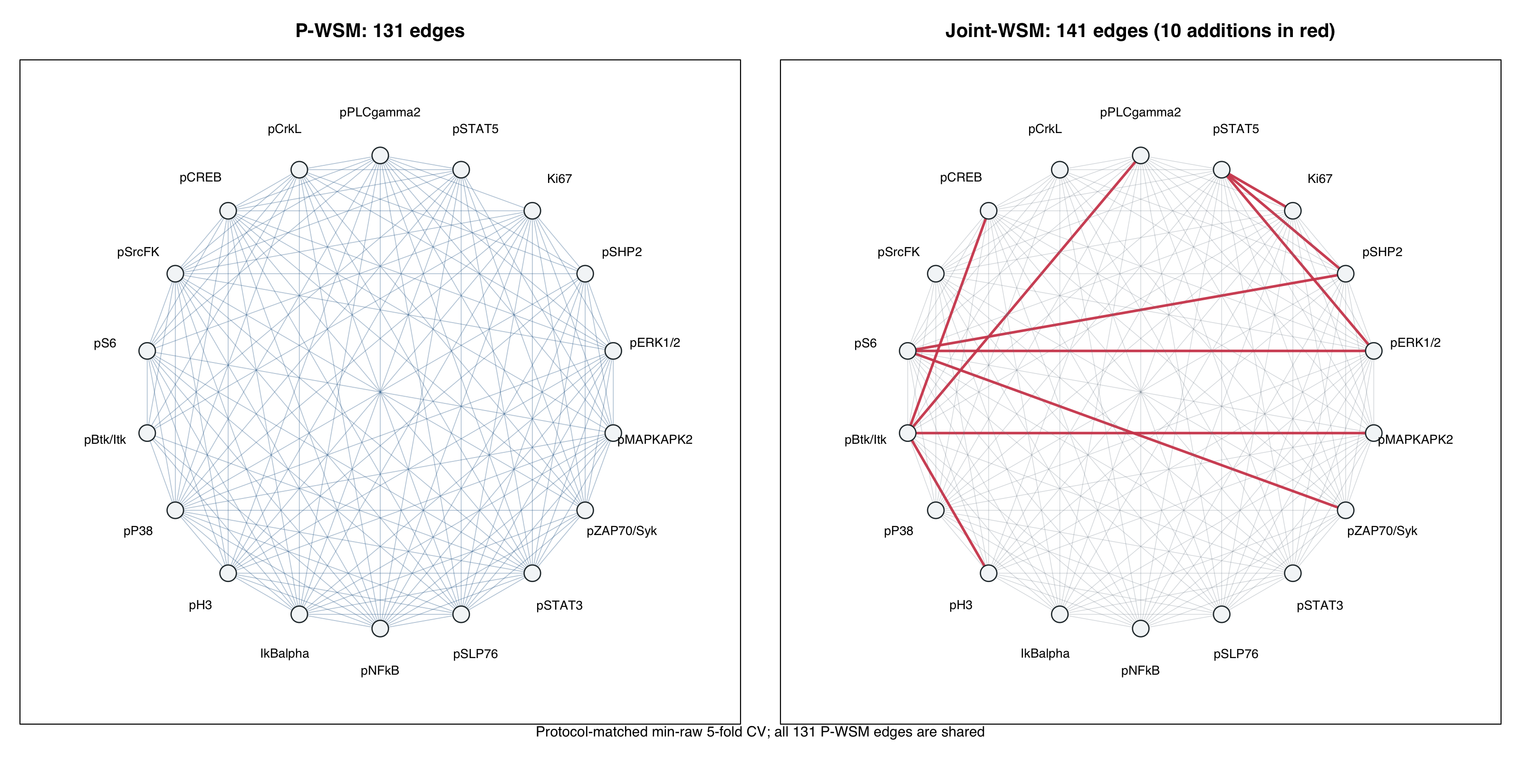}
			\caption{P-WSM and Joint-WSM interaction graphs for the
				Bendall LPS-monocyte data under the stated refitting protocol. The right panel highlights the 10
				Joint-WSM edges not selected by P-WSM.}
			\label{fig:bendall-support}
		\end{figure}
		
		At the fixed full-sample fraction, 50 stratified 80\% cell subsamples from
		this pooled one-donor dataset give mean graph size 131.96 (standard
		deviation 2.11; range 127--137). Of 131 full-sample edges, 127 have
		selection frequency at least 0.8, with mean frequency 0.979 across all 131.
		The strongest fully stable interactions by refitted block norm include
		Ki67--pH3, pP38--pSrcFK, pNFkB--pH3, pMAPKAPK2--pP38, and
		pERK1/2--pNFkB.
		
		With a cubic basis, linear boundary weight, or logistic scale 1.5, P-WSM
		selects 116, 136, and 130 edges versus Joint-WSM's 150, 145, and 143.
		Joint-WSM adds 34, 9, and 14 edges; only the last setting has one P-WSM-only
		edge. Raw one-standard-error tuning gives 90 and 109 edges, respectively.
		Full tuning, stability, overlap, and sensitivity results appear in Section~\ref{sec:supp-bendall-pwsm}
		in the Supplementary Material.
	}

\section{Conclusion}
\label{sec:conclusion}

Penalizing main effects can change the graph selected by
weighted score matching, even though those effects do not
define graph edges. P-WSM removes this direct shrinkage
contribution. Population preservation and a common-interval
separation construction establish consequences for edge
selection, while growing-dimensional recovery requires
restricted conditioning, sufficient signal, and an inactive
margin. The unrestricted estimator also admits weighted
Fisher-divergence rates and, under stronger conditions,
KL and Hellinger rates at fixed dimension.

Partial penalization does not remove finite-sample
complexity constraints: simulations show deterioration
as nuisance dimension or sieve size increases. 
The Bendall analysis gives a smaller P-WSM graph
with similar refitted validation scores, but does not
identify which discordant edges are correct. Future work
will study adaptive sieve selection and whether nuisance
regularization can control complexity while retaining
population subsystem preservation.

\textbf{Data availability.}
The bone-marrow signaling data are publicly available through
Cytobank \citep{bendall2011single}.

\appendix

\vspace{2em}
\setcounter{section}{0}
\setcounter{equation}{0}
\setcounter{theorem}{0}
\setcounter{lemma}{0}
\setcounter{corollary}{0}
\setcounter{assumption}{0}
\setcounter{definition}{0}
\setcounter{example}{0}
\setcounter{fact}{0}
\setcounter{remark}{0}
\setcounter{proposition}{0}
\setcounter{algorithm}{0}
\setcounter{table}{0}
\setcounter{figure}{0}

\renewcommand{\thesection}{\Alph{section}}
\renewcommand{\theequation}{S.\arabic{equation}}
\renewcommand{\thetheorem}{S.\arabic{theorem}}
\renewcommand{\thelemma}{S.\arabic{lemma}}
\renewcommand{\thecorollary}{S.\arabic{corollary}}
\renewcommand{\theassumption}{S.\arabic{assumption}}
\renewcommand{\thedefinition}{S.\arabic{definition}}
\renewcommand{\theexample}{S.\arabic{example}}
\renewcommand{\thefact}{S.\arabic{fact}}
\renewcommand{\theremark}{S.\arabic{remark}}
\renewcommand{\theproposition}{S.\arabic{proposition}}
\renewcommand{\thealgorithm}{S.\arabic{algorithm}}
\renewcommand{\thetable}{S.\arabic{table}}
\renewcommand{\thefigure}{S.\arabic{figure}}

\section{Notation and preliminary identities}
\label{supp:notation}\label{supp:problem_setup_details}
We retain the model, centered-basis, loss, and block notation of the main
paper. Matrix \(\|\cdot\|_\infty\) is the maximum absolute row sum,
\(\|\cdot\|_{\max}\) the largest absolute entry, and \(\|\cdot\|_2\)
the spectral norm (Euclidean norm for vectors). In the bounds below,
the notation \(X_n\le_p a_n\) designates \(X_n=O_p(a_n)\). Constants retain the dependence
specified in each result.

\subsection{Canonical projections and graph identification}
\label{supp:anova}
For an integrable function \(h\) on \(\Omega=[a,b]^p\), let \(A_i h\)
average its \(i\)th coordinate with respect to uniform probability.
These commuting averaging operators give
\[
\Pi_S=\prod_{i\in S}(I-A_i)\prod_{i\notin S}A_i,\qquad
h=\sum_{S\subseteq[p]}\Pi_Sh.
\]
Thus the canonical pairwise model is the restriction
\(\Pi_Sh_0=0\) for \(|S|\ge3\), with uniquely determined components
\(C=\Pi_\emptyset h_0\), \(f_i=\Pi_{\{i\}}h_0\), and
\(f_{ij}=\Pi_{\{i,j\}}h_0\).

\subsection{Bases and the score-matching identity}
\label{supp:methodology_details}\label{supp:basis_examples}
The bases in Theorem~\ref{th:density-unrestricted} are canonically
centered. For monomials, centering changes only the constant, so
nonconstant coefficients remain Taylor coefficients in the scaled coordinate
\(x/\pi\) when termwise differentiation is valid. Assumption~\ref{ass:expandability} and
uniformly bounded basis functions give
\[
\sup_i\|f_i-\tilde f_i\|_\infty=O(d_n^{1-\nu_1}),\qquad
\sup_{i<j}\|f_{ij}-\tilde f_{ij}\|_\infty=O(d_n^{1-\nu_2}).
\]
Here truncation retains \(r\le d_n\) for main effects and \(r,s\le d_n\)
for interactions; the latter tail includes both off-rectangle strips.

\label{supp:wsm_identity_methodology}
\begin{lemma}[Population score identity]\label{lem:1}
	Under Assumptions~\ref{ass:expandability} and~\ref{ass:1}, for every
	potential \(\eta_\theta\) with the stated coefficient decay, the loss
	\(\ell\) in \eqref{eq:empirical_wsm_loss} satisfies
	\begin{equation}\label{eqn:21}
		M(\theta):=E_0\ell(\theta,X)=\tfrac12J_g(p_0\|p_\theta)
		-\tfrac12E_0\sum_k g_ks_{0,k}^2.
	\end{equation}
	The identity also holds for any finite sieve potential with finite
	coefficients.
\end{lemma}
\begin{proof}
	Write \(s_\theta=\nabla\eta_\theta\) and \(s_0=\nabla\log p_0\).
	It suffices to justify coordinatewise integration by parts. Put
	\(\nu_{(1)}=\nu_4,\nu_{(2)}=\nu_5\). For \(a=1,2\), the coefficient-decay
	and basis-derivative bounds give
	\begin{equation}\label{eqn:uniform-score-derivatives}
		|\partial_k^a\eta_\theta|
		\lesssim\sum_{r\ge1}r^{-\nu_1+\nu_{(a)}}
		+(p-1)\Big(\sum_{r\ge1}r^{-\nu_2+\nu_{(a)}}\Big)
		\Big(\sum_{s\ge1}s^{-\nu_2}\Big)<\infty.
	\end{equation}
	Uniform convergence justifies termwise differentiation and bounds the true density
	and score components for each finite \(p\); for a finite sieve potential, the series reduce to finite sums.
	For fixed \(x_{-k}\), \(g_ks_{\theta,k}\) is absolutely continuous,
	and its boundary product with \(p_0\) vanishes because \(g_k(a)=g_k(b)=0\).
	Integration by parts and Fubini's theorem yield
	\begin{equation}\label{eqn:5}
		E_0\{g_k\partial_ks_{\theta,k}+g_k's_{\theta,k}\}
		=-E_0(g_ks_{\theta,k}s_{0,k}).
	\end{equation}
	Substituting \eqref{eqn:5} into \(M(\theta)\) and completing the square yields \eqref{eqn:21}. In particular, the
	population gradient at the truth vanishes along every finite sieve direction.
\end{proof}

\subsection{Block update and bounded restriction}
\label{supp:block_thresholding}
\begin{lemma}[Elliptic group soft-thresholding]
	\label{lem:elliptic_threshold}
	For \(H\succ0\), \(b\in\mathbb R^q\) and \(\lambda>0\), the unique
	minimizer of \(u^\top Hu/2+b^\top u+\lambda\|u\|_H\) is zero when
	\(\|b\|_{H^{-1}}\le\lambda\), and otherwise equals
	\(-\{1-\lambda/\|b\|_{H^{-1}}\}H^{-1}b\).
\end{lemma}
\begin{proof}
	With \(v=H^{1/2}u\) and \(c=H^{-1/2}b\), the subgradient equation is
	\(0\in v+c+\lambda\partial\|v\|_2\).
	At zero this is equivalent to \(\|c\|_2\le\lambda\); otherwise it gives
	\(v=-(1-\lambda/\|c\|_2)c\). Transforming back proves the formula;
	strict convexity gives uniqueness. In Algorithm~\ref{alg:1}, take
	\(H=H_G\) and \(b=\nabla_{\theta_G}Q_{n,\varepsilon}(\theta)-H_G\theta_G\).
\end{proof}

\label{supp:bounded_restriction}
If an original density \(p_0\) has support larger than \(\Omega\),
its conditional density is \(p_\Omega=p_0/\mathcal Z\), where
\(\mathcal Z=\int_\Omega p_0\in(0,1]\). The two log densities share the same score on
\(\Omega\). For known \(\mathcal Z\), multiplying the normalized estimate
\(p_{u_n}\) by \(\mathcal Z\) estimates the
restriction of \(p_0\). Rates are unchanged for fixed \(\mathcal Z>0\)
when the assumptions hold for \(p_\Omega\), whereas an estimated \(\mathcal Z\)
requires a separate error bound.

\section{Joint-WSM conditions and the scaling corollary}
\label{supp:theoretical-results}

\label{supp:support-recovery-conditions}
\label{supp:support-condition-interpretation}
Use the block and error-scale notation of
Section~\ref{sec:structure-recovery-theory}. The coordinate counts are
\begin{equation}\label{eqn:group-dimensions}
	q_j=d_n,\qquad q_{(j,k)}=d_n^2,\qquad
	q_S=\sum_{G\in S}q_G\le |S|d_n^2.
\end{equation}
A sufficient population version of the inactive Gram condition is
\[
\min_{G\in S^c}\lambda_{\min}(E\Phi_G)\ge2c_\Phi\Lambda_{\min},
\qquad
\max_{G\in S^c}\|\Phi_G-E\Phi_G\|_2=o_p(\Lambda_{\min}).
\]
Weyl's inequality then gives Assumption~\ref{ass:2}(1).
Bounds \eqref{lem:4}, \eqref{lem:5}, and
\eqref{lem:uniform-inactive-concentration} establish the three error
scales, with \(\log(|S|d_n^2)\le2\log(pd_n)\): coefficient decay
controls truncation, while derivative growth enlarges concentration error.

\label{supp:transparent-scaling-derivation}
\begin{proof}[Proof of Corollary~\ref{cor:support-transparent}]
	Only the error conditions in Assumption~\ref{ass:2} remain to be
	verified. With \(s_n=|S|\), the bounded spectra and inverse norms reduce
	the Hessian and retention requirements to \(r_{H,n}\to0\) and
	\((\lambda_n+r_{A,n})/\rho_n^*\to0\). Incoherence gives
	\(1\vee\sqrt{s_n}\mu_n/\Lambda_{\min}=O(1)\), so exclusion follows from
	\[
	\frac{pd_n}{\lambda_n}
	\left\{d_n^2\sqrt{\ell_n/n}
	+d_n(d_n^{-\nu_1+2}+d_n^{-\nu_2+2})\right\}
	\lesssim r_{\lambda,n}/\lambda_n=o(1).
	\]
	For the polynomial-rate formulation, the same three errors are, up to
	logarithmic factors,
	\[
	\begin{split}
		r_{H,n}&\lesssim n^{\sigma+4\alpha-1/2},\\
		r_{A,n}&\lesssim n^{\sigma+2\alpha-1/2}
		+n^{\zeta-\alpha(\nu_\star-3)},\\
		r_{\lambda,n}&\lesssim n^{\zeta+3\alpha-1/2}
		+n^{\zeta-\alpha(\nu_\star-4)}.
	\end{split}
	\]
	The corollary's strict inequalities give \(r_{H,n}=o(1)\) and
	\(r_{A,n},r_{\lambda,n}=o(\rho_n^*)\): the tail requirement with
	\(\nu_\star-4\) implies the weaker one with \(\nu_\star-3\).
	For \(\lambda_n=n^{-c}\), retention and exclusion require
	\(b<c<\min\{1/2-\zeta-3\alpha,\alpha(\nu_\star-4)-\zeta\}\).
	Strictness absorbs the logarithmic factors, and Theorem~\ref{th:2}
	yields recovery.
\end{proof}

\section{Special cases of the interaction graph model}
\label{app:examples}
\begin{example}[Gaussian graphical model]
	\label{exam:gaussian_graphical_model}
	For \(X\sim N_p(\mu,K^{-1})\), \(K\succ0\), conditioned on
	\(\Omega=[a,b]^p\), put \(m_{\mathcal X}=(a+b)/2\) and
	\(v_{\mathcal X}=(b-a)^2/12\). Expanding
	\(\log p_\Omega(x)=-(x-\mu)^\top K(x-\mu)/2+C_\Omega\) gives
	\[
	\begin{split}
		f_i(x_i)&=-\tfrac12K_{ii}\{(x_i-m_{\mathcal X})^2-v_{\mathcal X}\}
		+\{(K\mu)_i-m_{\mathcal X}\textstyle\sum_jK_{ij}\}(x_i-m_{\mathcal X}),\\
		f_{ij}(x_i,x_j)&=-K_{ij}(x_i-m_{\mathcal X})(x_j-m_{\mathcal X}).
	\end{split}
	\]
	These components are canonically centered, and \(f_{ij}=0\) if and
	only if \(K_{ij}=0\). Thus truncation to the rectangle preserves
	the Gaussian graph within this nonparametric model.
\end{example}

\begin{example}[Forest-structured densities]\label{exam:forest_density}
	A positive forest-structured density
	\(p_G(x)=\prod_i p_i(x_i)\prod_{\{i,j\}\in E}
	\allowbreak\{p_{ij}(x_i,x_j)/\allowbreak(p_i(x_i)p_j(x_j))\}\),
	where each \(p_{ij}\) has marginals \(p_i\) and \(p_j\), has only univariate and bivariate terms
	in its log density. Applying the canonical projections consequently
	gives \(h_0=C+\sum_i f_i+\sum_{\{i,j\}\in E}f_{ij}\), with every
	nonedge component zero. This includes Chow--Liu trees.
\end{example}
\section{Proof of Theorem~\ref{th:2}}

Write $\lambda=\lambda_n$ and $\tilde\theta^*=\pi_n\theta^*$.
Use the block dimensions in \eqref{eqn:group-dimensions}, the norms in
Section~\ref{supp:notation}, and the error scales before
Assumption~\ref{ass:2}; $S_n=S\ne\varnothing$ includes main effects
and interactions. Tensor entries are indexed by $(r-1)d_n+s$.
Choose $C^*\ge1$ in Assumption~\ref{ass:1} so that
$\|L_r\|_\infty<C^*$, $\|L_r'\|_\infty\le C^*r^{\nu_4}$, and
$\|L_r''\|_\infty\le C^*r^{\nu_5}$ for all $r$.

\subsection{Hessian and concentration bounds}

\begin{lemma}[Quadratic loss and principal blocks]
	\label{lem:3}\label{lem:6}\label{lem:7}
	For a group $G$, set
	$B_G^{(m)}=\partial_{\theta_G}\partial_m\eta_\theta$ and
	$\varphi_G^{(m)}=g_m^{1/2}B_G^{(m)}$. The nonzero derivative features are
	\begin{gather*}
		B_j^{(j)}=(L_r'(x_j))_{r\le d_n},\\
		(B_{jk}^{(j)})_{(r-1)d_n+s}=L_r'(x_j)L_s(x_k),\qquad
		(B_{jk}^{(k)})_{(r-1)d_n+s}=L_r(x_j)L_s'(x_k).
	\end{gather*}
	All Hessian blocks, including $\Sigma_{SS}$ and $\Sigma^c_{S^cS}$, are
	submatrices of the same positive semidefinite matrix, independent of $\theta$:
	\begin{equation}
		\label{eqn:lemma3-master-hessian}
		\Sigma_{G,H}=P_n\sum_{m=1}^p
		\varphi_G^{(m)}\{\varphi_H^{(m)}\}^{\top}.
	\end{equation}
	Disjoint groups have zero cross blocks, and for all $u,v$,
	$\nabla M_n(u)-\nabla M_n(v)=\Sigma(u-v)$.
	For any real symmetric block matrix $A$ with diagonal blocks $B_{ii}$,
	\[
	\lambda_{\max}(A)\ge\max_i\lambda_{\max}(B_{ii}),\qquad
	\lambda_{\min}(A)\le\min_i\lambda_{\min}(B_{ii}).
	\]
\end{lemma}
\begin{proof}
	Differentiate the quadratic part $\frac12P_n\sum_m
	g_m(\partial_m\eta_\theta)^2$; the divergence term is linear in $\theta$.
	The resulting formula includes both endpoints of an interaction diagonal block,
	$\Sigma_{jk,jk}=P_n\{\varphi_{jk}^{(j)}\varphi_{jk}^{(j)\top}
	+\varphi_{jk}^{(k)}\varphi_{jk}^{(k)\top}\}$.
	For example, when $j<k<l$, the shared-endpoint block is
	\[
	(\Sigma_{jk,kl})_{(r-1)d_n+s,(t-1)d_n+u}
	=P_n\{g_k L_r(x_j)L_s'(x_k)L_t'(x_k)L_u(x_l)\}.
	\]
	Main--interaction blocks obey the same rule, and symmetry gives reverse orderings.
	The gradient identity follows because $\Sigma$ is constant in $\theta$.
	The eigenvalue inequalities follow by padding a unit block eigenvector
	with zeros in the Rayleigh quotient, and then applying the bound to $-A$.
\end{proof}

\begin{lemma}[Uniform bounded-coordinate concentration]
	\label{lem:8}
	If $(q_{i1},\ldots,q_{iN})$ are i.i.d. over $i$, $|q_{ij}|\le C$, and
	$Q_j=n^{-1/2}\sum_iq_{ij}$, where the deterministic $C$ may depend on $n$,
	then $\max_{j\le N}|Q_j-EQ_j|=O_p(C\sqrt{\log N})$ as $N\to\infty$.
	For $C>0$, finite $N$ and $t>0$,
	\begin{equation}\label{eq:common-hoeffding}
		P\{\max_j|Q_j-EQ_j|\ge t\}
		\le2N e^{-t^2/(2C^2)}\le2N e^{-t^2/(8C^2)}.
	\end{equation}
\end{lemma}
\begin{proof}
	Hoeffding's inequality uses the uncentered range length $2C$; a union bound gives
	\eqref{eq:common-hoeffding} without independence across coordinates.
	Take $t=KC\sqrt{\log N}$, $K>\sqrt8$, for the order statement;
	$C=0$ makes every coordinate zero.
\end{proof}

\paragraph{Active Hessian.}
Under Assumption~\ref{ass:1}, put $A=E\Sigma_{SS}$ and
$\Delta_n=\Sigma_{SS}-A$.
The master Hessian formula bounds each centered summand by
$4(C^*)^4d_n^{2\nu_4}$. Lemma~\ref{lem:8} over $q_S^2$ entries and
$\|\Delta_n\|_\infty\le q_S\|\Delta_n\|_{\max}$ give
\begin{equation}\label{lem:4}
	\|\Delta_n\|_\infty
	=O_p\left((C^*)^4\frac{|S|d_n^{2+2\nu_4}}{\sqrt n}
	\sqrt{\log(|S|d_n^2)}\right).
\end{equation}
With Assumption~\ref{ass:2} as well, part~(2) implies
$\|A^{-1}\Delta_n\|_\infty=o_p(1)$, so the Neumann inverse expansion gives
\[
\|\Sigma_{SS}^{-1}\|_\infty
\le\frac{\|A^{-1}\|_\infty}{1-\|A^{-1}\Delta_n\|_\infty}
\le(1+\delta)\|A^{-1}\|_\infty
\]
with probability tending to one for every fixed $\delta>0$. Symmetry gives
$\|\Delta_n\|_2\le\|\Delta_n\|_\infty=O_p(R_{H,n})
=o_p(\Lambda_{\min})$; Weyl's inequality yields $\Sigma_{SS}\succ0$ and,
for every fixed $\varepsilon>0$,
\begin{equation}
	\label{eqn:sample-spectral-bounds}
	\lambda_{\max}(\Sigma_{SS})\le(1+\varepsilon)\Lambda_{\max},\qquad
	\|\Sigma_{SS}^{-1}\|_2\le(1+\varepsilon)/\Lambda_{\min}
\end{equation}
with probability tending to one.

\paragraph{Score fluctuation and truncation.}
For any sieve
coefficient $\theta$ and scalar feature $\psi_a$, write
\begin{equation}\label{eq:common-scalar-score}
	Y_a(\theta)=\sum_m\{\partial_m(g_m\partial_m\psi_a)
	+g_m(\partial_m\psi_a)\partial_m\eta_\theta\},
	\qquad \{\nabla M_n(\theta)\}_a=P_nY_a(\theta).
\end{equation}
At $\tilde\theta^*$, the coefficient-decay sums in
Assumption~\ref{ass:1} bound centered main-effect contributions by
$K_n=K_n(d_n)$, where
\begin{align*}
	K_n(t)&=2C^*(t^{\nu_5}+t^{\nu_4})
	+2(C^*)^2Bt^{\nu_4}\frac{\nu_1-\nu_4}{\nu_1-\nu_4-1}
	\\&\quad+2|S|(C^*)^3Bt^{\nu_4}
	\frac{\nu_2(\nu_2-\nu_4)}{(\nu_2-1)(\nu_2-\nu_4-1)}.
\end{align*}
An interaction coordinate has two endpoint contributions and one additional
basis factor, so its centered bound is $2C^*K_n$.
Set $C_{con}=(1\vee2C^*)K_n(1)$.
Lemma~\ref{lem:8} applied to $q_S\le |S|d_n^2$ coordinates yields
\begin{equation}
	\label{eqn:25}
	\|\nabla_S(M_n-M)(\tilde\theta^*)\|_\infty
	\le_p\frac{(1\vee2C^*)K_n}{\sqrt n}\sqrt{\log(|S|d_n^2)}
	\le\frac{C_{con}d_n^{\nu_4\vee\nu_5}}{\sqrt n}
	\sqrt{\log(|S|d_n^2)}.
\end{equation}
Here $M$ is the full population loss, evaluated at zero-padded sieve
parameters or at $\theta^*$, not a projected loss. The convergent
differentiated series and Lemma~\ref{lem:1} give
\begin{equation}
	\label{eqn:27}
	\nabla M(\theta^*)=0.
\end{equation}
Put $R_n(x)=\eta_{\tilde\theta^*}(x)-\eta_{\theta^*}(x)$.
The omitted interaction indices form the rectangle complement, whose
two disjoint parts satisfy the elementary power-sum bound
\begin{align}
	&\sum_{r>d_n,s\ge1}r^{-\nu_2+\nu_4}s^{-\nu_2}
	+\sum_{r\le d_n,s>d_n}r^{-\nu_2+\nu_4}s^{-\nu_2}\notag\\
	&\hspace{2em}\le
	\frac{2\nu_2-\nu_4}{(\nu_2-1)(\nu_2-\nu_4-1)}
	d_n^{-\nu_2+\nu_4+1}.\label{eq:common-rectangular-tail}
\end{align}
The main-coordinate bias is $E\{g_mL_q'\partial_mR_n\}$.
Its main-effect tail has coefficient $B(C^*)^2/(\nu_1-\nu_4-1)$, and
its at most $p-1$ interaction tails have the coefficient in
\eqref{eq:common-rectangular-tail} times $B(C^*)^3$.
An interaction coordinate adds two endpoint terms and one basis factor.
Thus, with
\[
C_{con1}=B(C^*)^2(1\vee2C^*)\left\{
\frac1{\nu_1-\nu_4-1}\vee
pC^*\frac{2\nu_2-\nu_4}{(\nu_2-1)(\nu_2-\nu_4-1)}\right\},
\]
we obtain
\begin{equation}
	\label{eqn:26}
	\|\nabla_SM(\tilde\theta^*)-\nabla_SM(\theta^*)\|_\infty
	\le C_{con1}d_n^{\nu_4}
	(d_n^{-\nu_1+\nu_4+1}+d_n^{-\nu_2+\nu_4+1}).
\end{equation}
Adding \eqref{eqn:25} and~\eqref{eqn:26}, using \eqref{eqn:27}, gives
the active score bound with their displayed constants. In the notation
of Assumption~\ref{ass:2}, since $C_{con}=O(|S|)$ and $C_{con1}=O(p)$,
\begin{equation}\label{lem:5}
	\|\nabla_SM_n(\tilde\theta^*)\|_\infty=O_p(R_{A,n}).
\end{equation}

\paragraph{Simultaneous inactive bounds.}
Only Assumption~\ref{ass:1}, $S\ne\varnothing$, and $S_n=S$ are needed
here. Write
\begin{align*}
	r_n&=4(C^*)^4\sqrt{|S|}\frac{d_n^{2+2\nu_4}}{\sqrt n}
	\sqrt{\log(p^2|S|d_n^4)},\\
	e_n&=\frac{d_n^{\nu_4\vee\nu_5}}{\sqrt n}\sqrt{\log(p^2d_n^2)}
	+d_n^{\nu_4}(d_n^{-\nu_1+\nu_4+1}+d_n^{-\nu_2+\nu_4+1}).
\end{align*}
Apply the entrywise argument for \eqref{lem:4} to at most
$p^2|S|d_n^4$ cross-Hessian entries, giving maximum error
$Cd_n^{2\nu_4}\sqrt{\log(p^2|S|d_n^4)/n}$. Convert this to operator norm via
\[
\|\Sigma^c_{G,S}-E\Sigma^c_{G,S}\|_2
\le\sqrt{q_Gq_S}\|\Sigma^c_{G,S}-E\Sigma^c_{G,S}\|_{\max},
\qquad \sqrt{q_Gq_S}\le\sqrt{|S|}d_n^2.
\]
Unlike the active bound, this uses $\sqrt{q_Gq_S}$ rather than $q_S$.
For the scores, at most $p-1$ convergent interaction series contribute to
each derivative, so $\max_m\|\partial_m\eta_{\tilde\theta^*}\|_\infty\le Cp$
and $|Y_a|\le Cpd_n^{\nu_4\vee\nu_5}$. Apply Lemma~\ref{lem:8} over
$p^2d_n^2$ coordinates and use \eqref{eqn:26} and~\eqref{eqn:27},
whose coordinatewise bias bound also holds off $S$. This gives
$\max_a|\{\nabla M_n(\tilde\theta^*)\}_a|\le Cpe_n$ on a common
high-probability event. Multiplication by $\sqrt{q_S}\le\sqrt{|S|}d_n$
or $\sqrt{q_G}\le d_n$ therefore gives a common event $\mathcal E_n$
of probability tending to one on which
\begin{align}
	\max_{G\in S^c}\|\Sigma^c_{G,S}-E\Sigma^c_{G,S}\|_2
	&\le Cr_n,\label{eqn:uniform-inactive-bounds}\\
	\|\nabla_SM_n(\tilde\theta^*)\|_2&\le Cp\sqrt{|S|}d_ne_n,\qquad
	\max_{G\in S^c}\|\nabla_GM_n(\tilde\theta^*)\|_2\le Cpd_ne_n.
	\label{lem:uniform-inactive-concentration}
\end{align}
Here $C$ is independent of $n,p,d_n$, and maxima over empty $S^c$ are zero.

\begin{proposition}[Strict dual feasibility excludes a block in every optimum]
	\label{prop:pwsm-strict-dual}
	Consider a convex differentiable quadratic $Q$ and a penalty
	$\lambda\sum_e\|\gamma_e\|_{\Phi_e}$, $\lambda>0$, with unpenalized
	coordinates also permitted. Suppose $(\check\beta,\check\gamma)$ satisfies
	KKT, and let $F$ be a collection of zero blocks for which
	$\Phi_e\succ0$ and $\|z_e\|_{\Phi_e^{-1}}<1$. Then every minimizer has
	zero coefficients on $F$, whether or not the full Hessian is positive
	definite.
\end{proposition}
\begin{proof}
	The quadratic expansion at $\check\theta$ has a nonnegative second-order remainder.
	The KKT relation cancels linear terms after applying the subgradient inequality
	outside $F$, leaving an objective difference at least
	\[
	\lambda\sum_{e\in F}
	\{\|u_e\|_{\Phi_e}-z_e^\top u_e\}
	\ge\lambda\sum_{e\in F}
	(1-\|z_e\|_{\Phi_e^{-1}})\|u_e\|_{\Phi_e}.
	\]
	This is strictly positive if any $u_e\ne0$ on $F$; no uniqueness outside $F$ is asserted.
\end{proof}

\subsection{Active retention and inactive exclusion}

\begin{proof}[Proof of Theorem~\ref{th:2}]
	\label{pf:Proof of Theorem2}
	We construct a primal--dual witness
	\citep{wainwright2006sharp,ravikumar2009sparse}.
	On $\{\Sigma_{SS}\succ0\}$, define
	\begin{equation}
		\label{eqn:restricted-estimator}
		\check\theta_S=\operatorname*{arg\,min}_{u_S}
		\{M_n(u_S,0)+\lambda J_{n,S}(u_S)\},\qquad
		\check\theta=(\check\theta_S,0).
	\end{equation}
	Positive restricted curvature gives coercivity and strict convexity,
	so the unique restricted minimizer admits
	$\check z_S\in\partial J_{n,S}(\check\theta_S)$ with
	\begin{equation}
		\label{eqn:22}
		\nabla_SM_n(\check\theta)+\lambda\check z_S=0.
	\end{equation}
	For $G\in S$, the penalty matrix $\Phi_G$ is a principal block of
	$\Sigma_{SS}$. The actual subgradient at $\check\theta_G$, including
	when that block is zero, satisfies
	\begin{equation}
		\label{eqn:active-dual-bound}
		\|\check z_G\|_{\Phi_G^{-1}}\le1,\qquad
		\|\check z_G\|_2\le\sqrt{\lambda_{\max}(\Phi_G)}.
	\end{equation}
	It suffices to establish nonzero active blocks and strictly feasible inactive dual blocks.
	
	\emph{Step 1: retain every active group.}
	The common Hessian identity gives exactly
	\begin{equation}
		\label{eqn:restricted-error-identity}
		\check\theta_S-\tilde\theta_S^*
		=-\Sigma_{SS}^{-1}\{\lambda\check z_S+
		\nabla_SM_n(\tilde\theta^*)\}.
	\end{equation}
	By the active bounds \eqref{lem:4} and~\eqref{lem:5},
	\eqref{eqn:sample-spectral-bounds}, and \eqref{eqn:active-dual-bound},
	\[
	\|\check\theta_S-\tilde\theta_S^*\|_\infty
	\lesssim_p\|(E\Sigma_{SS})^{-1}\|_\infty
	\{\lambda\sqrt{\Lambda_{\max}}+R_{A,n}\}
	=o_p(\rho_n^*)
	\]
	by Assumption~\ref{ass:2}(3). Since $S_n=S$ and
	$\min_{G\in S}\|\tilde\theta_G^*\|_\infty\ge\rho_n^*>0$, it follows that
	\begin{equation}
		\label{eqn:23}
		\min_{G\in S}\|\check\theta_G\|_\infty
		\ge\rho_n^*-\|\check\theta_S-\tilde\theta_S^*\|_\infty>0
	\end{equation}
	with probability tending to one.
	
	\emph{Step 2: exclude every inactive group.}
	If $S^c$ is empty this step is vacuous. Otherwise, define
	\begin{equation}
		\label{eqn:inactive-dual-definition}
		\check z_G=-\lambda^{-1}\nabla_GM_n(\check\theta),\qquad G\in S^c.
	\end{equation}
	Combining Lemma~\ref{lem:6} with \eqref{eqn:restricted-error-identity}
	reduces the required dual bound to three terms:
	\begin{equation}
		\label{eqn:inactive-witness-expansion}
		\check z_G=\Sigma^c_{G,S}\Sigma_{SS}^{-1}\check z_S
		+\lambda^{-1}\Sigma^c_{G,S}\Sigma_{SS}^{-1}\nabla_SM_n(\tilde\theta^*)
		-\lambda^{-1}\nabla_GM_n(\tilde\theta^*).
	\end{equation}
	Put $\mu_G=\|E\Sigma^c_{G,S}\|_2\le\mu_n$. On the common event of
	the preceding lemmas,
	$\|\Sigma^c_{G,S}\|_2\le\mu_G+Cr_n$ and
	$\|\check z_S\|_2^2\le|S|(1+\varepsilon)\Lambda_{\max}$.
	Using Euclidean operator norms in \eqref{eqn:inactive-witness-expansion}
	therefore gives, simultaneously over $G\in S^c$,
	\begin{align}
		\label{eqn:uniform-inactive-dual-bound}
		\|\check z_G\|_2
		&\le(1+\varepsilon)^{3/2}(\mu_G+Cr_n)
		\frac{\sqrt{|S|\Lambda_{\max}}}{\Lambda_{\min}}
		\notag\\
		&\quad+\frac{Cpd_ne_n}{\lambda}
		\left\{1+\frac{(1+\varepsilon)\sqrt{|S|}(\mu_G+Cr_n)}
		{\Lambda_{\min}}\right\}.
	\end{align}
	It remains to compare this Euclidean bound with the inactive dual radius
	$\sqrt{c_\Phi\Lambda_{\min}}$. Choose fixed $\varepsilon>0$ so that
	$(1+\varepsilon)^{3/2}(1-\delta_0)\le1-\delta_0/2$.
	Assumption~\ref{ass:2}(5) bounds the population part of the first line,
	after division by $\sqrt{\Lambda_{\min}}$, by
	$(1-\delta_0/2)\sqrt{c_\Phi}$.
	The remaining part of that line vanishes because
	\[
	\frac{r_n\sqrt{|S|\Lambda_{\max}}}{\Lambda_{\min}^{3/2}}
	\lesssim R_{H,n}\sqrt{\frac{\Lambda_{\max}}{\Lambda_{\min}^3}}\to0,
	\qquad
	\frac{\sqrt{|S|}r_n}{\Lambda_{\min}}\to0.
	\]
	Here $|S|\le p(p+1)/2$ controls the logarithm, and the second limit follows
	from the first since $\Lambda_{\max}\ge\Lambda_{\min}$.
	Also $e_n\lesssim R_{I,n}$, so the normalized second line is at most a
	constant times
	\[
	\frac{pd_nR_{I,n}}{\lambda\sqrt{\Lambda_{\min}}}
	\left(1\vee\frac{\sqrt{|S|}\mu_n}{\Lambda_{\min}}\right)=o(1)
	\]
	by Assumption~\ref{ass:2}(4). Consequently,
	\begin{equation}
		\label{eqn:uniform-strict-dual-euclidean}
		\max_{G\in S^c}\frac{\|\check z_G\|_2}{\sqrt{\Lambda_{\min}}}
		\le(1-\delta_0/2)\sqrt{c_\Phi}+o(1)<\sqrt{c_\Phi}
	\end{equation}
	with probability tending to one. The inactive penalty lower bound in
	Assumption~\ref{ass:2}(1) then yields
	\begin{equation}
		\label{eqn:24}
		\max_{G\in S^c}\|\check z_G\|_{\Phi_G^{-1}}
		\le\max_{G\in S^c}
		\frac{\|\check z_G\|_2}{\sqrt{\lambda_{\min}(\Phi_G)}}<1.
	\end{equation}
	
	Thus $\check z\in\partial J_n(\check\theta)$ and
	\begin{equation}
		\label{eqn:full-penalized-kkt}
		\nabla M_n(\check\theta)+\lambda\check z=0.
	\end{equation}
	Full-Gram invertibility is unnecessary: KKT gives a minimizer, and the
	general convex argument in Proposition~\ref{prop:pwsm-strict-dual}
	applied to $F_n=M_n+\lambda J_n$ gives, for every $u$,
	\begin{equation}
		\label{eqn:strict-inactive-optimality}
		F_n(u)-F_n(\check\theta)
		\ge\lambda\sum_{G\in S^c}
		(1-\|\check z_G\|_{\Phi_G^{-1}})\|u_G\|_{\Phi_G}.
	\end{equation}
	Hence every minimizer has zero inactive groups. Strict convexity on the
	active subspace then makes it equal to $\check\theta$.
	Together with \eqref{eqn:23}, this proves
	$\hat\theta_n^{\mathrm J}=\check\theta$ and
	$P(\widehat S^{\mathrm J}=S)\to1$. Canonical identifiability converts
	interaction-support recovery into the graph conclusion of the theorem.
\end{proof}

\section{Population preservation and profiled edge recovery}

We use the loss, block partition, nuisance solution, and profile in
\eqref{def:pwsm}--\eqref{eq:pwsm-profile}; write $q$ for $q_{\!P}$ and
$F_\lambda,Q_\lambda$ for the original and profiled objectives.
Hats denote empirical moments, and $\widehat\Phi_e=\widehat C_{ee}$.
The population profile differs from $\inf_\beta M$ by
$a^\top A^{-1}a/2$. We allow coordinate-specific vectors
$L_i=(L_{i1},\ldots,L_{id})^\top$ and compatible tensors $L_i\otimes L_j$,
ordered by $(r-1)d+s$. Let $\mathcal E=\{(i,j):i<j\}$ and $m=|\mathcal E|$.
Derivative supports imply $A=\operatorname{diag}(A_i)$,
$B_{i,e}=0$ for $i\notin e$, and $C_{ef}=0$ for disjoint edges,
without independence or correct specification. Coefficient spaces are
unconstrained, and P-WSM uses neither a main-effect penalty nor ridge.

\begin{proposition}[Population identities and their extensions]
	\label{prop:pwsm-propagation}\label{prop:pwsm-product}
	\label{prop:pwsm-independent}
	The conclusions of Propositions~\ref{prop:pwsm-propagation-main}
	and~\ref{prop:pwsm-preservation-main} hold for these coordinate-specific
	bases, with the following scopes.
	
	\textnormal{(i)} Under correct finite-dimensional specification, valid
	integration by parts, and $A\succ0$, the shrinkage identity holds at every
	main-effect stationary point for a convex penalty $J_\beta$ and
	$z_\beta\in\partial J_\beta(\beta)$.
	
	\textnormal{(ii)} For a product law, coordinatewise weights, finite
	moments, and $A_i\succ0$, the stated identities for $q,S$ hold without
	requiring the marginal log densities to lie in the sieve. Here
	$\mu_i=EL_i$, $V_i=\operatorname{Cov}(L_i)$,
	$A_i=E[g_iL_i'L_i'^\top]$, and $a_i=E[g_i'L_i'+g_iL_i'']$.
	If $\lambda>0$ and all $\Phi_e=C_{ee}\succ0$, the minimizer is unique
	with $\gamma=0$ and $\beta=-A^{-1}a$. At $\lambda=0$, zero interactions
	still minimize; $S\succ0$ additionally ensures their uniqueness.
	
	\textnormal{(iii)} For independent subsystems, compatible bases,
	coordinatewise weights, and finite moments, every minimizer restricted
	to within-subsystem edges extends by zeros to a full minimizer.
	This requires restricted existence, but not $A\succ0$.
	If $\lambda>0$, only cross-subsystem $\Phi_e$ need be positive definite
	to exclude those edges from every full minimizer. A positive-definite
	full population Gram gives uniqueness.
\end{proposition}
\begin{proof}[Proof of the population propositions]
	For (i), $\Gamma\theta^*+\rho=0$ by integration by parts. Solving
	$A(\beta-\beta^*)+B(\gamma-\gamma^*)+\lambda_\beta z_\beta=0$ for
	$\beta-\beta^*$ and substituting into
	$B^\top(\beta-\beta^*)+C(\gamma-\gamma^*)$ gives
	Proposition~\ref{prop:pwsm-propagation-main}.
	
	For (ii), profiling centers the other endpoint's basis. Indeed,
	for $e=(i,j)$,
	\[
	(B_{i,e})_{\alpha,(r,s)}=(A_i)_{\alpha r}\mu_{j,s},\qquad
	(B_{j,e})_{\alpha,(r,s)}=\mu_{i,r}(A_j)_{\alpha s}.
	\]
	Hence $b_{ij}=a_i\otimes\mu_j+\mu_i\otimes a_j
	=(B^\top A^{-1}a)_{ij}$ and $q=0$. For a coefficient matrix $H$,
	the two nonzero entries of the nuisance-projected derivative field are
	\begin{equation}\label{eq:pwsm-product-residual}
		L_i'^\top H(L_j-\mu_j),\qquad (L_i-\mu_i)^\top H L_j'.
	\end{equation}
	Independence makes this field orthogonal to every main-effect field,
	so $A\succ0$ identifies it as the projection residual. Its weighted
	second moment has coefficient matrix $A_i\otimes V_j+V_i\otimes A_j$.
	Two distinct edges sharing an endpoint have zero residual inner product:
	condition on that endpoint and integrate over the two independent centered
	factors at the other endpoints. This covers all three endpoint orderings
	by transposing $H$ when the edge orientation is reversed; disjoint edges
	have disjoint derivative supports. Thus $S$ has the stated blocks. Finally,
	the objective difference from $\gamma=0$ is
	$\gamma^\top S\gamma/2+\lambda\sum_e\|\gamma_e\|_{\Phi_e}$.
	Since $S\succeq0$, it is nonnegative, and strictly positive for
	$\gamma\ne0$ when $\lambda>0$. The claimed uniqueness at $\lambda=0$
	follows if $S\succ0$; in either case $A\succ0$ gives the unique nuisance
	minimizer.
	For (iii), at a restricted solution $\partial_i\eta$ depends only on the subsystem
	containing $i$. For a cross-subsystem edge $(i,j)$, factorization and the
	coordinatewise weights give
	\[
	\nabla_{\gamma_{ij}}M
	=\nabla_{\beta_i}M\otimes\mu_j
	+\mu_i\otimes\nabla_{\beta_j}M=0.
	\]
	Nuisance stationarity gives the last equality; zero cross-subsystem dual
	blocks therefore complete the restricted KKT system. Convexity gives
	optimality, and
	Proposition~\ref{prop:pwsm-strict-dual} excludes cross-subsystem blocks from
	every optimum when their penalty matrices are positive definite and
	$\lambda>0$. A positive-definite full Gram gives uniqueness.
\end{proof}
In particular, positive density on a connected rectangle interior,
positive interior weights and linearly independent canonically centered features
give a positive-definite population Gram: zero energy forces a constant
potential, centering removes the constant, and feature independence
removes its coefficients. This also gives restricted existence.

\subsection{A deterministic edge-recovery certificate}

For edge vectors use Euclidean block norms. Let $E\subseteq\mathcal E$,
$s=|E|>0$, and let $\gamma_d$ have exactly this support. Suppose
$\widehat A\succ0$, $\widehat\Phi_e\succ0$ for every edge, and
$\widehat S\succeq0$. Define
\[
r=\widehat S\gamma_d+\widehat q,\quad
k=\lambda_{\min}(\widehat S_{EE}),\quad
M=\max_{e\in E}\lambda_{\max}(\widehat\Phi_e),\quad
\phi=\min_e\lambda_{\min}(\widehat\Phi_e),
\]
\[
u=\max_{e\notin E}\|\widehat S_{eE}\|_2,\quad
R=\max_e\|r_e\|_2,\quad
\rho=\min_{e\in E}\|\gamma_{d,e}\|_2.
\]
When $E=\mathcal E$, all requirements involving inactive blocks below are
omitted and $u=0$.

\begin{theorem}[Deterministic certificate]
	\label{th:pwsm-deterministic}
	Suppose $k>0$, $\lambda>0$, and
	\[
	\rho>\frac{\sqrt{s}}{k}(R+\lambda\sqrt M),
	\qquad
	\frac{u\sqrt{sM}}{k\sqrt\phi}
	+\frac{R}{\lambda\sqrt\phi}
	\left(1+\frac{u\sqrt s}{k}\right)<1.
	\]
	Then the profiled penalized problem has a unique minimizer with edge support
	exactly $E$. The original partially penalized problem has the same unique
	interaction solution and the unique reconstructed main effects.
\end{theorem}
\begin{proof}
	We verify active retention and strict inactive dual feasibility.
	Since $k>0$, the criterion restricted to
	$\gamma_{E^c}=0$ is coercive and strictly convex. Its unique minimizer
	$\check\gamma$ satisfies, for active norm subgradients $z_E$,
	\begin{equation}\label{eq:pwsm-restricted-error}
		\check\gamma_E-\gamma_{d,E}
		=-\widehat S_{EE}^{-1}(r_E+\lambda z_E),\qquad
		\|\check\gamma_E-\gamma_{d,E}\|_2
		\le\frac{\sqrt{s}}k(R+\lambda\sqrt M)<\rho.
	\end{equation}
	Here $\|z_e\|_2\le\sqrt M$: write $z_e=\widehat\Phi_e^{1/2}v_e$
	with $\|v_e\|_2\le1$, also at a zero block. Thus every active block is
	nonzero. The required inactive KKT block is
	\begin{equation}\label{eq:pwsm-inactive-dual}
		z_e=-\lambda^{-1}(\widehat S\check\gamma+\widehat q)_e
		=\widehat S_{eE}\widehat S_{EE}^{-1}z_E
		+\lambda^{-1}
		(\widehat S_{eE}\widehat S_{EE}^{-1}r_E-r_e).
	\end{equation}
	Its $\widehat\Phi_e^{-1}$ norm is at most the second left-hand side
	in the theorem, hence is strictly below one. The full KKT equations give
	an attained global minimum even if $\widehat S$ is singular.
	Proposition~\ref{prop:pwsm-strict-dual} and restricted strict convexity
	give uniqueness; completing the square reconstructs the unique main effects.
	For $E=\mathcal E$ the restricted problem is the full strongly convex
	problem, and no inactive bound is used.
\end{proof}
If $E$ is empty, no restricted eigenvalue or signal condition is needed:
$\max_e\|\widehat q_e\|_{\widehat\Phi_e^{-1}}<\lambda$ suffices for a
unique zero interaction solution and a unique beta solution.

\subsection{Explicit simultaneous probability bounds}
\label{sec:pwsm-common-bounds}

Use the sampling, compatible features, deterministic $\theta_d$,
score envelopes and dimensions of Assumption~\ref{ass:pwsm-edge}(i),
but require only a common positive density and featurewise integration
by parts, not its canonical model or rectangular support. Weights remain
coordinatewise.
Let $E=\operatorname{supp}(\gamma_d)$; all population bounds may depend
on $n$. Moments are given by \eqref{eqn:lemma3-master-hessian} and
\eqref{eq:common-scalar-score}. The canonical special case is as follows.
\begin{assumption}[Sampling, approximation and envelopes]
	\label{ass:ercv2-model}
	The preceding conditions hold on a rectangle, and $p_0$ has canonical
	pairwise edge set $E$. Twice differentiable features may be replaced by
	piecewise versions when, for every feature $j$,
	\[
	E\sum_i(g_i'\partial_i\psi_j+g_i\partial_i^2\psi_j)
	=-E\sum_i g_i\partial_i\psi_j s_{0,i}.
	\]
	All expectations exist; essential supremum bounds are allowed with
	values at null-set cusps fixed. The equality of true and sieve edge sets
	is assumed, not inferred from the score approximation bound.
\end{assumption}
Set $L_\Gamma=2GU^2$, $L_\xi=2(GUT+G_1U+GV)$ and $b_0=2GUH$.
Use only the numerical definitions in Assumption~\ref{ass:pwsm-edge}(ii)--(iii),
with arbitrary $\delta\in(0,1)$ replacing $\delta_n$; positivity, recovery
and $\delta_n\to0$ are not imposed by reference.
Write $a_-=\underline a$, $\phi_-=\underline\phi$,
$M_+=\overline M$, $R=R_n$, and $k_-=\underline\kappa$ when $s=|E|>0$.
Then $U_\lambda=(k_-\rho_d/\sqrt{s}-R)/\sqrt{M_+}$ is the upper tuning
bound. For $0<s<m$, let $u_+=\overline\mu$ and define
\[
\chi=\frac{u_+\sqrt{sM_+}}{k_-\sqrt{\phi_-}},\qquad
V_P=\frac R{\sqrt{\phi_-}}\left(1+\frac{u_+\sqrt{s}}{k_-}\right),
\qquad L_\lambda=\frac{V_P}{1-\chi}.
\]
\begin{assumption}[Population stability and tuning interval]
	\label{ass:ercv2-stability}
	The quantities above satisfy $a_0,\phi_0,a_-,\phi_->0$.
	If $0<s<m$, require $k_->0$, $\chi<1$ and $L_\lambda<U_\lambda$,
	and let $I=(L_\lambda,U_\lambda)$.
	If $s=m$, require $k_->0$ and $U_\lambda>0$, and let
	$I=(0,U_\lambda)$; no inactive quantity is required.
	If $s=0$, let $I=(R/\sqrt{\phi_-},\infty)$, without defining active
	eigenvalue, signal or irrepresentability quantities.
	The penalty $\lambda$ is deterministic and belongs to $I$.
\end{assumption}

\begin{theorem}[Finite-sample recovery and growing sequences]
	\label{th:pwsm-finite}\label{th:ercv2-recovery}\label{cor:pwsm-growing}
	Under the common conditions above, the following conclusions hold.
	
	\textnormal{(i)} Under Assumption~\ref{ass:ercv2-stability}, with probability
	at least $1-\delta$, unstabilized P-WSM exists uniquely and has interaction
	support $E$, simultaneously for every $\lambda\in I$. For $s>0$,
	\[
	\|\widehat\gamma_E-\gamma_{d,E}\|_2
	\le\frac{\sqrt{s}}{k_-}(R+\lambda\sqrt{M_+}),\qquad
	\widehat\gamma_{E^c}=0.
	\]
	This includes Assumption~\ref{ass:ercv2-model}, but does not require a
	canonical model, a population sieve minimizer $\theta_d$, full empirical
	Gram positive definiteness, or main-effect sparsity or beta-min.
	For any random choice $\widetilde\lambda$, the same event gives
	\[
	P\{\text{unique exact recovery at }\widetilde\lambda\}
	\ge1-\delta-P\{\widetilde\lambda\notin I\}.
	\]
	Failure of existence outside $I$ is counted as failure; the last
	probability is not controlled for CV or 1SE.
	
	\textnormal{(ii)} Allow $p,d,s$ and all population constants to vary with
	$n$, and form the bounds with $\delta_n\to0$. Suppose sieve edge
	visibility holds, $a_-,\phi_->0$, and $k_->0$ when $s>0$.
	For deterministic $\lambda_n>0$, sufficient conditions for
	$P(\widehat E=E)\to1$ are, for some fixed $\eta>0$,
	\[
	\chi\le1-\eta,\quad V_P/\lambda_n\to0\quad(0<s<m),\qquad
	\frac{\sqrt{s}(R+\lambda_n\sqrt{M_+})}{k_-\rho_d}\to0\quad(s>0).
	\]
	For $s=0$, replace these by $R/(\lambda_n\sqrt{\phi_-})<1$ eventually,
	without active eigenvalue, signal or incoherence quantities. More generally,
	eventual nonemptiness of $I_n$ and any deterministic $\lambda_n\in I_n$
	suffice, including full graphs with no inactive conditions.
\end{theorem}
\begin{proof}
	It suffices to establish the two deterministic recovery inequalities on one event
	independent of $\lambda$. Put $t_\varepsilon=t_w-b_0$, and use
	\eqref{eq:common-scalar-score} at $\theta_d$:
	$\xi_a=Y_a(\theta_d)$, $h=E\xi$, and $\varepsilon=P_n\xi-E\xi$.
	The feature support gives $|\xi_a|\le L_\xi$ and an absolute Gram-entry
	bound $L_\Gamma$. Integration by parts gives
	$h_a=E\sum_i g_i\partial_i\psi_a(s_{d,i}-s_{0,i})$, so $\|h\|_\infty\le b_0$.
	The $2C^2$ bound in \eqref{eq:common-hoeffding}, with $C=L_\Gamma$
	or $L_\xi$, assigns failure budget $\delta/2$ to $D^2$ Gram entries
	and to $D$ score entries. It yields an event $\mathcal H$ of
	probability at least $1-\delta$ on which
	\begin{equation}\label{eq:pwsm-common-event}
		\|\widehat\Gamma-\Gamma\|_{\max}\le t_\Gamma,\qquad
		\|\varepsilon\|_\infty\le t_\varepsilon,\qquad
		w:=\widehat\Gamma\theta_d+\widehat\rho=h+\varepsilon,
		\quad\|w\|_\infty\le t_w.
	\end{equation}
	A zero envelope makes the corresponding variable identically zero.
	The following estimates are deterministic on $\mathcal H$.
	
	The block dimensions give errors $e_A,e_B,e_\Phi$ for $A_i,B_{i,e},C_{ef}$.
	If $\underline a>0$, Weyl's inequality and
	$\widehat A_i^{-1}-A_i^{-1}=\widehat A_i^{-1}(A_i-\widehat A_i)A_i^{-1}$
	give
	\begin{equation}\label{eq:pwsm-common-matrices}
		\widehat A_i\succeq\underline a I,\qquad
		\|\widehat A_i^{-1}-A_i^{-1}\|_2\le\frac{e_A}{a_0\underline a},
		\qquad \underline\phi I\preceq\widehat\Phi_e\preceq\overline M I.
	\end{equation}
	The empirical Gram is positive semidefinite, hence so is its Schur
	complement. For a shared endpoint $i$ of edges $e,f$, the projection error is
	\begin{align*}
		&\widehat B_{i,e}^\top\widehat A_i^{-1}\widehat B_{i,f}
		-B_{i,e}^\top A_i^{-1}B_{i,f}\\
		&=(\widehat B_{i,e}-B_{i,e})^\top\widehat A_i^{-1}\widehat B_{i,f}
		+B_{i,e}^\top\widehat A_i^{-1}(\widehat B_{i,f}-B_{i,f})
		+B_{i,e}^\top(\widehat A_i^{-1}-A_i^{-1})B_{i,f}.
	\end{align*}
	Its norm is at most
	$e_B(2B_0+e_B)/\underline a+B_0^2e_A/(a_0\underline a)$.
	At most two endpoints contribute. Summing the resulting block errors by
	Cauchy--Schwarz gives, whenever the relevant blocks exist,
	\begin{equation}\label{eq:pwsm-common-schur}
		\max_{e,f}\|\widehat S_{ef}-S_{ef}\|_2\le e_S,\qquad
		\lambda_{\min}(\widehat S_{EE})\ge\underline\kappa,\qquad
		\max_{e\notin E}\|\widehat S_{eE}\|_2\le\overline\mu.
	\end{equation}
	The active block error is at most $se_S$, and the inactive block-row error
	at most $\sqrt{s}e_S$; no additional factor $d$ enters these two bounds.
	
	Each edge has $d^2$ coordinates and touches two $d$-dimensional nuisance
	blocks. Thus any vector $v=(v_\beta,v_\gamma)$ satisfies
	\begin{equation}\label{eq:pwsm-common-profile}
		\max_e\|(v_\gamma-\widehat B^\top\widehat A^{-1}v_\beta)_e\|_2
		\le K_P\|v\|_\infty,\qquad
		K_P=d+\frac{2(B_0+e_B)\sqrt d}{\underline a}.
	\end{equation}
	In particular, the exact identity
	\begin{equation}\label{eq:pwsm-common-residual}
		r:=\widehat S\gamma_d+\widehat q
		=w_\gamma-\widehat B^\top\widehat A^{-1}w_\beta
	\end{equation}
	gives $\max_e\|r_e\|_2\le K_Pt_w=R_n$ without sample splitting or
	conditional independence. Profiling $w=h+\varepsilon$ gives separate bounds
	$R_\varepsilon=K_Pt_\varepsilon$ and $R_H=K_Pb_0$, with
	$R=R_\varepsilon+R_H$. The empirical projection of $h$ differs from
	its population projection by at most
	\[
	2\sqrt d\,b_0\left\{\frac{e_B}{a_-}
	+\frac{B_0e_A}{a_0a_-}\right\}
	\]
	on each edge: expand the endpoint projection error as
	$(\widehat B_{i,e}-B_{i,e})^\top\widehat A_i^{-1}
	+B_{i,e}^\top(\widehat A_i^{-1}-A_i^{-1})$ and use
	$\|h_{\beta_i}\|_2\le\sqrt d\,b_0$.
	
	It remains to apply Theorem~\ref{th:pwsm-deterministic} with
	$k\ge k_-$, $M\le M_+$, $\phi\ge\phi_-$, $u\le u_+$ and residual
	bound $R_n=R$. For $0<s<m$, the actual random-metric operator satisfies
	\[
	\widehat{\mathcal I}:=
	\max_{e\notin E}\sup_{\|v_f\|_2\le1,\ f\in E}
	\left\|\widehat\Phi_e^{-1/2}\widehat S_{eE}
	\widehat S_{EE}^{-1}
	\operatorname{diag}_{f\in E}(\widehat\Phi_f^{1/2})v_E\right\|_2
	\le\chi
	\]
	because the diagonal product has norm at most $\sqrt{sM_+}$.
	Since $\lambda<U_\lambda$, \eqref{eq:pwsm-restricted-error} gives the
	stated error bound strictly below $\rho_d$. For inactive edges,
	\eqref{eq:pwsm-inactive-dual} gives
	$\max_{e\notin E}\|z_e\|_{\widehat\Phi_e^{-1}}
	\le\chi+V_P/\lambda<1$, equivalently $\chi<1$ and
	$\lambda>L_\lambda$. These sufficient conditions give existence and
	uniqueness. For $s=m$ only the active inequality is used. For $s=0$,
	$r=\widehat q$ and
	$\max_e\|\widehat q_e\|_{\widehat\Phi_e^{-1}}
	\le R/\sqrt{\phi_-}<\lambda$ gives its empty-graph certificate.
	The event $\mathcal H$ is independent of $\lambda$ and has probability
	at least $1-\delta$, proving (i), including its random-choice bound.
	In (ii), the displayed ratios put $\lambda_n$ in $I_n$ eventually;
	the failure probability is then at most $\delta_n\to0$.
\end{proof}

\paragraph{Connecting the envelopes to coefficient decay.}
If $\sup_r\|L_r\|_\infty\le K_0$,
$\|L_r'\|_\infty\le K_1r^{\nu_4}$ and
$\|L_r''\|_\infty\le K_2r^{\nu_5}$, with $\nu_4\ge0$, one may use
$U=K_1\max(1,K_0)d^{\nu_4}$ and
$V=K_2\max(1,K_0)d^{\nu_5}$. Suppose the canonical expansion represents
the true score, truncation retains every true edge, and uniformly
$|\beta_{ir}|\le B r^{-\nu_1}$ and
$|\gamma_{ij,rs}|\le B r^{-\nu_2}s^{-\nu_2}$, with
$\nu_1,\nu_2>\nu_4+1$. Summing differentiated tails gives
\[
H\le C\{d^{-\nu_1+\nu_4+1}
+\Delta_n d^{-\nu_2+\nu_4+1}\},\qquad
T\le C(1+\Delta_n),
\]
where $\Delta_n$ is the maximum true graph degree. Indeed, the main tail is
at most $BK_1d^{-\nu_1+\nu_4+1}/(\nu_1-\nu_4-1)$; the tail of each
incident interaction is $BK_0K_1$ times the rectangular sum in
\eqref{eq:common-rectangular-tail}. Summing over at most $\Delta_n$
neighbors gives $H$, and the full convergent first-derivative series
gives $T$. It is $R_H=K_P\,2GUH$, not $H$ alone, that must fit below the
tuning and signal budgets. These rates establish verifiable sufficient conditions for recovery, while population conditioning and incoherence are governed by the empirical and approximation bounds developed above.

	\begin{proposition}[A nonempty growing-graph regime]
		\label{prop:pwsm-nonvacuity}
		\label{prop:ercv2-nonvacuity}
		For sufficiently large $n$, set
		\[
		s_n=\lfloor n^{1/10}\rfloor,\qquad p_n=2s_n,\qquad
		d_n=\lfloor n^{1/40}\rfloor,\qquad
		E_n=\{(2k-1,2k):1\le k\le s_n\}.
		\]
		On $[0,1]$, use $L_r(x)=\sqrt2\cos(\pi r x)$ and
		$g(x)=x^2(1-x)^2/4$. Consider the density on $[0,1]^{p_n}$
		\[
		p_{0,n}(x)=\prod_{k=1}^{s_n} Z_n^{-1}
		\exp\!\left\{bL_1(x_{2k-1})+bL_1(x_{2k})
		+\tau_nL_1(x_{2k-1})L_1(x_{2k})\right\},
		\]
		where $b=1/10$, $\tau_n=n^{-1/20}/10$, and $Z_n$ normalizes each
		two-coordinate factor. The canonical edge set
		is the nonempty matching $E_n$, its maximum degree is $\Delta_n=1$, and
		every node has a nonzero main effect. For the exact sieve coefficient
		$\theta_d$, the score approximation error is $H=0$ and the minimum edge
		signal is $\rho_d=\tau_n$ for all $d\ge1$. The sufficient conditions of
		Theorem~\ref{th:pwsm-finite}, equivalently the open-interval conditions
		in Assumption~\ref{ass:ercv2-stability}, hold eventually with
		\[
		\delta_n=n^{-2},\qquad \lambda_n=n^{-1/5}.
		\]
		Consequently, the unstabilized P-WSM estimator exists uniquely and recovers
		$E_n$ with probability at least $1-n^{-2}$ for all sufficiently large $n$.
	\end{proposition}
	
	\begin{proof}
		It suffices to verify the population lower bounds and the three tuning
		ratios. The centered orthonormal cosine basis has derivative exponents
		$\nu_4=1$, $\nu_5=2$; $g$ vanishes at both endpoints and
		$|g'|^2/g=(1-2x)^2\le1$. Since the true potential uses only $L_1$,
		$H=0$ and the coefficient-decay bounds hold for any fixed decay exponents.
		
		Put $K=2\sqrt2/10+2/10$ and $R=e^{4K}$. Each pair potential is bounded
		by $K$, so its normalized density and marginals lie in
		$[e^{-2K},e^{2K}]$. Every one- or two-coordinate marginal, including
		coordinates from different subsystems, therefore lies in $[R^{-1},R]$
		relative to uniform measure. No full-dimensional comparison is needed.
		
		The scalar Hardy argument in Lemma~\ref{r2-lem:curvature}, independent of
		its product-density assumptions, gives
		$\int_0^1g(f')^2\ge\int_0^1 f^2/64$ for every centered finite cosine sum.
		Orthonormality gives $A_i\succeq I/64$ under uniform measure. Applying this
		in each centered coordinate of an interaction and adding gives
		$C_{ee}\succeq I/32$. Local density comparison therefore yields
		\[
		a_0\ge\frac1{64R},\qquad
		\phi_0\ge\frac1{32R},\qquad
		\kappa_0\ge\frac1{32R}.
		\]
		For the Schur bound, minimize the two-node energy over the main effects.
		Their derivative fields are orthogonal to interaction fields under
		uniform measure, so the infimum is at least $\|\gamma_e\|_2^2/32$, and
		the density lower bound survives minimization. Since matching edges have
		disjoint derivative supports and nuisance blocks, $S_{EE}$ is block
		diagonal with the same lower bound independently of $s_n$.
		
		The bound $g\le1/64$ and cosine derivative orthogonality further give
		\[
		\max_i\|A_i\|_2\le\frac{R\pi^2d^2}{64},\qquad
		M_0\le\frac{R\pi^2d^2}{32},\qquad B_0\le R\pi^2d^2.
		\]
		The bound for $B_0$ covers all candidate edges by the Gram inequality
		$\|B_{i,e}\|_2\le\|A_i\|_2^{1/2}\|C_{ee}\|_2^{1/2}$.
		
		Every inactive edge joins independent pair subsystems, so its
		nuisance-projected field is \eqref{eq:pwsm-product-residual}.
		An active-edge residual is supported in its subsystem. Their inner
		product is zero: a shared endpoint leaves an independent centered factor,
		and disjoint derivative supports give zero pointwise. Hence
		\[
		S_{eE}=0\quad(e\notin E),\qquad \mu_0=0.
		\]
		
		Let $\ell_n=\sqrt{\log n/n}$. The feature dimension grows polynomially,
		and valid envelopes in the finite-sample theorem are
		\[
		G=\frac1{64},\quad G_1=\frac18,\quad
		U=2\pi d,\quad V=2\pi^2d^2,\quad
		T\le\sqrt2\pi b+\frac{2\pi}{10},\quad H=0.
		\]
		Consequently, with constants independent of $n$,
		\begin{align*}
			t_\Gamma&=O(d^2\ell_n),& t_w&=O(d^2\ell_n),\\
			e_A&=O(d^3\ell_n),& e_B&=O(d^{7/2}\ell_n),&
			e_\Phi&=O(d^4\ell_n),\\
			e_S&=O(d^7\ell_n),& R_n&=O(d^{9/2}\ell_n).
		\end{align*}
		The largest first-order Schur term is $B_0^2e_A=O(d^7\ell_n)$.
		With the proposed exponents,
		\[
		e_A=O(n^{-17/40}\sqrt{\log n}),\quad
		e_\Phi=O(n^{-2/5}\sqrt{\log n}),\quad
		se_S=O(n^{-9/40}\sqrt{\log n}),
		\]
		so the nuisance, penalty, and active Schur lower bounds remain positive.
		Since $\mu_0=0$, $\chi=O(sd\,e_S)=O(n^{-1/5}\sqrt{\log n})\to0$
		and $1+u_+\sqrt{s}/k_-=1+O(se_S)=1+o(1)$. The remaining ratios are
		\[
		\frac{R_n}{\lambda_n}
		=O(n^{-3/16}\sqrt{\log n}),\qquad
		\frac{\sqrt{s}R_n}{\rho_d}
		=O(n^{-23/80}\sqrt{\log n}),\qquad
		\frac{\sqrt{s}\lambda_n\sqrt{\overline M}}{\rho_d}
		=O(n^{-3/40}).
		\]
		Thus the two deterministic recovery inequalities hold strictly, or
		equivalently $L_\lambda<\lambda_n<U_\lambda$, proving the assertion.
	\end{proof}

\subsection{Proof of Proposition~\ref{prop:pwsm-existence}}
\label{supp:pwsm-existence-proof}

\begin{proof}
	Take \(p=2\), \(d=n=2\), basis \((x,x^2-1/3)\) on \([-1,1]\),
	observations \((0,0)\) and \((a,b)=(1/2,1/4)\), and weight
	\(g(u)=(1-u^2)^2/16\).
	Here \(P_n\) averages over the two observations, and \(X,Y\) denote
	their first and second coordinates.
	The two derivative rows of each main-effect design are \((1,0)\)
	and \((1,2a)\) or \((1,2b)\), so \(\widehat A\succ0\).
	The four interaction features are
	\[
	xy,\qquad x(y^2-1/3),\qquad (x^2-1/3)y,\qquad
	(x^2-1/3)(y^2-1/3).
	\]
	Their four derivative rows, two coordinates at each observation, form
	a square matrix. At \((0,0)\), the first two rows isolate the second
	and third columns with entries \(-1/3\). After eliminating these
	columns, the remaining two-by-two determinant is
	\[
	\det\begin{pmatrix}
		b & 2a(b^2-1/3)\\
		a & 2b(a^2-1/3)
	\end{pmatrix}
	=\frac23(a^2-b^2)=\frac18\ne0.
	\]
	Thus the raw empirical edge Gram
	\(\widehat\Phi=\widehat C_{(1,2),(1,2)}\) is positive definite.
	The main-effect derivative design is already an invertible
	\(4\times4\) matrix spanning all derivative observations, so
	\(\widehat S=0\).
	
	For the interaction direction \(xy\), its unique score-canceling
	nuisance adjustment gives, up to an irrelevant constant,
	\[
	h(x,y)=xy-x^2/4-y^2.
	\]
	Its gradient is zero at both sample points, but its linear
	score-matching term is
	\[
	P_n\sum_i\partial_i(g_i\partial_i h)
	=-P_n\{g(X)/2+2g(Y)\}<0.
	\]
	Consequently, \(\widehat q_{\!P}\ne0\), and the profiled objective is,
	up to an additive constant,
	\(\widehat q_{\!P}^{\top}\gamma+\lambda\|\gamma\|_{\widehat\Phi}\).
	For
	\[
	0<\lambda<\|\widehat q_{\!P}\|_{\widehat\Phi^{-1}}
	=\lambda_{\max}^{\rm P},
	\]
	its value along
	\(\gamma=-t\widehat\Phi^{-1}\widehat q_{\!P}\), as \(t\to\infty\),
	diverges to \(-\infty\).
	All full-rank and nonzero conditions persist on an open neighborhood
	of these observations; \(\widehat S=0\) persists whenever the
	\(4\times4\) main-effect derivative design is invertible.
	Hence a positive continuous sampling density assigns positive
	probability to this nonexistence phenomenon.
\end{proof}

\begingroup
{
	\section{Regularity for identification and optimization}
	\label{supp:new-theory-acceptance}
	We first state regularity conditions for identification and define the
	optimizer outside the recovery event. We then give a P-WSM/Joint-WSM
	comparison and a growing-graph recovery result. Each section states its
	additional assumptions. The results concern exact support without ridge
	regularization at the specified deterministic tuning values.
}
{
	\begin{lemma}[Measurable minimum-norm selection or failure]
		\label{iface-lem:selector}
		Fix finite \(D\), deterministic \(\lambda>0\), and disjoint penalized
		blocks \(G\), permitting unpenalized coordinates and an empty block set.
		Let \(\mathcal T\) be the Borel domain of \(t=(Q,b,\{\Phi_G\})\) with
		\(Q\succeq0\), \(b\in\mathbb R^D\), and \(\Phi_G\succ0\), all matrices
		symmetric. Define
		\[
		f_t(u)=\tfrac12u^\top Qu+b^\top u+
		\lambda\sum_G(u_G^\top\Phi_Gu_G)^{1/2},\qquad u\in\mathbb R^D.
		\]
		For integers \(j\ge1\), the minimizer \(u_j(t)\) of
		\(f_t(u)+\|u\|_2^2/(2j)\) exists uniquely and is continuous on
		\(\mathcal T\). Its finite limit exists exactly when \(f_t\) attains its
		minimum, and is then the unique minimum-Euclidean-norm minimizer.
		Returning this limit, or the isolated symbol \(\dagger\) when it does
		not exist or \(t\notin\mathcal T\), defines a Borel map into
		\(\mathbb R^D\sqcup\{\dagger\}\).
	\end{lemma}
	
	\begin{proof}
		Strong convexity and the lower bound
		\(-\|b\|_2\|u\|_2+\|u\|_2^2/(2j)\) give a unique minimizer with
		\(\|u_j(t)\|_2\le2j\|b\|_2\). For fixed \(j\), this locally uniform
		bound, uniform convergence of the objectives on compact sets, and
		uniqueness prove continuity.
		
		If \(\operatorname*{arg\,min}f_t\ne\varnothing\), this closed convex set
		has a unique minimum-norm member \(v\). Optimality gives
		\[
		f_t(u_j)+\frac{\|u_j\|_2^2}{2j}
		\le f_t(v)+\frac{\|v\|_2^2}{2j}.
		\]
		Thus \(\|u_j\|_2\le\|v\|_2\) and
		\(0\le f_t(u_j)-f_t(v)\le\|v\|_2^2/(2j)\), making every cluster point
		equal to \(v\). Conversely, a finite limit \(u\) satisfies
		\(f_t(u)\le f_t(z)\) for every fixed \(z\), by the same inequality.
		
		The finite-limit set is Borel:
		\(\bigcap_{m\ge1}\bigcup_{N\ge1}\bigcap_{j,k\ge N}
		\{t:\|u_j(t)-u_k(t)\|_2<1/m\}\).
		Coordinatewise limits there and extension by \(\dagger\) are Borel.
	\end{proof}
	
	\begin{remark}[Uniform spectra and unscaled trigonometric blocks]
		\label{iface-rem:trig-spectrum}
		For unscaled, equally normalized sine/cosine pairs, \(\sin^2+\cos^2=1\)
		makes the summed squared derivatives at frequency \(r\) equal \(r^2\)
		times the frequency-one sum. The four interaction tensors at \((r,r)\)
		have the same property, also after weighting and integration.
		If the active Hessian has minimum eigenvalue at least \(c>0\), its
		frequency-one principal trace is at least \(mc\), where \(m=2\) for a
		main effect and \(m=4\) for an interaction. The frequency-\(r\) trace is
		then at least \(mcr^2\), so the full maximum eigenvalue is at least
		\(cr^2\). Growing complete blocks with a nonempty active set therefore
		cannot have constant upper and lower spectral bounds simultaneously.
		This limits the conditional Joint corollary's applicability, not its
		implication or the numerical basis.
	\end{remark}
}

\newcommand{\E}{\mathbb E}
\newcommand{\R}{\mathbb R}
\newcommand{\op}{\mathrm{op}}
\newcommand{\supp}{\operatorname{supp}}
\newcommand{\osc}{\operatorname{osc}}
\newcommand{\diag}{\operatorname{diag}}
\newcommand{\norm}[1]{\left\lVert#1\right\rVert}
\newcommand{\argmin}{\operatorname*{arg\,min}}

{
	
	\section{Comparison of partial and joint penalization}\label{supp:comparison}
	
	We compare exact support recovery by P-WSM and Joint-WSM over a common
	deterministic tuning interval.
	
	\subsection{Density and estimators}
	
	Let $L_r(x)=\sqrt2\cos(\pi r x)$, $r\ge1$, and
	$g(x)=x^2(1-x)^2/4$ on $[0,1]$. Define
	\begin{align}
		Z(a)&=\int_0^1e^{aL_1(x)}\,dx,&
		q(x)&=e^{2L_1(x)}/Z(2),& Q&=q^{\otimes3},\label{r2-eq:q}\\
		h(x,y)&=L_1(x)L_1(y)+\sum_{r=2}^\infty2^{-r}L_r(x)L_r(y),&
		F(x)&=h(x_1,x_2)+h(x_2,x_3).\label{r2-eq:h}
	\end{align}
	The fixed true distribution is
	\begin{equation}\label{r2-eq:true}
		\frac{dP_\tau}{dQ}(x)=
		\frac{\exp\{\tau F(x)\}}{\E_Q\exp\{\tau F(X)\}},\qquad \tau>0.
	\end{equation}
	Put $\ell=e^{-4\sqrt2}$, $u=e^{4\sqrt2}$,
	$c_0=\ell/u$, $c=c_0/7$, and $\eta_0=c_0/64000$.
	Assume
	\begin{equation}\label{r2-eq:tau}
		0<\tau\le\frac1{12}\log(1+\eta_0/12).
	\end{equation}
	The positive $\tau$ is fixed in $n,d$; subsequent constants may depend on it.
	
	Use $L=(L_1,\ldots,L_d)^\top$ at each node and $L(x_i)\otimes L(x_j)$
	at every pair,
	ordered $(r-1)d+s$ within each pair; thus $D=3d+3d^2$ and
	$\eta_\theta=\theta^\top\psi$. For iid data from $P_\tau$, write
	$\widehat M=M_n$, $\widehat\Gamma=\widehat\Gamma_n$, and
	$\widehat\rho=\widehat\rho_n$ as in \eqref{def:loss_function}.
	The estimators are exactly P-WSM \eqref{def:pwsm} and Joint-WSM
	\eqref{def:loss_function_with_penalization}, without ridge or coefficient
	constraints. Both use the raw empirical blocks
	$\widehat A_i,\widehat\Phi_e=\widehat C_{ee}$ of
	\eqref{eq:block_score_gram}, and Joint uses the same $\lambda$ for all
	groups. Support means exact nonzero edge coefficients. Subscripts $R$
	denote population moments; unqualified norms are Euclidean or spectral.
	
	\begin{theorem}[A common-interval support separation]\label{r2-th:separation}
		The density \eqref{r2-eq:true}--\eqref{r2-eq:tau} has conditional-independence
		graph $E=\{12,23\}$ and infinitely many nonzero coefficients on both edges.
		Let $d_n=\lceil\log n\rceil$ and $I_n=[n^{-1/4},n^{-1/8}]$.
		For all sufficiently large $n$, one event of probability at least $1-n^{-2}$
		has positive definite full and raw-block empirical Grams and, simultaneously
		for every $\lambda\in I_n$, attained unique optima satisfying
		$\supp(\widehat\gamma^P_\lambda)=E$ and
		$\widehat\gamma^J_{\lambda,13}\ne0$. The Joint-WSM conclusion follows from a violation of necessary KKT dual feasibility on edge $13$. The threshold in $n$ depends only on the fixed density and weight;
		logarithms are natural.
	\end{theorem}
	
	\subsection{Product geometry and a scalar witness}
	
	All fixed-order derivatives of \eqref{r2-eq:h} converge uniformly, since
	$\sum r^k2^{-r}<\infty$ for each fixed $k$. In particular $P_\tau$ is a
	positive smooth density. Since $\norm{h}_\infty\le3$ and
	$\norm{F}_\infty\le6$,
	\begin{equation}\label{r2-eq:densityratio}
		\ell\le q\le u,\qquad
		e^{-12\tau}\le dP_\tau/dQ\le e^{12\tau}.
	\end{equation}
	The canonical mains are $2L_1$ and the only interactions are $\tau h$
	on 12 and 23. The density factorizes in $x_1,x_3$ given $x_2$;
	the first nonzero sine--sine coefficient of each true edge's mixed
	derivative excludes either remaining conditional independence.
	
	Write $A_{0,i}=\E_Q[g_i L_i'L_i'^\top]$, $\Phi_{0,e}=C_{Q,ee}$, and
	\begin{equation}\label{r2-eq:d0}
		D_0=\diag(A_{0,1},A_{0,2},A_{0,3},\Phi_{0,12},\Phi_{0,13},\Phi_{0,23}).
	\end{equation}
	Put $A_0=\diag_iA_{0,i}$. These deterministic metrics are used only to
	express the proof in normalized coordinates.
	
	\begin{lemma}[Uniform reference coercivity]\label{r2-lem:curvature}
		For every $d$,
		\[
		D_0\succeq k_D I,\quad k_D=\min(\ell/64,\ell^2/32)>0,
		\qquad cI\preceq G_0:=D_0^{-1/2}\Gamma_QD_0^{-1/2}\preceq6I.
		\]
		In the main/edge partition $G_0=[I,B_0;B_0^\top,C_0]$,
		$\norm{B_{0,e}}_\op\le1$, $\norm{B_0}_\op\le\sqrt3$, and the Schur
		complement $S_0$ is edge-block diagonal with $S_{0,ee}\succeq c_0I$.
	\end{lemma}
	\begin{proof}
		We bound the raw metrics and profiled Schur complement, then complete
		the square to obtain full coercivity.
		
		For a smooth mean-zero $f$, set $H(x)=\int_0^x f(t)\,dt$.
		Then $H(0)=H(1)=0$. Integration by parts in $\int H^2/x^2$ followed by
		Cauchy--Schwarz gives $\int H^2/x^2\le4\int f^2$; the endpoint term
		$H(x)^2/x$ tends to zero at 0. Reflection gives the corresponding bound
		with $(1-x)^{-2}$. Since $1/g\le8\{x^{-2}+(1-x)^{-2}\}$,
		\[
		\int H^2/g\le64\int f^2,\qquad
		\int f^2=-\int Hf'\le(64\int f^2\int g(f')^2)^{1/2}.
		\]
		Consequently $\int g(f')^2\ge\int f^2/64$. The cosine basis is
		Lebesgue-centered and orthonormal, so $A_{0,i}\succeq\ell I/64$.
		Writing $W=\E_q LL^\top$, we have $W\succeq\ell I$ and
		$\Phi_{0,e}=A_{0,i}\otimes W+W\otimes A_{0,j}\succeq\ell^2I/32$.
		For every centered cosine combination $f$,
		\[
		\operatorname{Var}_q(f)=\inf_b\E_q(f-b)^2
		\ge\ell\inf_b\int(f-b)^2=\ell\int f^2
		\ge c_0\E_q f^2.
		\]
		Proposition~\ref{prop:pwsm-product} gives zero off-diagonal Schur blocks
		and raw diagonal
		$A_{0,i}\otimes\operatorname{Cov}_q(L)+
		\operatorname{Cov}_q(L)\otimes A_{0,j}$.
		The variance bound therefore gives $S_{0,ee}\succeq c_0I$.
		The main projection of each normalized edge field is a contraction,
		giving $\norm{B_{0,e}}\le1$ and $\norm{B_0}\le\sqrt3$.
		All six diagonal group blocks of $G_0$ are $I$; its Gram representation
		and Cauchy--Schwarz give $G_0\preceq6I$. For the full lower bound, complete
		the square and compare the resulting coordinates with the original ones:
		\[
		(x,y)^\top G_0(x,y)=\norm{x+B_0y}^2+y^\top S_0y,
		\quad \norm{x}^2+\norm{y}^2\le2\norm{x+B_0y}^2+7\norm{y}^2.
		\]
		As $c_0\le1$, this proves the lower bound $c_0/7$.
	\end{proof}
	
	Put $A_{11}=\E_q[g(L_1')^2]$, $\mu=\E_qL_1$, and $\nu=\E_qL_1^2$.
	\begin{lemma}[A strict product witness]\label{r2-lem:witness}
		For every $d\ge1$, $R_0=\sqrt2\mu/\sqrt\nu>9/8$.
		Let $v=e_1\otimes e_1$ in the original missing-edge block and
		$v_0=\Phi_{0,13}^{1/2}v/\norm{v}_{\Phi_{0,13}}$.
		If $z_{\beta,0}$ denotes the concatenated normalized directions of
		$A_{0,i}^{1/2}(2e_1)$, then
		$v_0^\top B_{0,13}^\top z_{\beta,0}=R_0$.
	\end{lemma}
	\begin{proof}
		Termwise integration of the absolutely convergent exponential series gives
		$Z(a)=\sum_{k\ge0}(a^2/2)^k/(k!)^2$; odd cosine moments vanish and
		$\int\cos^{2k}(\pi x)\,dx=(2k)!/\{2^{2k}(k!)^2\}$.
		Thus $Z'(2)=\sum_{k\ge0}k2^k/(k!)^2$. The sum through $k=5$ of
		$8Z'(2)-9Z(2)=\sum_{k\ge0}(8k-9)2^k/(k!)^2$ is $37/900>0$;
		all later terms are positive. Hence $\mu=Z'(2)/Z(2)>9/8$,
		while $\nu\le2$, proving the first claim.
		The raw main covectors are
		$z_{\beta,0,i}^{\rm raw}=A_{0,i}e_1/\sqrt{A_{11}}$.
		The two endpoints contribute $2\mu\sqrt{A_{11}}$ to
		$v^\top B_{Q,13}^\top A_0^{-1}z_{\beta,0}^{\rm raw}$, whereas
		$\|v\|_{\Phi_{0,13}}=\sqrt{2A_{11}\nu}$.
		Transforming covectors by $A_0^{-1/2}$ and the edge by
		$\Phi_{0,13}^{1/2}$ gives $R_0$, independently of higher sieve coordinates.
	\end{proof}
	
	\subsection{KKT conditions for the empirical penalties}
	
	It suffices to show that the minimizer restricted to $\gamma_{13}=0$
	satisfies the full KKT system for P-WSM but violates the missing-edge KKT condition for Joint-WSM.
	
	Set $t=D_0^{1/2}\theta$,
	$G=D_0^{-1/2}\widehat\Gamma D_0^{-1/2}=[A,B;B^\top,C]$, and
	$S=C-B^\top A^{-1}B$. Assume $\|G-G_0\|\le\eta\le\eta_0$.
	Then $G\succeq cI/2$, and each actual penalty is $\|t_g\|_{K_g}$
	with $K_g=G_{gg}$ and $\|K_g-I\|\le\eta$.
	The following bounds control its KKT equations:
	\begin{align}
		\norm{A^{-1}}&\le2,& \norm{A^{-1}-I}&\le2\eta,
		&\norm{S-S_0}&\le20\eta,\label{r2-eq:perturb1}\\
		\norm{S_{13,E}}&\le20\eta,& \norm{S_{EE}^{-1}}&\le2/c_0,
		&\norm{B_E^\top A^{-1}}&\le6,\label{r2-eq:perturb2}\\
		\norm{B_{13}^\top A^{-1}-B_{0,13}^\top}&\le4\eta.
		&&&&\label{r2-eq:perturb3}
	\end{align}
	The Schur difference is bounded by $\norm{C-C_0}$ and the terms in
	\begin{equation}\label{r2-eq:schurexp}
		\begin{aligned}
			B^\top A^{-1}B-B_0^\top B_0
			={}&(B-B_0)^\top A^{-1}B\\
			&+B_0^\top(A^{-1}-I)B+B_0^\top(B-B_0).
		\end{aligned}
	\end{equation}
	Here $\norm{B}\le\sqrt3+\eta<2$; the resulting bound is less than
	$20\eta$. As $20\eta<c_0/2$, the active inverse bound follows.
	The other inequalities use $\norm{B_{0,13}}\le1$.
	
	Let $\theta_d$ be the literal truncation: its mains are $2e_1$, its
	two true-edge diagonals are $\tau(1,2^{-2},\ldots,2^{-d})$, and
	$\gamma_{d,13}=0$. Put $t^*=D_0^{1/2}\theta_d$,
	$w=D_0^{-1/2}(\widehat\Gamma\theta_d+\widehat\rho)$,
	$W=\|w\|$, and $r=w_\gamma-B^\top A^{-1}w_\beta$.
	Thus $\|r\|\le7W$; no finite-sieve population optimality of $\theta_d$
	is assumed.
	
	For either procedure restrict to $\gamma_{13}=0$, with unique optimum
	$\check t$. Every $K_g$-norm subgradient has Euclidean norm at most
	$\sqrt{1+\eta}$, including at zero. Restricted stationarity, the inverse
	Gram bound $2/c$, and at most six penalized groups give
	\begin{equation}\label{r2-eq:restrictederror}
		\|\check t-t^*\|\le R(\lambda):=2(W+3\lambda)/c.
	\end{equation}
	The true main norms are $b=2\sqrt{A_{11}}$, and true-edge norms are at
	least $\sigma=\tau\ell/\sqrt{32}$. Hence $R(\lambda)<\min(b,\sigma)$
	retains all these groups. For nonzero $t,t^*$ and $\|K-I\|\le\eta<1/2$,
	\[
	\left\|\frac{Kt}{\sqrt{t^\top Kt}}-\frac{t^*}{\|t^*\|}\right\|
	\le4\eta+2\|t-t^*\|/\|t^*\|.
	\]
	This follows by inserting $t/\|t\|$ and bounding the two differences.
	Applied to Joint's main covectors, it gives
	\begin{equation}\label{r2-eq:maincontinuity}
		\|z_\beta-z_{\beta,0}\|\le7\eta+4R(\lambda)/b,\qquad
		\|z_\beta\|<3,\quad\|z_E\|<2.
	\end{equation}
	Only the main directions need this continuity bound.
	
	\paragraph{Deterministic separation.}
	Suppose $R(\lambda)<\min(b,\sigma)$ and
	\begin{align}
		\frac{80\eta/c_0+14W/\lambda}{\sqrt{1-\eta}}&<1,
		\label{r2-eq:psuccess}\\
		1000\eta/c_0+4R(\lambda)/b+14W/\lambda
		&<R_0-\sqrt{1+\eta}.\label{r2-eq:jfail}
	\end{align}
	We show that these conditions give P recovery and a nonzero Joint edge 13.
	The error bound already retains both true edges; only missing-edge KKT remains.
	
	\emph{P-WSM.}
	Eliminating main stationarity at the restricted P solution, its required
	missing-edge covector is exactly
	\begin{equation}\label{r2-eq:pdual}
		z_{13}^P=S_{13,E}S_{EE}^{-1}z_E+
		(S_{13,E}S_{EE}^{-1}r_E-r_{13})/\lambda.
	\end{equation}
	It has norm at most $80\eta/c_0+14W/\lambda$ by
	\eqref{r2-eq:perturb1}--\eqref{r2-eq:perturb2}. Since $K_{13}\succeq(1-\eta)I$,
	\eqref{r2-eq:psuccess} makes its actual dual norm
	$\norm{z_{13}^P}_{K_{13}^{-1}}<1$. All full KKT conditions now hold.
	Positive definiteness makes this the unique full optimizer.
	
	\emph{Joint-WSM.}
	Joint's main stationarity is
	$A(\check t_\beta-t^*_\beta)+B_E(\check t_E-t_E^*)+
	w_\beta+\lambda z_\beta=0$.
	Substitution into edge stationarity yields the exact required covector
	\begin{align}\label{r2-eq:jdual}
		z_{13}^J={}&S_{13,E}S_{EE}^{-1}z_E+
		(B_{13}^\top-S_{13,E}S_{EE}^{-1}B_E^\top)A^{-1}z_\beta\\
		&\quad +(S_{13,E}S_{EE}^{-1}r_E-r_{13})/\lambda.\nonumber
	\end{align}
	Here the active-edge term has norm at most $80\eta/c_0$, and the indirect Schur--main correction
	has norm at most $(20\eta)(2/c_0)6\cdot3=720\eta/c_0$.
	The direct main-operator perturbation contributes at most $12\eta$, while its
	direction perturbation contributes at most $7\eta+4R(\lambda)/b$.
	Lemma \ref{r2-lem:witness} and the residual bound therefore imply
	\begin{equation}\label{r2-eq:witnesserror}
		|v_0^\top z_{13}^J-R_0|
		\le1000\eta/c_0+4R(\lambda)/b+14W/\lambda.
	\end{equation}
	At a zero edge, dual feasibility requires
	$|v_0^\top z_{13}|\le\sqrt{v_0^\top K_{13}v_0}\le\sqrt{1+\eta}$,
	contradicting \eqref{r2-eq:jfail}. A full optimizer with edge 13 zero
	would equal the unique restricted optimizer, so none exists.
	
	\subsection{Simultaneous empirical and approximation bounds}
	
	\begin{proof}[Proof of Theorem \ref{r2-th:separation}]
		First, we establish an event of probability at least $1-n^{-2}$,
		independent of $\lambda$, on which, for $d=d_n$ and fixed $C<\infty$,
		\begin{align}
			\norm{D_0^{-1/2}(\widehat\Gamma-\Gamma_{P_\tau})D_0^{-1/2}}
			&\le\epsilon_n=C d^4\sqrt{\log n/n},\label{r2-eq:gramrate}\\
			W&\le W_n=C\{d^3\sqrt{\log n/n}+d2^{-d}\}.
			\label{r2-eq:noiserate}
		\end{align}
		The feature derivative bounds are $Cd,Cd^2$. Because
		$\sum r2^{-r}<\infty$, the truncated score $s_d$ is uniformly bounded
		and $\|s_d-s_0\|_\infty\le Cd2^{-d}$. Thus every Gram entry and residual
		coordinate
		$\xi_a=\sum_i\{g_i\partial_i\psi_a s_{d,i}
		+g_i'\partial_i\psi_a+g_i\partial_i^2\psi_a\}$
		is bounded by $Cd^2$.
		Hoeffding and a union bound over $D^2+D$ coordinates, with failure
		probability $n^{-2}$, give deviations $Cd^2\sqrt{\log n/n}$.
		Multiplying by $D=O(d^2)$ for the matrix norm and $\sqrt D$ for the
		residual norm, then using $D_0\succeq k_DI$, proves the stochastic bounds.
		
		For the bias, integration by parts is valid because $g$ vanishes at both
		endpoints and all finite-sieve derivatives are bounded. It yields
		$\E\xi_a=\E\sum_i g_i\partial_i\psi_a(s_{d,i}-s_{0,i})$.
		For $\|v\|=1$, let $f_v=v^\top D_0^{-1/2}\psi$. Cauchy--Schwarz gives
		\[
		\begin{aligned}
			|v^\top D_0^{-1/2}\E\xi|
			&\le
			\left(\E\sum_i g_i(\partial_i f_v)^2\right)^{1/2}
			\left(\E\sum_i g_i(s_{d,i}-s_{0,i})^2\right)^{1/2}\\
			&\le\sqrt{6e^{12\tau}}\,Cd2^{-d},
		\end{aligned}
		\]
		by \eqref{r2-eq:densityratio} and Lemma~\ref{r2-lem:curvature}.
		Taking the supremum proves the bias bound without a feature-count factor.
		
		It remains to verify the deterministic separation conditions on this event.
		Density comparison gives
		$\|D_0^{-1/2}(\Gamma_{P_\tau}-\Gamma_Q)D_0^{-1/2}\|
		\le6(e^{12\tau}-1)\le\eta_0/2$.
		Since $\epsilon_n\to0$, eventually $\|G-G_0\|\le\eta_0$, proving
		positive definiteness and unique attained optima.
		Uniformly for $\lambda\in I_n$,
		\[
		W/\lambda\le W_n n^{1/4}\to0,\qquad
		R(\lambda)\le2(W_n+3n^{-1/8})/c\to0.
		\]
		Indeed the first bound has terms
		$C(\log n)^{7/2}n^{-1/4}$ and $C(\log n)n^{1/4-\log2}$.
		Thus $R<\min(b,\sigma)$ eventually. The limiting bounds in
		\eqref{r2-eq:psuccess} and \eqref{r2-eq:jfail} are respectively
		$\{80/64000\}/\sqrt{1-\eta_0}<1$ and $1/64$; the latter is smaller than
		$R_0-\sqrt{1+\eta_0}>9/8-1-1/64000$.
		Thus the deterministic separation holds throughout $I_n$.
	\end{proof}
	
}

{
	
	\section{Graph recovery under local weak interactions}\label{supp:local-recovery}
	
	To prove recovery on a growing graph, it suffices to control the profiled
	inverse, propagation to inactive edges, and the profiled residual on one
	sample event. We derive these bounds from local density conditions, then
	verify them for a connected path with infinite interaction expansions.
	
	\subsection{P-WSM under local density conditions}
	
	Let $p\ge2$, $d\ge1$, $\mathcal M=\{(i,j):i<j\}$, and
	$D=pd+\binom p2d^2$. Use the cosine basis and weight of
	Section~\ref{supp:comparison} on $[0,1]^p$, with all candidate pairs.
	We retain the iid empirical loss and block notation of
	\eqref{def:loss_function}, and study P-WSM \eqref{def:pwsm} with
	$\vartheta=(\beta,\gamma)$, $\widehat\Phi_e=\widehat C_{ee}$.
	No stabilization or sample splitting is used.
	
	\begin{assumption}[Local weak interactions]\label{r1-ass:density}
		For positive $C^2$ reference densities $q_i$ and constants
		$0<\ell\le u<\infty$, $\Delta\ge1$, $v\ge0$,
		\begin{equation}\label{r1-eq:density}
			P(dx)=Z^{-1}Q(dx)\exp\!\left\{\sum_{e\in E}h_e(x_e)\right\},
			\qquad Q=\bigotimes_{i=1}^p q_i,\qquad
			\ell\le q_i\le u.
		\end{equation}
		The graph has maximum degree $\Delta$ or less. Each $h_e$ is nonzero,
		$C^2$, centered separately in each Lebesgue coordinate, and has
		$\osc(h_e)\le v$. Integration by parts is valid, and constants are uniform
		over the sequence of $p,d,n$. The $q_i$ need not be marginals of $P$.
	\end{assumption}
	
	\begin{assumption}[Visible literal truncation and score envelopes]\label{r1-ass:tail}
		The deterministic literal canonical truncation $\vartheta_d$ of $\log P$
		has edge support exactly $E$. Write $s_{d,i}=\partial_i(\vartheta_d^\top\psi)$,
		$s_{0,i}=\partial_i\log P$. For every $i,a$, deterministic envelopes satisfy
		\begin{equation}\label{r1-eq:envelopes}
			\|s_{d,i}\|_\infty\le T,\quad
			\|s_{d,i}-s_{0,i}\|_\infty\le H,\quad
			0\le g\le G,\quad |g'|\le G_1,\quad
			|\partial_i\psi_a|\le U,\quad |\partial_i^2\psi_a|\le V.
		\end{equation}
		The envelopes may depend on $d,n$, and finite-sieve boundary terms vanish.
		Small oscillation alone does not imply these derivative bounds.
	\end{assumption}
	
	Only $h_{ij}$ contributes to the mixed $ij$ derivative; its centering
	makes that derivative nonzero whenever $h_{ij}\ne0$. Positivity therefore
	identifies $E$ with the conditional-independence graph.
	
	Write $\|x\|_{2,\infty}=\max_e\|x_e\|_2$.
	Reference blocks are $A_{0,i}=A_{Q,i}$ and $\Phi_{0,e}=C_{Q,ee}$, with
	block diagonals $A_0,\Phi_0$. Their lower bounds below use
	\begin{equation}\label{r1-eq:reference}
		a_0=\ell/64,\qquad \phi_0=\ell^2/32,\qquad c_0=\ell/u.
	\end{equation}
	
	\subsection{Geometry derived from the density}
	
	\begin{lemma}[Local and product geometry]
		\label{r1-lem:local}\label{r1-lem:product}\label{r1-lem:population}
		Under Assumption~\ref{r1-ass:density}, a $j$-coordinate marginal satisfies
		\begin{equation}\label{r1-eq:local}
			e^{-j\Delta v}\le dP_J/dQ_J\le e^{j\Delta v},\qquad
			\epsilon_j:=e^{j\Delta v}-1.
		\end{equation}
		The reference blocks satisfy $A_{0,i}\succeq a_0I$,
		$\Phi_{0,e}\succeq\phi_0I$. The product Schur complement is edge-block
		diagonal, with
		\begin{equation}\label{r1-eq:product}
			S_{Q,ee}\succeq c_0\Phi_{0,e},\qquad
			\|A_{0,i}^{-1/2}B_{Q,i,e}\Phi_{0,e}^{-1/2}\|\le1\quad(i\in e).
		\end{equation}
		After normalization by these reference inverse square roots, the
		$P$-versus-$Q$ errors for $A_i$, $B_{i,e}$, $C_{ee}$, and distinct
		incident $C_{ef}$ are at most $\epsilon_1,\epsilon_2,\epsilon_2,\epsilon_3$,
		respectively. Disjoint-edge and nonincident main--edge blocks vanish
		pointwise, hence also in the sample.
	\end{lemma}
	\begin{proof}
		\emph{Local comparison.}
		Conditional on $x_{-J}$, at most $j\Delta$ edge potentials depend on $x_J$.
		Their total oscillation is at most $j\Delta v$, giving the conditional
		ratio bounds in \eqref{r1-eq:local}; integrate against $P_{-J}$.
		
		\emph{Product geometry.}
		Lemma~\ref{r2-lem:curvature}'s scalar Hardy bound is
		\begin{equation}\label{r1-eq:hardy}
			\int_0^1g(f')^2\ge\frac1{64}\int_0^1 f^2
		\end{equation}
		for every centered finite cosine combination.
		For $W_i=\E_{q_i}LL^\top$, $V_i=\operatorname{Cov}_{q_i}(L)$,
		density comparison gives $W_i\succeq\ell I$,
		$A_{0,i}\succeq\ell I/64$, and $V_i\succeq c_0W_i$.
		The product projection of Proposition~\ref{prop:pwsm-product} gives
		$\Phi_{0,ij}=A_{0,i}\otimes W_j+W_i\otimes A_{0,j}$ and
		$S_{Q,ij,ij}=A_{0,i}\otimes V_j+V_i\otimes A_{0,j}$, with zero off-diagonal
		Schur blocks. These identities prove the raw and Schur lower bounds.
		Cauchy--Schwarz between a unit-energy main field and an endpoint part
		of a unit-energy edge field proves the per-endpoint bound.
		
		\emph{Perturbations.}
		Each energy product uses one, two, or three coordinates. Its absolute
		$Q$ expectation in normalized unit directions is at most one by
		Cauchy--Schwarz. Multiplying by the local density-ratio error gives the
		corresponding $\epsilon_j$ bound. Disjoint derivative supports give the zeros.
	\end{proof}
	
	\subsection{A single event and actual empirical profiling}
	
	For $0<\delta<1$, take $t_\Gamma,t_w$ from
	Assumption~\ref{ass:pwsm-edge}(ii), replacing $\delta_n$ by $\delta$;
	only these two definitions are imported. Normalize their block errors as
	\begin{align}\label{r1-eq:budgets}
		e_A&=dt_\Gamma/a_0,\quad
		e_B=d^{3/2}t_\Gamma/\sqrt{a_0\phi_0},\quad
		e_C=d^2t_\Gamma/\phi_0,\nonumber\\
		\eta&=\epsilon_3+\max(e_A,e_B,e_C).
	\end{align}
	\begin{lemma}[Common event]\label{r1-lem:event}
		The event $\mathcal H$ of \eqref{eq:pwsm-common-event} has probability
		at least $1-\delta$, independently of $\lambda$. On it, the empirical
		reference-normalized block errors relative to $Q$ are at most $\eta$.
	\end{lemma}
	\begin{proof}
		The iid sampling, finite-sieve integration by parts, and six envelopes
		required for \eqref{eq:pwsm-common-event} are precisely those imposed here.
		In particular, for $\xi_a=Y_a(\vartheta_d)$ in \eqref{eq:common-scalar-score},
		the same identity retains the bias
		\begin{equation}\label{r1-eq:bias}
			\E_P\xi_a=\E_P\sum_i g_i\partial_i\psi_a(s_{d,i}-s_{0,i}),
			\qquad |\E_P\xi_a|\le2GUH.
		\end{equation}
		The event's $\delta/2$ budgets for $D^2$ Gram and $D$ residual entries
		give the chosen $t_\Gamma,t_w$. Operator errors for blocks of sizes
		$d\times d$, $d\times d^2$, $d^2\times d^2$ are at most
		$d,d^{3/2},d^2$ times the entrywise error. Normalizing by the reference
		lower bounds and adding Lemma~\ref{r1-lem:population}'s local error proves
		the claim.
	\end{proof}
	
	Bars henceforth denote empirical blocks multiplied on each side by the
	corresponding reference inverse square root. If $\eta\le1/10$,
	$\|\bar A_i-I\|\le\eta$ implies $\widehat A\succ0$, and
	\begin{equation}\label{r1-eq:schur}
		\bar S=\bar C-\bar B^\top\bar A^{-1}\bar B,\qquad K_e=\bar C_{ee}.
	\end{equation}
	For $x=\Phi_0^{1/2}\gamma$, the actual penalty is
	$\sum_e\|x_e\|_{K_e}$ with $(1-\eta)I\preceq K_e\preceq(1+\eta)I$.
	
	\begin{lemma}[Schur localization and inverse]\label{r1-lem:schur}
		Suppose $\eta\le1/10$ and put
		\begin{equation}\label{r1-eq:kappa}
			k=c_0-(10\Delta-2)\eta,\qquad \kappa=10\Delta\eta/k.
		\end{equation}
		If $k>0$ and $E$ is nonempty, then
		\begin{align}\label{r1-eq:inverse}
			\bar S_{EE}&\succeq kI,\qquad
			\|\bar S_{EE}^{-1}\|_{2,\infty\to2,\infty}\le1/k,\\
			\|\bar S_{eE}\bar S_{EE}^{-1}\|_{2,\infty\to2}
			&\le\kappa\qquad(e\notin E).\nonumber
		\end{align}
	\end{lemma}
	\begin{proof}
		Apply \eqref{r2-eq:schurexp} at a common endpoint, where the reference
		main--edge block has norm at most one. Since
		$\|\bar A_i^{-1}\|\le(1-\eta)^{-1}$ and
		$\|\bar A_i^{-1}-I\|\le\eta/(1-\eta)$, the three terms contribute
		$\eta(1+\eta)/(1-\eta)$, $\eta(1+\eta)/(1-\eta)$, and $\eta$.
		Their sum is $J(\eta)=(3\eta+\eta^2)/(1-\eta)\le31\eta/9$.
		Disjoint edges have no common endpoint, distinct incident edges have one,
		and diagonal blocks have two. Product Schur orthogonality yields
		\begin{equation}\label{r1-eq:schurblocks}
			\|\bar S_{ef}\|\le\eta+J(\eta)<5\eta\quad(e\ne f,\ e\cap f\ne\varnothing),
			\qquad \bar S_{ee}\succeq(c_0-8\eta)I.
		\end{equation}
		An active row has at most $2\Delta-2$ off-diagonal blocks.
		Writing $\bar S_{EE}=J_E+R_E$, with $J_E$ block diagonal and
		$\alpha=c_0-8\eta$, gives
		$\|J_E^{-1}R_E\|_{2,\infty\to2,\infty}\le(2\Delta-2)5\eta/\alpha<1$.
		Its Neumann series bounds the inverse by
		$\{\alpha-(2\Delta-2)5\eta\}^{-1}=1/k$.
		Symmetry and $2ab\le a^2+b^2$ bound the off-diagonal quadratic form by
		$(2\Delta-2)5\eta\sum_{e\in E}\|x_e\|^2$, giving the same lower eigenvalue $k$.
		An inactive edge has at most $2\Delta$ active neighbors, giving row norm
		$10\Delta\eta$ and hence the propagation bound $\kappa$, without
		$\sqrt{|E|}$.
	\end{proof}
	
	\paragraph{Profiled residual.}
	Let $w^{\rm raw}=\widehat\Gamma\vartheta_d+\widehat\rho$,
	$w_{\beta,i}=A_{0,i}^{-1/2}w^{\rm raw}_{\beta,i}$, and
	$w_{\gamma,e}=\Phi_{0,e}^{-1/2}w^{\rm raw}_{\gamma,e}$.
	The exact identity \eqref{eq:pwsm-common-residual} becomes
	$r_e=w_{\gamma,e}-\sum_{i\in e}\bar B_{i,e}^\top\bar A_i^{-1}w_{\beta,i}$.
	On $\mathcal H$, the normalized residual blocks are bounded by
	$\sqrt d\,t_w/\sqrt{a_0}$ and $dt_w/\sqrt{\phi_0}$.
	For $\eta\le1/10$, each of the two endpoint operators has norm at most
	$(1+\eta)/(1-\eta)$, so
	\begin{equation}\label{r1-eq:R}
		\max_e\|r_e\|\le R_n:=
		\left\{\frac d{\sqrt{\phi_0}}
		+\frac{2(1+\eta)\sqrt d}{(1-\eta)\sqrt{a_0}}\right\}t_w.
	\end{equation}
	This is a same-sample bound and includes the bias $2GUH$.
	
	\subsection{Recovery despite a potentially singular full Gram}
	
	\begin{theorem}[Finite-sample, degree-localized recovery]\label{r1-thm:general}
		Under Assumptions~\ref{r1-ass:density}--\ref{r1-ass:tail}, let $E$ be nonempty,
		$\eta\le1/10$, and $k>0$, with the budgets above. Define
		$\rho_{\rm ref}=\min_{e\in E}\|\Phi_{0,e}^{1/2}\gamma_{d,e}\|$.
		For any $\lambda>0$ such that
		\begin{equation}\label{r1-eq:signal}
			\rho_{\rm ref}>\frac{R_n+\lambda\sqrt{1+\eta}}k
		\end{equation}
		and, when inactive candidates exist,
		\begin{equation}\label{r1-eq:dual}
			\frac{\kappa\sqrt{1+\eta}+(1+\kappa)R_n/\lambda}{\sqrt{1-\eta}}<1,
		\end{equation}
		the original estimator \eqref{def:pwsm} has an attained unique full
		parameter minimizer on $\mathcal H$, and its edge support is exactly $E$.
		Hence the conclusion has probability at least $1-\delta$. The same event
		works simultaneously for every $\lambda$ satisfying these deterministic
		conditions. Full empirical Gram positive definiteness is not required.
	\end{theorem}
	\begin{proof}
		We use the restricted KKT identities
		\eqref{eq:pwsm-restricted-error}--\eqref{eq:pwsm-inactive-dual}, but
		bound them in block-maximum rather than Euclidean norm.
		
		\emph{Active retention.}
		Profiling by \eqref{eq:pwsm-profile} and setting
		$x=\Phi_0^{1/2}\gamma$ gives Hessian $\bar S\succeq0$, residual $r$ at
		$x_d=\Phi_0^{1/2}\gamma_d$, and penalty $\lambda\sum_e\|x_e\|_{K_e}$.
		On $x_{E^c}=0$, curvature $\bar S_{EE}\succeq kI$ ensures a unique
		attained optimum $\check x$.
		The exact restricted equation is
		$\check x_E-x_{d,E}=-\bar S_{EE}^{-1}(r_E+\lambda z_E)$.
		Since each subgradient has $\|z_e\|\le\sqrt{1+\eta}$ even at zero,
		Lemma~\ref{r1-lem:schur} gives
		$\|\check x_E-x_{d,E}\|_{2,\infty}\le(R_n+\lambda\sqrt{1+\eta})/k$.
		Condition \eqref{r1-eq:signal} therefore retains every active group.
		
		\emph{Inactive feasibility and uniqueness.}
		The required inactive covector is
		\begin{equation}\label{r1-eq:requireddual}
			z_e=\bar S_{eE}\bar S_{EE}^{-1}z_E+
			\lambda^{-1}(\bar S_{eE}\bar S_{EE}^{-1}r_E-r_e).
		\end{equation}
		Its Euclidean norm is at most
		$\kappa\sqrt{1+\eta}+(1+\kappa)R_n/\lambda$; divide by
		$\sqrt{1-\eta}$ for its $K_e^{-1}$ norm.
		Condition \eqref{r1-eq:dual} makes this strictly below one, completing
		KKT for the convex full objective.
		Proposition~\ref{prop:pwsm-strict-dual} forces every minimum to have zero
		inactive groups, and restricted strict convexity makes it equal to
		$\check x$. The unique main vector is reconstructed using $\widehat A\succ0$.
		If there are no inactive groups, the full profiled problem is already
		strongly convex.
	\end{proof}
	
	For completeness, if $E$ is empty, the direct condition
	$R_n<\lambda\sqrt{1-\eta}$ proves the unique zero-edge solution by the same
	strict-dual argument, without $k$ or a minimum-signal condition.
	Set
	\begin{equation}\label{r1-eq:etastar}
		\eta_*=\frac{c_0}{1000(\Delta+1)}.
	\end{equation}
	When $\eta\le\eta_*$, $k\ge0.99c_0$ and $\kappa<1/99$.
	In particular
	$\kappa\sqrt{(1+\eta)/(1-\eta)}<(1/99)\sqrt{11/9}<1/80$.
	Thus the density condition $\epsilon_3\le\eta_*/2$, together with
	$\max(e_A,e_B,e_C)\le\eta_*/2$, produces a fixed strict dual margin.
	
	\subsection{A growing, connected, genuinely infinite-dimensional example}
	
	\begin{corollary}[One connected path with growing $p$ and $d$]\label{r1-cor:path}
		Let $\Delta=2$, $p_n=\lceil\exp(n^{1/3})\rceil$, $d_n=\lceil\log n\rceil$,
		and $E_n=\{\{i,i+1\}:1\le i<p_n\}$, with natural logarithms.
		Use $q$ from \eqref{r2-eq:q} and $h_0=h$ from \eqref{r2-eq:h}, and set
		\begin{equation}\label{r1-eq:path}
			P_{0,n}(dx)\propto\left\{\prod_{i=1}^{p_n}q(x_i)dx_i\right\}
			\exp\!\left\{\tau\sum_{i=1}^{p_n-1}h_0(x_i,x_{i+1})\right\}.
		\end{equation}
		Take $\ell=e^{-4\sqrt2}$, $u=e^{4\sqrt2}$, and $\eta_*$ from
		\eqref{r1-eq:etastar}. Fix once and for all
		\begin{equation}\label{r1-eq:tau}
			0<\tau\le\frac{\log(1+\eta_*/2)}{18\Delta}.
		\end{equation}
		For all sufficiently large $n$, with threshold depending only on the fixed
		constants, one event of probability at least $1-n^{-2}$ gives a unique
		attained full minimizer of \eqref{def:pwsm} and exact recovery of all path
		edges for every $\lambda\in[n^{-1/4},n^{-1/8}]$.
		Each true interaction has infinitely many canonical coefficients, although
		the full empirical Gram is singular on every sample eventually.
	\end{corollary}
	\begin{proof}
		Theorem~\ref{r1-thm:general} applies once the signal stays positive and
		the block errors and $R_n/\lambda$ vanish uniformly on the interval.
		
		\emph{Density and signal.}
		The construction \eqref{r2-eq:h} has the required regularity and centering,
		$\|h_0\|_\infty\le3$, and a nonzero mixed sine mode on every edge.
		Its truncation tail has squared norm
		$\|\tau(h_0-h_{0,d})\|_2^2=\tau^2\sum_{r>d}4^{-r}=\tau^2 4^{-d}/3>0$.
		The mains are $2L_1$ and each true-edge diagonal is
		$\tau(1,2^{-2},\ldots,2^{-d})$, so truncation has exact support $E_n$ and
		\begin{equation}\label{r1-eq:signalpath}
			\rho_{\rm ref}\ge\tau\sqrt{\phi_0}>0.
		\end{equation}
		Equation \eqref{r2-eq:densityratio} bounds $q$, while $v=6\tau$ and
		\eqref{r1-eq:tau} give $\epsilon_3=e^{18\Delta\tau}-1\le\eta_*/2$.
		Envelopes uniform in $p_n$ are
		\begin{equation}\label{r1-eq:envelopepath}
			\begin{gathered}
				G=1/64,\quad G_1=1/8,\quad U=2\pi d,\quad V=2\pi^2d^2,\\
				T=2\sqrt2\pi+5\pi\Delta\tau,\qquad
				H=2\pi\Delta\tau(d+2)2^{-d}.
			\end{gathered}
		\end{equation}
		For $T,H$, use $\sum_{r\ge2}r2^{-r}=3/2$,
		$\sum_{r>d}r2^{-r}=(d+2)2^{-d}$ and at most $\Delta$ incident edges.
		Smoothness and the endpoint zeros of $g$ justify integration by parts.
		
		\emph{Recovery inequalities.}
		With $\delta=n^{-2}$ and $D=pd+\binom p2d^2$,
		$\log(4D^2/\delta)=O(n^{1/3})$. The residual logarithm is no larger.
		Consequently, with constants independent of $n,p_n,d_n$,
		\begin{align}\label{r1-eq:rates}
			t_\Gamma&=O(d^2n^{-1/3}),&
			t_w&=O(d^2n^{-1/3}+d^22^{-d}),\\
			e_A&=O(d^3n^{-1/3}),&e_B&=O(d^{7/2}n^{-1/3}),\nonumber\\
			e_C&=O(d^4n^{-1/3}),&R_n&=O(d^3n^{-1/3}+d^32^{-d}).\nonumber
		\end{align}
		Thus $\eta\le\eta_*$ eventually, and uniformly on the tuning interval,
		\begin{equation}\label{r1-eq:window}
			R_n/\lambda\le C\{(\log n)^3n^{-1/12}
			+(\log n)^3n^{1/4-\log2}\}\longrightarrow0.
		\end{equation}
		Together with $\lambda\le n^{-1/8}\to0$,
		\eqref{r1-eq:signalpath}, and $k\ge0.99c_0$, this verifies
		\eqref{r1-eq:signal}--\eqref{r1-eq:dual} on the common event.
		
		\emph{Full-Gram singularity.}
		Each observation contributes rank at most $p_n$, whereas
		$D=p_nd+\binom{p_n}2d^2>np_n$ eventually; hence the full Gram is always
		singular. The restricted dimension $O(p_nd^2)$ is compatible with rank
		$np_n$ since $d^2=o(n)$. Uniqueness follows from restricted curvature and
		strict dual feasibility, not full-Gram curvature.
	\end{proof}
	
	The same proof covers $p_n=\lceil n\rceil$, with smaller concentration logarithms.
}

\endgroup
 
\begingroup 
{
	\section{Proofs for unrestricted density estimation and normalization}
	\label{supp:density-inverse-closure}
	
	\subsection{Score-risk bound}
	To prove Theorem~\ref{th:density-unrestricted}, we control truncation,
	the Gram matrix, and residual variance, then apply the normal equations.
	Use the fixed-\(p\) setting, bases, and weights of
	Theorem~\ref{th:density-unrestricted}, writing \(d=d_n\), \(D_d=D_n\),
	\(\eta_0\) for the true canonical potential, and \(\mu\) for uniform
	product probability on \(\Omega\); take \(g_i'(0)=0\). Constants may depend on
	\(p,B,\nu_1,\nu_2\), not on \(n,d\). Let \(\psi_d\) denote the truncated feature vector collecting the
	centered main-effect basis functions and tensor-product interaction
	basis functions up to level \(d\). Put 
	\(A_d=(\partial_i\psi_d^\top)_{i=1}^p\), \(G=\operatorname{diag}(g_i)\),
	and define the summands of the empirical normal equations by
	\[
	K_d=A_d^\top GA_d,\quad \Gamma_d=\mathbb E_0K_d,\quad
	b_d=\sum_i(g_i'\partial_i\psi_d+g_i\partial_{ii}\psi_d),\quad
	\widehat\Gamma_d=P_nK_d,\quad \widehat\rho_d=P_nb_d.
	\]
	For literal truncation \(\theta_d^{\rm tr}\), put
	\(s_d=A_d\theta_d^{\rm tr}\), \(s_0=\nabla\log p_0\), and
	\[
	a_d^2=\mathbb E_0\|G^{1/2}(s_d-s_0)\|^2,\quad
	\zeta_d=A_d^\top Gs_d+b_d,\quad h_d=\mathbb E_0\zeta_d,\quad
	\varepsilon_d=(P_n-P_0)\zeta_d.
	\]
	\paragraph{Truncation and the residual mean.}
	The bounds \(\|L_r\|_\infty\le2\), \(\|L_r'\|_\infty\le Cr\), and
	\(\|L_r''\|_\infty\le Cr^2\) give uniform convergence through second
	derivatives, bounded \(s_d\), and \(m\le dp_0/d\mu\le M\), where
	\(m=e^{-2\|\eta_0\|_\infty}>0\) and \(M=m^{-1}\).
	The absolutely continuous product \(g_i\partial_i\psi_dp_0\) vanishes
	at both endpoints, so integration by parts gives
	\begin{equation}
		\mathbb E_0b_d=-\mathbb E_0A_d^\top Gs_0,\qquad
		h_d=\mathbb E_0A_d^\top G(s_d-s_0).
		\label{di:mean}
	\end{equation}
	The cusp introduces no boundary flux. The main derivative tail is
	\(O(\sum_{r>d}r^{1-\nu_1})\); splitting omitted interaction indices
	into \(\{r>d,s\ge1\}\) and \(\{r\le d,s>d\}\) gives
	\(O(d^{2-\nu_2}+d^{1-\nu_2})\). Summing components, with and without
	differentiation, respectively, gives
	\begin{align}
		a_d^2&\le C_A\{d^{-2(\nu_1-2)}+d^{-2(\nu_2-2)}\},\label{di:tail}\\
		\|\eta_{\theta_d^{\rm tr}}-\eta_0\|_\infty
		&\le C\{d^{1-\nu_1}+d^{1-\nu_2}\}.\label{di:log-tail}
	\end{align}
	
	\paragraph{Relative Gram concentration.}
	For a polynomial \(Q\) of degree at most \(k\), orthogonal Legendre
	expansion, \(|P_j|\le1\), and \(\int_{-1}^1P_j^2=2/(2j+1)\) give,
	after affine rescaling,
	\begin{equation}
		|Q(t)|^2\le (k+1)^2|I|^{-1}\int_IQ^2,\qquad t\in I.
		\label{di:polynomial-evaluation}
	\end{equation}
	For a degree-\(N\) trigonometric polynomial on \(t=x/\pi\in[-1,1]\),
	Parseval gives \(\|T\|_\infty^2\le(2N+1)\int T^2dt/2\).
	A boundary strip of width \(\epsilon=\{2(2N+1)\}^{-1}\) contains at
	most half this integral, so, for \(w=(1-|t|)^2\),
	\(wT^2\le8(2N+1)^3\int wT^2dt/2\).
	For monomials, \(\sqrt wT\) has degree at most \(d\) on each half interval;
	\eqref{di:polynomial-evaluation} instead gives
	\(wT^2\le2(d+1)^2\int wT^2dt/2\).
	
	To lift these bounds,
	\(F_i=\sqrt{g_i}\partial_i(\psi_d^\top v)\) lies in the tensor product
	of the weighted derivative space and the orthogonal sum of the constant
	and the centered one-coordinate spaces for \(j\ne i\).
	The latter evaluation kernel is at most \(1+2(p-1)d\) for trig and
	\(1+(p-1)\{(d+1)^2-1\}\) for monomials.
	Multiplying the kernels, summing over \(i\), and using \(p_0\ge m\mu\)
	gives \(K_d(x)\preceq R_d\Gamma_d\), where \(R_d=C_R(d+1)^4\ge1\).
	Zero \(\Gamma_d\)-energy forces a constant potential and hence zero
	coefficients by canonical centering and tensor-basis independence.
	Thus \(\Gamma_d\succ0\).
	
	The iid matrices \(Y_k=\Gamma_d^{-1/2}K_d(X_k)\Gamma_d^{-1/2}\)
	have mean \(I\) and satisfy \(0\preceq Y_k\preceq R_dI\).
	Matrix Chernoff yields, for \(0<t\le1\),
	\begin{equation}
		\mathbb P\{\|\Gamma_d^{-1/2}\widehat\Gamma_d\Gamma_d^{-1/2}-I\|_{\rm op}>t\}
		\le2D_de^{-nt^2/(3R_d)}.
		\label{di:relative-gram}
	\end{equation}
	In particular, \(\mathcal E_{n,d}=\{\Gamma_d/2\preceq\widehat\Gamma_d
	\preceq3\Gamma_d/2\}\) has failure probability at most
	\(2D_de^{-n/(12R_d)}\).
	
	\paragraph{Derivative inequality and residual variance.}
	For either basis we need
	\begin{equation}
		\int g^2(T')^2\le Cd^2\int gT^2.
		\label{di:trig-derivative}
	\end{equation}
	For trigonometric degree at most \(d\), set \(h=1+\cos x\). Since
	\(2g\le h\le(\pi^2/2)g\) and \(|h'|^2\le\pi^2g\), Parseval applied to
	\(hT\), followed by \(hT'=(hT)'-h'T\), proves the bound.
	For polynomials of degree at most \(d-1\), put \(H=1-t^2\), \(t=x/\pi\).
	The operator
	\(\mathcal LQ=-(1-t^2)Q''+6tQ'=-H^{-2}(H^3Q')'\)
	is symmetric in \(L_2(H^2dt)\), by integration by parts, and triangular
	on monomials with distinct eigenvalues \(r(r+5)\).
	Its polynomial eigenbasis therefore gives
	\(\int H^3(T')^2\le(d-1)(d+4)\int H^2T^2\).
	Now \(g^2\le H^3\) and \(H^2\le4g\) prove \eqref{di:trig-derivative};
	constants absorb the fixed rescaling. The degree-zero case is immediate.
	
	For any \(v\), set \(T_i=\partial_i(\psi_d^\top v)\). The identity
	\[
	v^\top\zeta_d=\sum_i\{g_iT_is_{d,i}+g_i'T_i+g_i\partial_iT_i\}
	\]
	with bounded \(s_d\), \(|g_i'|^2\le g_i\), and \(g_i^2\le g_i\), bounds
	its second moment by
	\(C\sum_i\mathbb E_0\{g_iT_i^2+g_i^2(\partial_iT_i)^2\}\).
	Apply \eqref{di:trig-derivative} with the other coordinates fixed,
	then integrate and compare \(p_0\) to \(\mu\). This gives
	\(\operatorname{Var}_0(v^\top\zeta_d)\le Cd^2v^\top\Gamma_dv\).
	Since \(v\) is unrestricted, whitening and taking the trace give
	\begin{equation}
		\operatorname{Cov}_0(\zeta_d)\preceq C_Vd^2\Gamma_d,
		\qquad \mathbb E_0\|\varepsilon_d\|_{\Gamma_d^{-1}}^2
		\le C_Vd^2D_d/n.
		\label{di:residual-variance}
	\end{equation}
	
	\paragraph{Normal equations and risk.}
	For \(u_d=u_n\) in \eqref{eq:density-unrestricted},
	\eqref{di:mean} gives \(\|h_d\|_{\Gamma_d^{-1}}\le a_d\). On
	\(\mathcal E_{n,d}\), the normal equations imply
	\[
	\Delta=u_d-\theta_d^{\rm tr}=-\widehat\Gamma_d^{-1}(h_d+\varepsilon_d),
	\qquad \|\Delta\|_{\Gamma_d}^2
	\le4\|h_d+\varepsilon_d\|_{\Gamma_d^{-1}}^2.
	\]
	Thus \(J_g\le2a_d^2+2\|\Delta\|_{\Gamma_d}^2\) yields
	\begin{equation}
		J_g(p_0\|p_{u_d})\le18a_d^2+16\|\varepsilon_d\|_{\Gamma_d^{-1}}^2.
		\label{di:inverse-oracle}
	\end{equation}
	If \(a_d=0\), the sharper bound is \(4\|\varepsilon_d\|_{\Gamma_d^{-1}}^2\).
	For any finite, possibly data-dependent \(\widetilde\theta\), define
	\(r_d=\widehat\Gamma_d\widetilde\theta+\widehat\rho_d\). The identity
	\(\widehat\Gamma_d(\widetilde\theta-\theta_d^{\rm tr})=r_d-h_d-\varepsilon_d\),
	three-term Cauchy--Schwarz, and
	\(\Gamma_d^{-1}\preceq(3/2)\widehat\Gamma_d^{-1}\) yield
	\begin{equation}
		J_g(p_0\|p_{\widetilde\theta})
		\le26a_d^2+24\|\varepsilon_d\|_{\Gamma_d^{-1}}^2
		+36r_d^\top\widehat\Gamma_d^{-1}r_d.
		\label{di:approximate-oracle}
	\end{equation}
	By \eqref{di:relative-gram}--\eqref{di:residual-variance} and Markov's
	inequality, the risk exceeds \(18a_d^2+16C_Vd^2D_d/(n\delta)\) with
	probability at most \(2D_de^{-n/(12R_d)}+\delta\).
	Under \(d^4\log(eD_d)=o(n)\), this proves
	Theorem~\ref{th:density-unrestricted}, including the smaller
	\(D_dd^2/n\) term when \(p=1\). In particular, if \(d\asymp n^q\),
	\(0<q<1/4\),
	\begin{equation}
		J_g(p_0\|p_{u_d})=O_p(R_{n,d}),\qquad
		R_{n,d}=d^4/n+d^{-2(\nu_1-2)}+d^{-2(\nu_2-2)}.
		\label{di:unrestricted-rate}
	\end{equation}
	The zero fallback covers singular samples; these are probability bounds,
	not inverse-risk expectations.
	
	\subsection{From score risk to normalized density risk}
	For \(f\in C^1[0,1]\), integration by parts with \(v=f-f(0)\) gives
	\(\int v^2=2\int(1-t)vv'\le
	2(\int v^2\int(1-t)^2(f')^2)^{1/2}\).
	Applying this on both halves of \([-1,1]\), rescaling, and tensorizing
	by conditional variance and Jensen's inequality gives
	\(\operatorname{Var}_\mu F\le4\pi^2\mathbb E_\mu\sum_i g_i(\partial_iF)^2\).
	Density comparison then yields
	\begin{equation}
		\operatorname{Var}_0F\le4\pi^2(M/m)\mathbb E_0\sum_i g_i(\partial_iF)^2.
		\label{di:weighted-poincare}
	\end{equation}
	For \(r=\eta_{u_d}-\eta_0\) and \(W=\sup r-\inf r\), exact normalization
	and Taylor's remainder give
	\begin{equation}
		\mathrm{KL}(p_0\|p_{u_d})
		=\log\mathbb E_0e^{r-\mathbb E_0r}
		\le\tfrac12e^W\operatorname{Var}_0r
		\le2\pi^2(M/m)e^WJ_g(p_0\|p_{u_d}).
		\label{di:kl-oracle}
	\end{equation}
	Here \(|r-\mathbb E_0r|\le W\), and Jensen's inequality gives
	\(H^2\le-2\log\int\sqrt{p_0p_{u_d}}\le\mathrm{KL}(p_0\|p_{u_d})\).
	It therefore remains to prove \(W=O_p(1)\).
	
	The centered main-plus-pair evaluation kernel under \(\mu\) is
	\(O(d^2)\) for trig and \(O(d^4)\) for monomials. Together with the
	uniform-measure version of \eqref{di:weighted-poincare} this implies
	\(\|\psi_d^\top v\|_\infty\le C_pd\|v\|_{\Gamma_d}\) and
	\(C_pd^2\|v\|_{\Gamma_d}\), respectively.
	Since \(\|\Delta\|_{\Gamma_d}=O_p(a_d+d^2/\sqrt n)\), adding
	\eqref{di:log-tail}, with \(\nu=\min(\nu_1,\nu_2)\), gives
	\begin{equation}
		\|r\|_\infty=
		\begin{cases}
			O_p(d^{3-\nu}+d^3/\sqrt n),&\text{trig},\\
			O_p(d^{4-\nu}+d^4/\sqrt n),&\text{monomials}.
		\end{cases}
		\label{di:log-oscillation}
	\end{equation}
	This proves the corollary for trig with \(q\le1/6\), and for monomials
	with \(\nu\ge4,q\le1/8\). To include the monomial choice \(q=1/7\),
	we sharpen the stochastic term.
	
	\paragraph{Monomials: leading fluctuation and inverse error.}
	\label{di:polynomial-supnorm-refinement}
	Set \(L_{n,d}(x)=\psi_d(x)^\top\Gamma_d^{-1}\varepsilon_d\).
	For deterministic \(v\), the tensor evaluation bounds imply
	\begin{equation}
		|v^\top\zeta_d(x)|\le C_pd^3\|v\|_{\Gamma_d}.
		\label{di:point-residual-envelope}
	\end{equation}
	The terms \(g_iT_is_{d,i}\) and \(g_i'T_i\) have squared bounds
	\(C_pd^4v^\top\Gamma_dv\).
	For the remaining term, \(g_i\partial_iT_i\) is degree at most \(d\)
	on each own-coordinate half interval. Its tensor evaluation kernel is
	\(O(d^4)\); applying \eqref{di:trig-derivative} after integration gives
	\[
	|g_i\partial_iT_i(x)|^2
	\le C_pd^4\int g_i^2(\partial_iT_i)^2d\mu
	\le C_pd^6\int g_iT_i^2d\mu
	\le C_pd^6v^\top\Gamma_dv.
	\]
	Summing over the \(3p\) terms proves \eqref{di:point-residual-envelope}.
	
	For every monomial sieve function \(F\), a deterministic product grid
	\(|\mathcal Q_d|\le C_p(d+1)^{3p}\) satisfies
	\begin{equation}
		\|F\|_\infty\le2\max_{x\in\mathcal Q_d}|F(x)|.
		\label{di:polynomial-norming-grid}
	\end{equation}
	To construct it, fix all but one coordinate and expand the resulting
	degree-\(d\) polynomial in Chebyshev polynomials.
	Their coefficients have magnitude at most \(2\|F\|_\infty\) and
	\(\|\mathsf T_j'\|_\infty\le j^2\), giving
	\(\|\partial_iF\|_\infty\le Cd^3\|F\|_\infty/\pi\).
	A mesh at most \(\pi/(2pCd^3)\) makes the error from any point to the
	nearest grid point at most \(\|F\|_\infty/2\).
	
	For fixed \(x\), put \(v_x=\Gamma_d^{-1}\psi_d(x)\).
	The log-sieve evaluation bound gives
	\(v_x^\top\Gamma_dv_x\le C_pd^4\). Consequently
	\(Z_k(x)=v_x^\top\{\zeta_d(X_k)-h_d\}\) is centered and satisfies
	\(\mathbb EZ_k(x)^2\le C_pd^6\), \(|Z_k(x)|\le C_pd^5\).
	The scalar Bernstein inequality therefore gives
	\[
	\mathbb P\{|L_{n,d}(x)|>C_p(d^3\sqrt{t/n}+d^5t/n)\}\le2e^{-t}.
	\]
	Take \(t=\log(2|\mathcal Q_d|/\delta)\), union over the grid, and use
	\eqref{di:polynomial-norming-grid}:
	\begin{equation}
		\|L_{n,d}\|_\infty
		=O_p\{d^3\sqrt{\log(ed)/n}+d^5\log(ed)/n\}.
		\label{di:leading-polynomial-supnorm}
	\end{equation}
	
	For the inverse remainder, put
	\(H_{n,d}=\Gamma_d^{-1/2}\widehat\Gamma_d\Gamma_d^{-1/2}\). The Chernoff
	bound and \(d^4\log(eD_d)=o(n)\) give
	\(\|H_{n,d}-I\|_{\rm op}=O_p(d^2\sqrt{\log(eD_d)/n})=o_p(1)\).
	When this norm is at most \(1/2\), \(\|H_{n,d}^{-1}-I\|_{\rm op}\)
	is at most twice it, and the exact decomposition is
	\[
	\Delta=-\widehat\Gamma_d^{-1}h_d-\Gamma_d^{-1}\varepsilon_d
	-\Gamma_d^{-1/2}(H_{n,d}^{-1}-I)\Gamma_d^{-1/2}\varepsilon_d.
	\]
	The three log-sieve supnorms are bounded by \(C_pd^2a_d\),
	\(\|L_{n,d}\|_\infty\), and
	\(C_pd^2\|H_{n,d}-I\|_{\rm op}\|\varepsilon_d\|_{\Gamma_d^{-1}}\),
	respectively. The last is
	\(O_p(d^6\sqrt{\log(eD_d)}/n)\); the bound is eventwise and needs no
	independence between the inverse and the residual. Adding the log tail gives
	\begin{equation}
		\|r\|_\infty=O_p\{d^{4-\nu}
		+d^3\sqrt{\log(ed)/n}+d^5\log(ed)/n
		+d^6\sqrt{\log(eD_d)}/n\}.
		\label{di:refined-polynomial-oscillation}
	\end{equation}
	For \(\nu\ge4\) and \(0<q<1/6\), the bias is bounded and the three
	stochastic terms vanish, completing Corollary~\ref{cor:density-normalization}
	by \eqref{di:kl-oracle}. If \(\nu>4\), the supnorm tends to zero.
	The monomial endpoint \(q=1/6\) and exponential-moment bounds are not asserted.
}

\endgroup

{
	\section{Additional simulation results for P-WSM}
	\label{sec:supp-pwsm-simulations}
	
	\subsection{Simulation design and tuning}
	\label{sec:supp-pwsm-design}
	
	The graph-recovery experiments in the main paper cover 28 settings, each with 100 independent replicates. P-WSM and Joint-WSM use the same observations, the same five cross-validation folds, the same basis functions, the same solver tolerance and support threshold, and the same 31-point relative penalty grid. Both methods use the raw empirical score-Gram block norm for the interaction penalty. P-WSM refits all main effects and selected interactions, whereas Joint-WSM refits selected main effects and interactions. This paired design compares the complete procedures. Because the methods have distinct path endpoints and calibration geometries, each procedure selects its regularization parameter adaptively along its own cross-validation path.
	
	For the primary analysis, we choose the penalty that minimizes the mean held-out weighted score after refitting the selected graph without penalization, namely the minimum raw five-fold cross-validation rule defined in the main paper. The basis dimension is fixed within each setting, so cross-validation selects only the penalty level; a separate fixed-design experiment compares \(d=2,3,4\). We also evaluate the raw one-standard-error rule, the dimension-adjusted minimum and one-standard-error rules, and the choice with the smallest structural Hamming distance along the already fitted path. We use these alternatives only as sensitivity analyses. Graphical lasso and \texttt{highscore} use EBIC with \(\gamma=0.5\); BEAM uses its Bayes-factor edge rule; and BGGM uses its posterior edge-selection rule after 5,000 iterations.
	
	\begin{table}[!htbp]
		\centering
		\caption{Summary of the formal graph-recovery settings. Every setting contains 100 independent replicates.}
		\label{tab:supp-pwsm-settings}
		\resizebox{0.96\textwidth}{!}{%
			\begin{tabular}{lllll}
				\hline
				Module & Model and graph & Main dimensions & Values varied & Purpose \\
				\hline
				Benchmarks & Gaussian and nonlinear chains & \((6,1500,2)\), \((8,2000,2)\) & model class & baseline recovery \\
				Main-effect amplitude & nonlinear, three disjoint pairs & \((6,1800,2)\) & \(a=0,0.25,\ldots,2\) & effect of main-effect penalization \\
				Preservation & empty graph; two independent chains & \((6,1800,2)\) & graph structure & product and subsystem preservation \\
				Difficulty & nonlinear chains & \(p=8,d=2\) or 3 & \(n,\rho,d\) & weak signal and overfitting \\
				Topology & Gaussian/nonlinear hub and sparse random & \((8,2000,2)\) & graph structure & robustness beyond chains \\
				Nuisance dimension & nonlinear disjoint pairs & \(d=2\) & \(pd/n=0.01\) to 0.30 & finite-sample scaling \\
				Basis dimension & nonlinear chain & \((8,1600,d)\) & \(d=2,3,4\) & sieve sensitivity \\
				\hline
		\end{tabular}}
	\end{table}
	
	For the Gaussian models, the precision matrix has unit diagonal and off-diagonal entry \(-\rho\) on every true edge, and the centered-monomial basis is used. For the nonlinear models,
	\[
	p(\tb x)\propto
	\exp\left\{
	a\sum_{j=1}^p\sin(x_j)+
	\rho\sum_{(j,k)\in E}\sin(x_j)\sin(x_k)
	\right\}
	\mathbbm 1\{\tb x\in[-\pi,\pi]^p\},
	\]
	and we use the trigonometric basis. The hub graph connects node 1 to each of the other seven nodes. The fixed sparse-random graph contains the edges
	\[
	(1,2),(1,4),(2,5),(3,5),(5,7),(5,8),(6,7).
	\]
	
	\subsection{Main-effect penalization and propagated interaction gradients}
	\label{sec:supp-pwsm-propagation}
	
	Table~\ref{tab:supp-pwsm-amplitude} gives the complete seven-point main-effect amplitude experiment. We keep the graph and interaction strength fixed and vary only the amplitude \(a\). As the main effects become stronger, Joint-WSM selects increasingly many false edges, whereas P-WSM remains accurate over most of the range. At \(a=2\), most of the remaining P-WSM error comes from missed interactions.
	
	\begin{table}[!htbp]
		\centering
		\caption{Main-effect amplitude experiment. SHD, false positives (FP), and false negatives (FN) are Monte Carlo means; values in parentheses are Monte Carlo standard errors.}
		\label{tab:supp-pwsm-amplitude}
		\resizebox{0.92\textwidth}{!}{%
			\begin{tabular}{cclcccc}
				\hline
				Amplitude & Method & SHD & FP & FN & Exact & Selected edges \\
				\hline
				0 & P-WSM & 0.010 (0.010) & 0.010 (0.010) & 0.000 (0.000) & 0.990 & 3.010 \\
				& Joint-WSM & 0.030 (0.022) & 0.030 (0.022) & 0.000 (0.000) & 0.980 & 3.030 \\
				0.25 & P-WSM & 0.020 (0.014) & 0.020 (0.014) & 0.000 (0.000) & 0.980 & 3.020 \\
				& Joint-WSM & 1.290 (0.186) & 1.290 (0.186) & 0.000 (0.000) & 0.470 & 4.290 \\
				0.50 & P-WSM & 0.130 (0.053) & 0.130 (0.053) & 0.000 (0.000) & 0.930 & 3.130 \\
				& Joint-WSM & 0.380 (0.084) & 0.380 (0.084) & 0.000 (0.000) & 0.760 & 3.380 \\
				0.75 & P-WSM & 0.050 (0.026) & 0.050 (0.026) & 0.000 (0.000) & 0.960 & 3.050 \\
				& Joint-WSM & 1.140 (0.125) & 1.140 (0.125) & 0.000 (0.000) & 0.370 & 4.140 \\
				1.00 & P-WSM & 0.060 (0.028) & 0.060 (0.028) & 0.000 (0.000) & 0.950 & 3.060 \\
				& Joint-WSM & 7.490 (0.165) & 7.490 (0.165) & 0.000 (0.000) & 0.000 & 10.490 \\
				1.50 & P-WSM & 0.860 (0.092) & 0.400 (0.079) & 0.460 (0.069) & 0.430 & 2.940 \\
				& Joint-WSM & 10.770 (0.102) & 10.770 (0.102) & 0.000 (0.000) & 0.000 & 13.770 \\
				2.00 & P-WSM & 2.640 (0.085) & 0.360 (0.086) & 2.280 (0.101) & 0.010 & 1.080 \\
				& Joint-WSM & 11.240 (0.090) & 11.240 (0.090) & 0.000 (0.000) & 0.000 & 14.240 \\
				\hline
		\end{tabular}}
	\end{table}
	
	To connect this empirical difference to Proposition~\ref{prop:pwsm-propagation-main}, we reconstruct the displacement caused by penalizing the main effects at every Joint-WSM path point. Holding the interaction coefficient fixed, we solve for the exact unpenalized main-effect estimate and compute the resulting contribution
	\[
	H_{\gamma,\beta}\bigl(\widehat\beta_{\rm Joint}-\widehat\beta_{\rm ref}\bigr)
	\]
	to the interaction gradient. We measure its norm in the raw score-Gram dual geometry. The path-averaged norm increases from 0.0016 at \(a=0\) to 0.2339 at \(a=2\). For P-WSM, the same quantity is numerically zero because the main effects satisfy the unpenalized stationarity equation at every interaction update.
	
	At \(a=1\), the paired mean difference in structural Hamming distance, P-WSM minus Joint-WSM, is \(-7.43\), with a 95\% Monte Carlo interval \([-7.756,-7.104]\). We also use a local gradient-threshold diagnostic at the selected Joint-WSM fit. We hold the entire fitted interaction coefficient \(\widehat\gamma\) fixed, replace the penalized main-effect estimate by its exact unpenalized refresh conditional on \(\widehat\gamma\), and recompute the loss-gradient norm for each fitted false-positive interaction block. The refreshed gradient norm is below the corresponding penalty level for 21.9\% of these blocks at \(a=1\) and 94.2\% at \(a=1.5\). This confirms that the false-positive interaction selections in Joint-WSM are locally driven by gradient distortion from the penalized main effects.
	
	\begin{figure}[!htbp]
		\centering
		\includegraphics[width=0.78\textwidth]{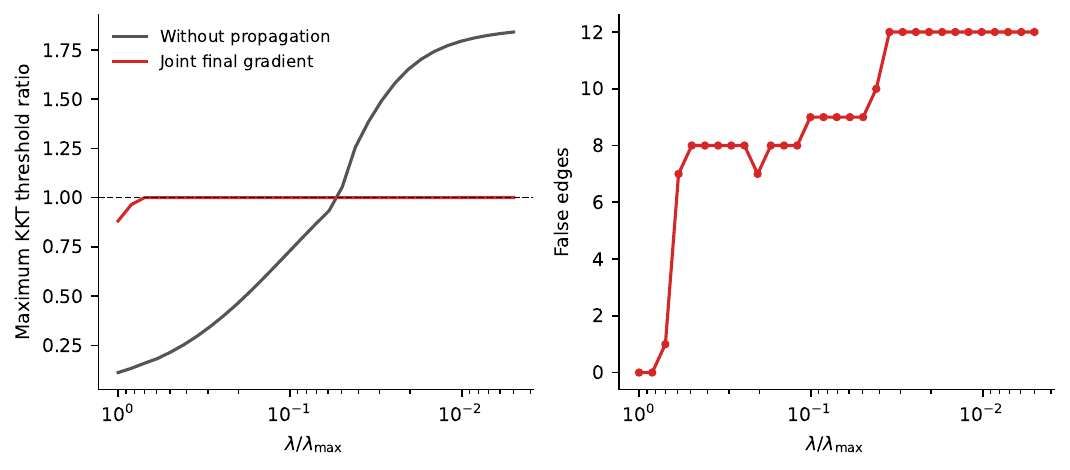}
		\caption{Local gradient-threshold diagnostic for fitted false-positive interaction blocks. All fitted interaction coefficients are held fixed while the penalized main-effect estimate is replaced by its exact unpenalized refresh; the diagnostic records whether the refreshed loss-gradient norm falls below the corresponding penalty level.}
		\label{fig:supp-pwsm-kkt-crossing}
	\end{figure}
	
	This diagnostic directly demonstrates that main-effect shrinkage materially alters the interaction gradients at fitted false-positive blocks.
	
	\subsection{Empty graphs and independent subsystems}
	\label{sec:supp-pwsm-preservation}
	
	In the empty-graph experiment, the six nonlinear variables are mutually independent. P-WSM selects an average of 0.07 edges and recovers the empty graph in 95\% of the replicates. Joint-WSM selects an average of 0.65 edges and recovers the empty graph in 61\% of the replicates.
	
	Table~\ref{tab:supp-pwsm-subsystems} gives the full comparison for two independent three-node chains. The columns ``Within A'' and ``Within B'' are the SHDs computed separately within the two true subsystems; ``Cross edges'' counts selected edges joining the two subsystems; and ``Any cross'' is the proportion of replicates containing at least one such edge.
	
	\begin{table}[!htbp]
		\centering
		\caption{Recovery for two independent three-node subsystems.}
		\label{tab:supp-pwsm-subsystems}
		\resizebox{0.95\textwidth}{!}{%
			\begin{tabular}{lcccccccc}
				\hline
				Method & Within A & Within B & Cross edges & Any cross & SHD & TPR & FDR & Exact \\
				\hline
				P-WSM & 0.100 & 0.050 & 0.220 & 0.130 & 0.370 & 0.975 & 0.044 & 0.760 \\
				Joint-WSM & 0.690 & 0.730 & 6.630 & 1.000 & 8.050 & 1.000 & 0.664 & 0.000 \\
				Graphical lasso & 0.280 & 0.330 & 0.300 & 0.290 & 0.910 & 0.912 & 0.103 & 0.380 \\
				\texttt{highscore} & 0.380 & 0.350 & 0.250 & 0.240 & 0.980 & 0.865 & 0.084 & 0.420 \\
				BEAM & 0.110 & 0.130 & 0.480 & 0.410 & 0.720 & 0.995 & 0.127 & 0.470 \\
				BGGM & 0.090 & 0.120 & 0.480 & 0.420 & 0.690 & 0.995 & 0.123 & 0.480 \\
				\hline
		\end{tabular}}
	\end{table}
	
	\begin{figure}[!htbp]
		\centering
		\subfigure[Probability of at least one cross-subsystem edge.]{\includegraphics[width=0.48\textwidth]{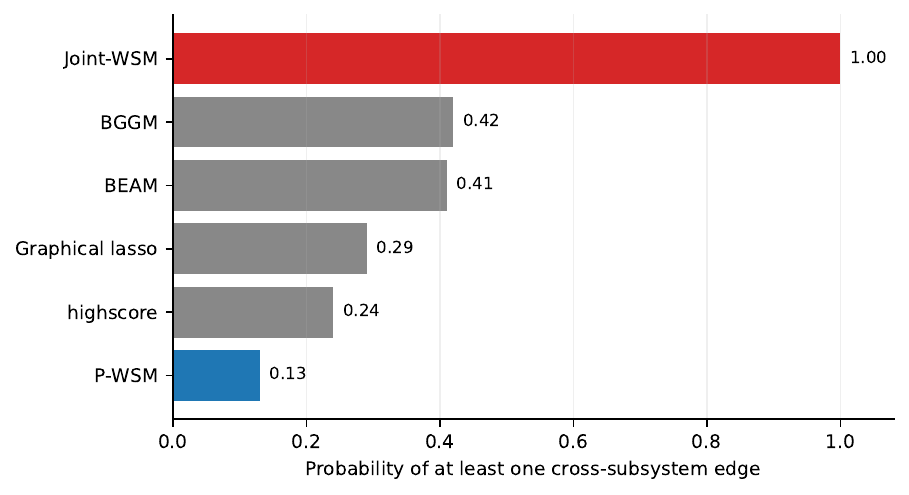}}
		\hfill
		\subfigure[Distribution of the number of cross-subsystem edges.]{\includegraphics[width=0.48\textwidth]{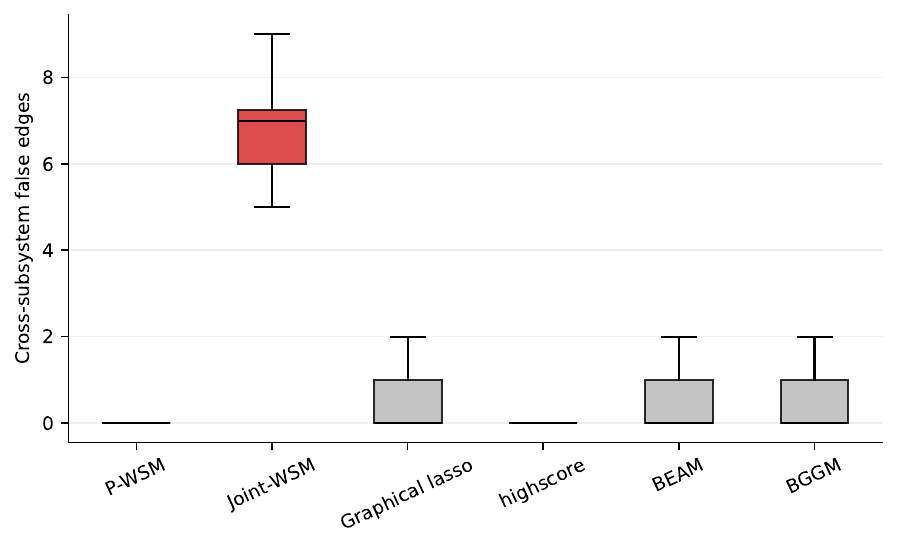}}
		\caption{Cross-subsystem errors in the independent-subsystem experiment.}
		\label{fig:supp-pwsm-cross-subsystem}
	\end{figure}
	
	\subsection{Additional graph topologies}
	\label{sec:supp-pwsm-topologies}
	
	For the topology experiment, we use \(p=8\), \(n=2000\), \(d=2\), and seven true edges in each graph. The Gaussian settings use \(\rho=0.25\), while the nonlinear settings use \(a=1\) and \(\rho=0.50\). Table~\ref{tab:supp-pwsm-topology} reports the complete comparison.
	
	\begin{table}[!htbp]
		\centering
		\caption{Graph-topology experiment. SHD is reported as mean (Monte Carlo standard error).}
		\label{tab:supp-pwsm-topology}
		\scriptsize
		\resizebox{\textwidth}{!}{%
			\begin{tabular}{llcccccc}
				\hline
				Setting & Method & SHD & Precision & Recall & F1 & Exact & Selected edges \\
				\hline
				Gaussian hub & P-WSM & {0.120 (0.043)} & {0.986} & {1.000} & {0.992} & {0.910} & {7.120} \\
				& Joint-WSM & {18.700 (0.131)} & {0.273} & {1.000} & {0.429} & {0.000} & {25.700} \\
				& Graphical lasso & {1.530 (0.132)} & {0.839} & {1.000} & {0.908} & {0.270} & {8.530} \\
				& \texttt{highscore} & {1.410 (0.127)} & {0.850} & {1.000} & {0.914} & {0.270} & {8.410} \\
				& BEAM & {1.030 (0.101)} & {0.884} & {1.000} & {0.935} & {0.350} & {8.030} \\
				& BGGM & {1.130 (0.117)} & {0.877} & {1.000} & {0.930} & {0.340} & {8.130} \\
				\hline
				Gaussian sparse random & P-WSM & {0.040 (0.020)} & {0.995} & {1.000} & {0.997} & {0.960} & {7.040} \\
				& Joint-WSM & {18.590 (0.132)} & {0.274} & {1.000} & {0.430} & {0.000} & {25.590} \\
				& Graphical lasso & {1.830 (0.136)} & {0.810} & {1.000} & {0.891} & {0.140} & {8.830} \\
				& \texttt{highscore} & {1.290 (0.118)} & {0.860} & {1.000} & {0.921} & {0.290} & {8.290} \\
				& BEAM & {1.140 (0.105)} & {0.873} & {1.000} & {0.929} & {0.320} & {8.140} \\
				& BGGM & {1.150 (0.103)} & {0.872} & {1.000} & {0.928} & {0.310} & {8.150} \\
				\hline
				Nonlinear hub & P-WSM & 5.850 (0.163) & 0.480 & 0.269 & 0.305 & 0.010 & 2.610 \\
				& Joint-WSM & 14.470 (0.256) & 0.331 & 1.000 & 0.496 & 0.000 & 21.470 \\
				& Graphical lasso & 7.500 (0.099) & 0.295 & 0.051 & 0.085 & 0.000 & 1.220 \\
				& \texttt{highscore} & 6.850 (0.039) & 0.138 & 0.023 & 0.038 & 0.000 & 0.170 \\
				& BEAM & 5.520 (0.175) & 0.685 & 0.373 & 0.463 & 0.000 & 3.740 \\
				& BGGM & 5.660 (0.177) & 0.679 & 0.357 & 0.447 & 0.000 & 3.660 \\
				\hline
				Nonlinear sparse random & P-WSM & {2.670 (0.165)} & {0.891} & {0.737} & {0.773} & {0.090} & {5.990} \\
				& Joint-WSM & {14.690 (0.201)} & {0.326} & {1.000} & {0.490} & {0.000} & {21.690} \\
				& Graphical lasso & {5.180 (0.136)} & {0.981} & {0.266} & {0.388} & {0.010} & {1.900} \\
				& \texttt{highscore} & {6.440 (0.146)} & {0.188} & {0.081} & {0.102} & {0.000} & {0.580} \\
				& BEAM & {2.770 (0.150)} & {0.848} & {0.750} & {0.785} & {0.030} & {6.270} \\
				& BGGM & {2.930 (0.148)} & {0.835} & {0.741} & {0.774} & {0.020} & {6.310} \\
				\hline
		\end{tabular}}
	\end{table}
	
	The Gaussian results show that the strong performance of P-WSM is not specific to chain graphs. The two nonlinear settings are more difficult and show that recovery depends on the graph structure and signal configuration. {In the nonlinear sparse-random setting, P-WSM has the smallest mean structural Hamming distance, while BEAM has the largest mean F1.}
	
	\subsection{Tuning and finite-sample limits}
	\label{sec:supp-pwsm-limits}
	
	Figure~\ref{fig:supp-pwsm-tuning} compares the primary minimum raw cross-validation rule with four alternatives evaluated on the same fitted paths. The primary rule performs well in the benchmark and main-effect settings, but the difficult settings retain a gap between the selected graph and the best graph available on the path. In the weak-signal setting, the P-WSM structural Hamming distance is 6.63 under the primary rule and 4.75 at the best point on the path. In the overfitted-sieve setting, the corresponding values are 6.21 and 3.17, illustrating that finite-sample estimation errors under overfitted sieves stem fundamentally from model complexity rather than suboptimality of the selection rule.
	
	\begin{figure}[!htbp]
		\centering
		\includegraphics[width=0.86\textwidth]{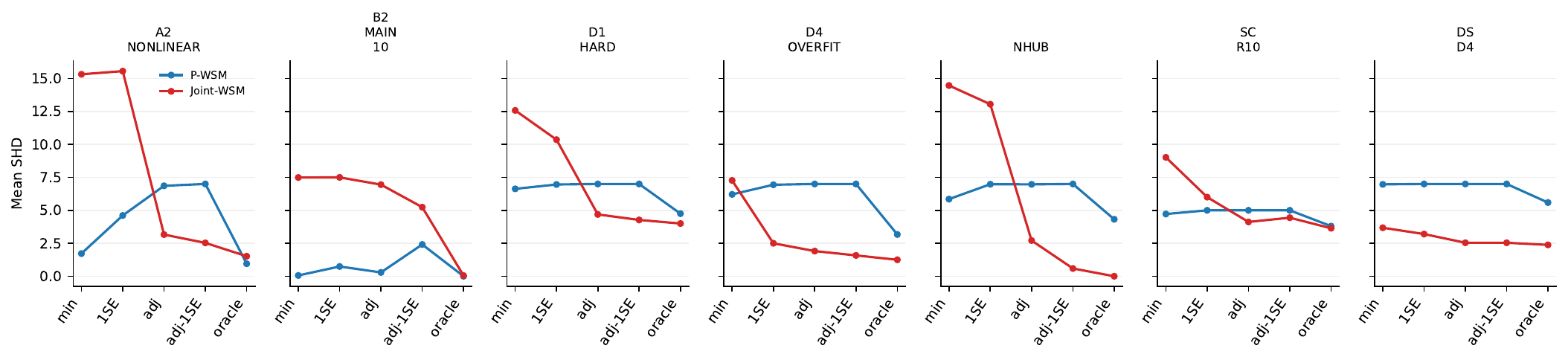}
		\caption{Sensitivity to the tuning rule. ``min'' denotes the primary minimum raw five-fold cross-validation rule; the remaining rules are sensitivity analyses evaluated on the same fitted paths.}
		\label{fig:supp-pwsm-tuning}
	\end{figure}
	
	We next vary the dimension of the unpenalized main-effect model relative to the sample size. The six settings use nonlinear disjoint-pair graphs with \(a=1\), \(\rho=0.75\), and \(d=2\). Table~\ref{tab:supp-pwsm-scaling} reports the P-WSM results together with the conditioning of the main-effect Hessian and the frequency with which the selected penalty is the zero-edge endpoint.
	
	\begin{table}[!htbp]
		\centering
		\caption{Nuisance-dimension scaling for P-WSM.}
		\label{tab:supp-pwsm-scaling}
			\begin{tabular}{cccccccc}
				\hline
				\(p\) & \(n\) & \(pd/n\) & \(\lambda_{\min}(H_{\beta\beta})\) & \(\kappa(H_{\beta\beta})\) & SHD & Exact & Endpoint \\
				\hline
				6 & 1200 & 0.010 & 0.105 & 2.01 & 0.30 & 0.76 & 0.00 \\
				6 & 400 & 0.030 & 0.100 & 2.19 & 2.13 & 0.05 & 0.43 \\
				10 & 400 & 0.050 & 0.098 & 2.26 & 3.84 & 0.01 & 0.43 \\
				10 & 200 & 0.100 & 0.093 & 2.47 & 4.72 & 0.00 & 0.80 \\
				6 & 60 & 0.200 & 0.082 & 3.04 & 2.99 & 0.00 & 0.98 \\
				6 & 40 & 0.300 & 0.078 & 3.41 & 2.99 & 0.00 & 0.97 \\
				\hline
		\end{tabular}
		\end{table}
	
	\begin{figure}[!htbp]
		\centering
		\subfigure[Nuisance-dimension scaling.]{\includegraphics[width=0.48\textwidth]{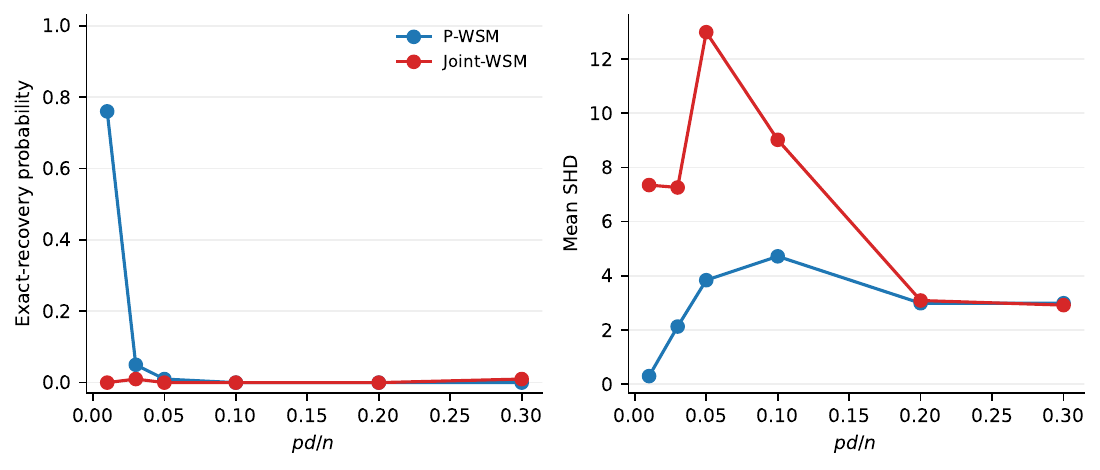}}
		\hfill
		\subfigure[Fixed-design basis-dimension sensitivity.]{\includegraphics[width=0.48\textwidth]{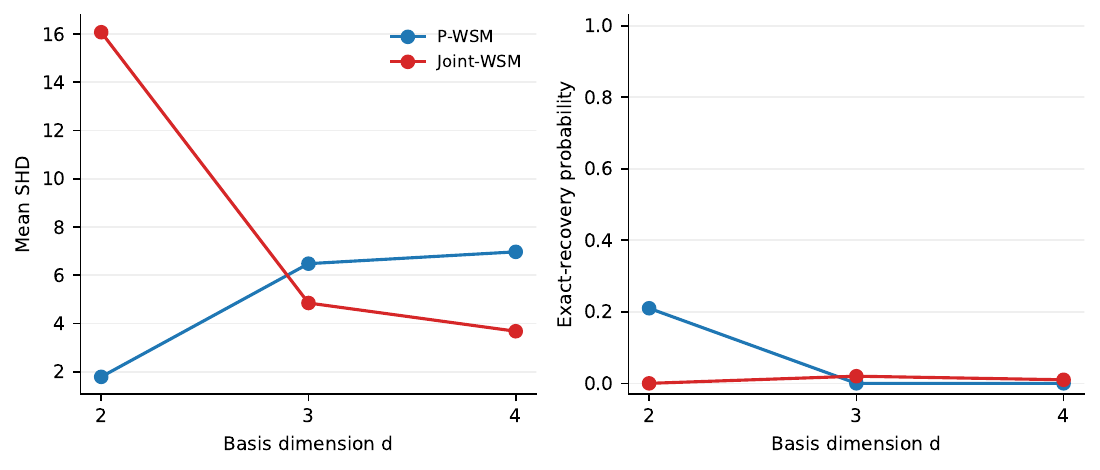}}
		\caption{Finite-sample limits of P-WSM. The left panel varies \(pd/n\); the right panel holds the data-generating model fixed and varies the basis dimension.}
		\label{fig:supp-pwsm-limits}
	\end{figure}
	
	As \(pd/n\) increases, exact recovery disappears and the selected model increasingly collapses to the zero-edge endpoint. The smaller structural Hamming distance observed at the two largest ratios reflects aggressive model sparsity rather than improved edge precision: because the true graph contains only three edges, an almost empty estimated graph trivially incurs an SHD near three. In the fixed-design basis experiment, the mean P-WSM structural Hamming distances for \(d=2,3,4\) are 1.79, 6.48, and 6.97, with exact-recovery probabilities 0.21, 0, and 0. These results reflect the increasing difficulty of estimating both the unpenalized main effects and the \(d^2\)-dimensional interaction blocks from a finite sample.
	
	\subsection{Numerical checks and computation}
	\label{sec:supp-pwsm-computation}
	
	{All P-WSM and Joint-WSM fits in the 2,800-replicate study satisfied the convergence and KKT criteria. The largest KKT residual was \(4.9999991\times10^{-6}\), below the tolerance \(5\times10^{-6}\), and no primary terminal fit failed. All \(2,800\times5\times31=434,000\) Joint-WSM cross-validation refits satisfied the normal-equation checks, with held-out scores agreeing with those used for selection. No spectral truncation was used.}
	
	The main-effect displacement reconstruction used in the propagation analysis covers 651,000 block/path records and {passed its stored convergence and KKT checks}. Because P-WSM refreshes the unpenalized main-effect coefficients exactly, it can require more linear-system work than Joint-WSM.

	\section{Bendall P-WSM analysis}
	\label{sec:supp-bendall-pwsm}
	
	\subsection{Data, representation, and tuning}
	
	The analysis uses 19,987 cells from the LPS-stimulated sample of the first
	donor in the Bendall bone-marrow data.  We retain 18 intracellular state
	markers and the three monocyte populations used in the main paper.  Within
	each training fold, every marker is centered by its median, scaled by its
	median absolute deviation, mapped through the logistic function with scale
	one, and rescaled to \(( -\pi,\pi )\).  Validation observations use the
	corresponding training-fold transformation.  The primary representation uses
	the centered-monomial basis with \(d=2\) and the squared coordinatewise
	boundary-distance weight.
	
	P-WSM leaves all 18 main-effect blocks unpenalized and penalizes only the 153
	pairwise interaction blocks.  Joint-WSM penalizes both block types.  Both
	methods use the same five folds and 31 relative penalty fractions from 1 to
	0.005; each fraction multiplies the method- and training-fold-specific
	\(\lambda_{\max}\). {At each path point, the unpenalized refit
		uses all main effects and selected interactions for P-WSM, and selected
		main effects and interactions for Joint-WSM. The primary rule minimizes
		the mean raw held-out score of this refit. This refitting protocol is
		distinct from the common all-main refit of the later four-route comparison.}
	
	Table~\ref{tab:supp-bendall-tuning} reports the primary rule and three
	prespecified alternatives.  The minimum-raw curve is flat near its minimum:
	the one-standard-error rule reduces the P-WSM graph from 131 to 90 edges.
	Across all four rules, P-WSM selects fewer interactions than Joint-WSM.
	
	\begin{table}[!htbp]
		\centering
		\caption{Bendall tuning-rule comparison.  The score columns report mean raw
			held-out loss after fixed-support refitting; smaller values are better.}
		\label{tab:supp-bendall-tuning}
		\resizebox{\textwidth}{!}{%
			\begin{tabular}{lrrrrrrrrrr}
				\hline
				Rule & P fraction & P edges & P score & J fraction & J edges & J score & Common & P only & J only \\
				\hline
				Minimum raw & 0.04967 & 131 & -1.30199 & 0.04967 & 141 & -1.30271 & 131 & 0 & 10 \\
				Raw 1-SE & 0.12011 & 90 & -1.29163 & 0.17100 & 109 & -1.29380 & 88 & 2 & 21 \\
				Minimum adjusted & 0.20403 & 57 & -1.24841 & 0.29046 & 81 & -1.26074 & 57 & 0 & 24 \\
				Adjusted 1-SE & 0.24344 & 48 & -1.22586 & 0.29046 & 81 & -1.26074 & 48 & 0 & 33 \\
				\hline
		\end{tabular}}
	\end{table}
	
	\begin{figure}[H]
		\centering
		\includegraphics[width=0.86\textwidth]{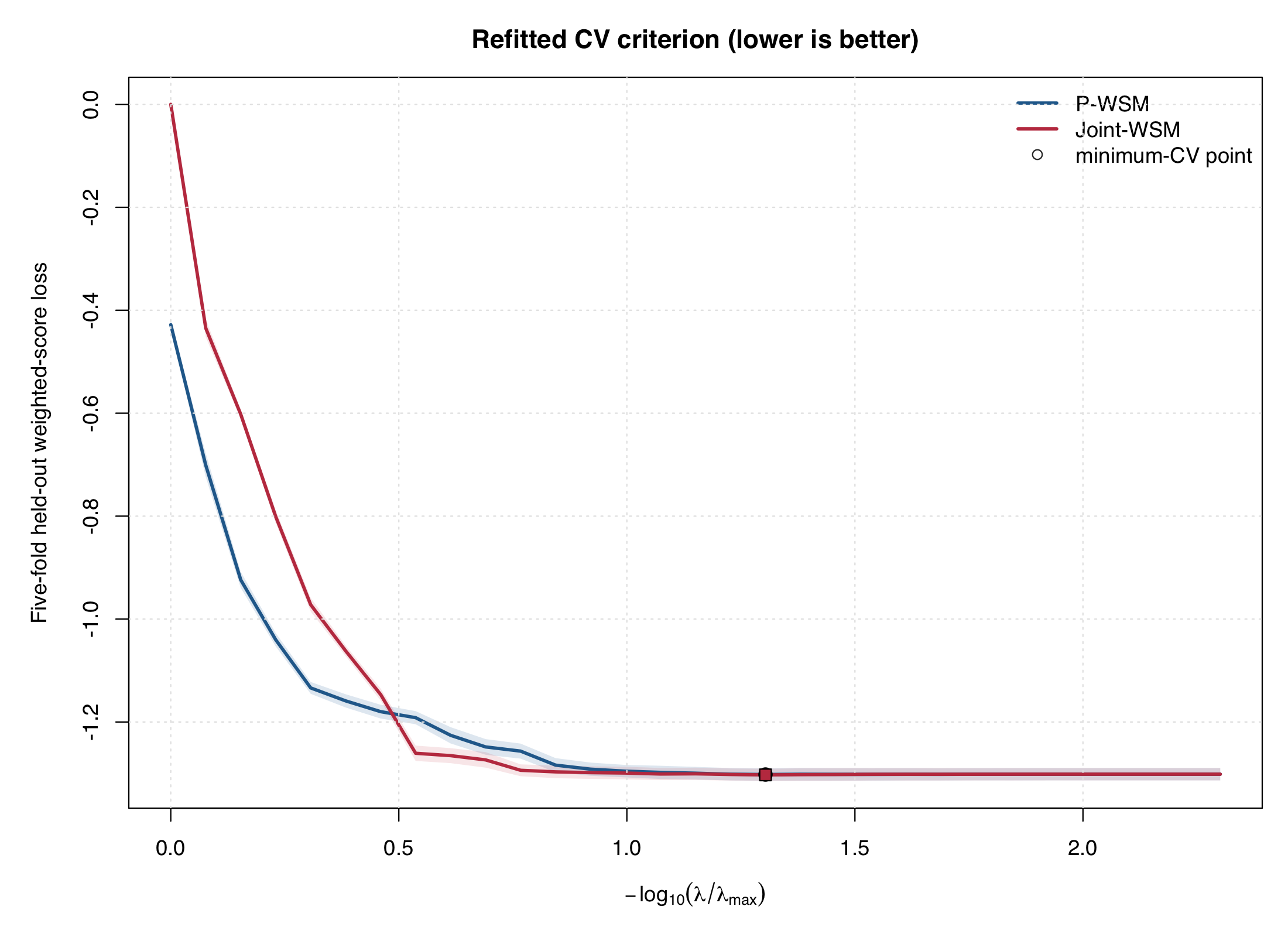}
		\caption{Five-fold raw held-out weighted-score curves after fixed-support
			refitting.  Filled symbols mark the minimum-raw choices.}
		\label{fig:supp-bendall-cv}
	\end{figure}
	
	\subsection{Primary fit and method comparison}
	
	Table~\ref{tab:supp-bendall-primary} gives the primary numerical results.  The
	P-WSM and Joint-WSM fits choose the same relative path position.  P-WSM selects
	131 interactions and Joint-WSM selects 141; all P-WSM edges are shared.  Both
	restricted refits are full rank and satisfy their normal equations to numerical
	precision.
	
	\begin{table}[!htbp]
		\centering
		\caption{Primary Bendall P-WSM and protocol-matched Joint-WSM fits.}
		\label{tab:supp-bendall-primary}
		\resizebox{0.98\textwidth}{!}{%
			\begin{tabular}{lrrrrrrrr}
				\hline
				Method & \(\lambda\) & Edges & Selection loss & Refit loss & CV refit (SE) & KKT max & Refit rank & Condition \\
				\hline
				P-WSM & 0.020955 & 131 & -1.315289 & -1.338636 & -1.301991 (0.012190) & \(3.79\times10^{-7}\) & 560 & 120.88 \\
				Joint-WSM & 0.021014 & 141 & -1.302897 & -1.340208 & -1.302713 (0.012081) & \(9.97\times10^{-7}\) & 600 & 121.12 \\
				\hline
		\end{tabular}}
	\end{table}
	
	\begin{sloppypar}
		{The observed foldwise paired P-WSM-minus-Joint-WSM difference
			in refitted validation loss is 0.000722, with a fold-based descriptive
			standard error of 0.000310 and 95\% interval \([-0.000138,0.001582]\).
			The 10 Joint-WSM interactions not selected by P-WSM are}
		pPLCgamma2--pBtk/Itk, pSTAT5--Ki67, pSTAT5--pSHP2,
		pSTAT5--pERK1/2, pSHP2--pS6, pERK1/2--pS6,
		pMAPKAPK2--pBtk/Itk, pZAP70/Syk--pS6, pH3--pBtk/Itk, and
		pBtk/Itk--pCREB.
	\end{sloppypar}
	
	The graphical-lasso and \texttt{highscore} analyses select 137 and 144
	edges.  P-WSM shares 123 edges with graphical lasso and 127 with
	\texttt{highscore}, corresponding to Jaccard similarities 0.848 and 0.858.
	Graphical lasso and \texttt{highscore} share 137 edges.  Their Gaussian
	conditional-dependence target on the standardized pre-logistic data differs
	from the bounded nonlinear interaction target of P-WSM, so these comparisons
	concern support overlap rather than a common score.
	
	\subsection{Stability}
	
	We draw 50 independent 80\% subsamples without replacement, stratified by the
	three monocyte populations.  The primary full-sample penalty fraction is held
	fixed and multiplied by each subsample's \(\lambda_{\max}\); the penalty is not
	retuned.  The selected graph contains 131.96 edges on average, with standard
	deviation 2.11 and range 127--137.  Among the 131 primary edges, 127 have
	selection frequency at least 0.8, the mean frequency is 0.979, and the minimum
	frequency is 0.70.
	These high subsampling frequencies confirm the pronounced algorithmic stability of the
	estimated conditional signaling network under cellular data perturbations.
	
	\begin{figure}[!htbp]
		\centering
		\includegraphics[width=0.92\textwidth]{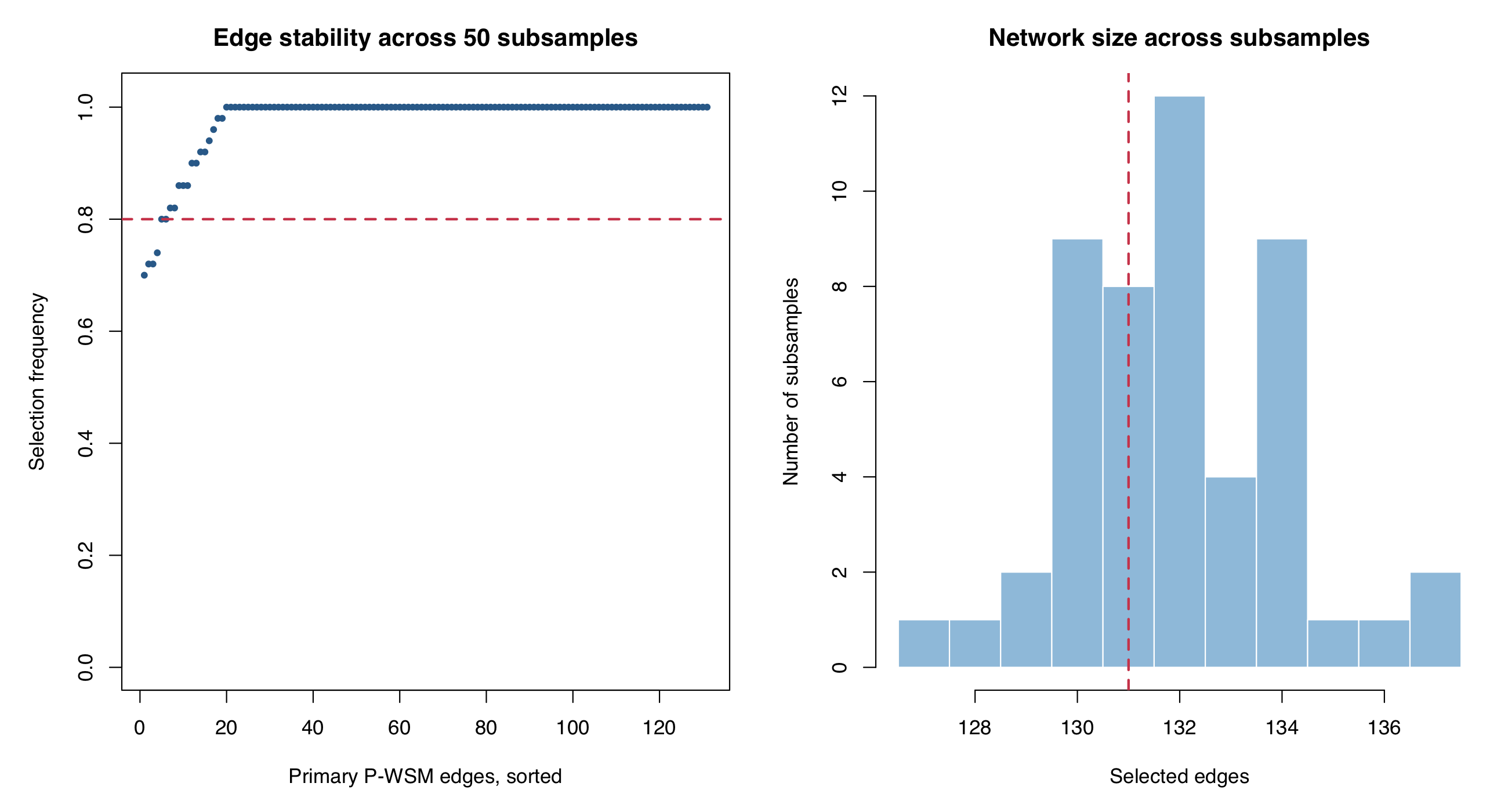}
		\caption{P-WSM stability across 50 stratified 80\% cell subsamples.  The left
			panel gives selection frequencies for the 131 primary edges; the right panel
			gives the resampled graph sizes.}
		\label{fig:supp-bendall-stability}
	\end{figure}
	
	\begin{table}[!htbp]
		\centering
		\caption{Ten fully stable P-WSM interactions with the largest refitted block
			norms.}
		\label{tab:supp-bendall-stable-edges}
		\begin{tabular}{lrr}
			\hline
			Interaction & Selection frequency & Refit block norm \\
			\hline
			Ki67--pH3 & 1.00 & 4.240 \\
			pP38--pSrcFK & 1.00 & 2.795 \\
			pNFkB--pH3 & 1.00 & 2.521 \\
			pMAPKAPK2--pP38 & 1.00 & 2.439 \\
			pERK1/2--pNFkB & 1.00 & 2.434 \\
			pH3--pSrcFK & 1.00 & 2.314 \\
			pH3--pP38 & 1.00 & 2.170 \\
			Ki67--pNFkB & 1.00 & 1.749 \\
			pNFkB--pP38 & 1.00 & 1.684 \\
			Ki67--pSrcFK & 1.00 & 1.645 \\
			\hline
		\end{tabular}
	\end{table}
	
	\subsection{Sensitivity and numerical checks}
	
	Table~\ref{tab:supp-bendall-sensitivity} varies one modeling choice at a time
	and repeats the complete five-fold path and refit for both methods.  Joint-WSM
	selects 9--34 more interactions in all three settings.  Increasing the basis
	dimension produces the largest support change and the largest separation
	between the two methods.
	
	\begin{table}[!htbp]
		\centering
		\caption{Bendall sensitivity analysis under minimum raw five-fold tuning.}
		\label{tab:supp-bendall-sensitivity}
		\resizebox{0.96\textwidth}{!}{%
			\begin{tabular}{lrrrrrrr}
				\hline
				Setting & P edges & J edges & Common & P only & J only & Jaccard & P-minus-J score \\
				\hline
				Monomial \(d=3\) & 116 & 150 & 116 & 0 & 34 & 0.773 & -0.00556 \\
				Linear boundary weight & 136 & 145 & 136 & 0 & 9 & 0.938 & 0.00007 \\
				Logistic scale 1.5 & 130 & 143 & 129 & 1 & 14 & 0.896 & -0.00094 \\
				\hline
		\end{tabular}}
	\end{table}
	
	All 31 full-data path fits and all 155 fold-path fits for each primary method
	converged and passed the \(5\times10^{-6}\) KKT tolerance.  The largest KKT
	residual was \(1.22\times10^{-6}\) on the full-data paths and
	\(1.37\times10^{-6}\) across the fold paths.  Every active interaction block
	matched a nonzero coefficient block, and both final refits preserved the
	selected support exactly.  All 50 stability fits and all six sensitivity
	analyses also passed their convergence, KKT, rank, and refit-stationarity
	checks.
}

\bibliographystyle{agsm}
\bibliography{references_7}

@misc{janofsky2018learning,
  author        = {Janofsky, Eric},
  title         = {Learning High-Dimensional Graphical Models with Regularized Quadratic Scoring},
  year          = {2018},
  note          = {arXiv:1809.05638},
  doi           = {10.48550/arXiv.1809.05638}
}

@article{yu2019generalized,
  author  = {Yu, Shiqing and Drton, Mathias and Shojaie, Ali},
  title   = {Generalized Score Matching for Non-Negative Data},
  journal = {Journal of Machine Learning Research},
  volume  = {20},
  number  = {76},
  pages   = {1--70},
  year    = {2019}
}

@article{yu2022generalized,
  author  = {Yu, Shiqing and Drton, Mathias and Shojaie, Ali},
  title   = {Generalized Score Matching for General Domains},
  journal = {Information and Inference: A Journal of the IMA},
  volume  = {11},
  number  = {2},
  pages   = {739--780},
  year    = {2022},
  doi     = {10.1093/imaiai/iaaa041}
}

@article{gu2013nonparametric,
  title={Nonparametric density estimation in high-dimensions},
  author={Gu, Chong and Jeon, Yongho and Lin, Yi},
  journal={Statistica Sinica},
  volume = {23},
  number = {3},
  pages={1131--1153},
  year={2013},
  publisher={JSTOR}
}

@article{leday2019fast,
	title={Fast Bayesian inference in large Gaussian graphical models},
	author={Leday, Gwena{\"e}l GR and Richardson, Sylvia},
	journal={Biometrics},
	volume={75},
	number={4},
	pages={1288--1298},
	year={2019},
	publisher={Oxford University Press}
}

@article{bendall2011single,
  title={Single-cell mass cytometry of differential immune and drug responses across a human hematopoietic continuum},
  author={Bendall, Sean C and Simonds, Erin F and Qiu, Peng and Amir, El-ad D and Krutzik, Peter O and Finck, Rachel and Bruggner, Robert V and Melamed, Rachel and Trejo, Angelica and Ornatsky, Olga I and others},
  journal={Science},
  volume={332},
  number={6030},
  pages={687--696},
  year={2011},
  publisher={American Association for the Advancement of Science}
}

@article{lin2016estimation,
	title={Estimation of high-dimensional graphical models using regularized score matching},
	author={Lin, Lina and Drton, Mathias and Shojaie, Ali},
	journal={Electronic journal of statistics},
	volume={10},
	number={1},
	pages={806--854},
	year={2016}
}

@article{tan2019layer,
	title={Layer-wise learning strategy for nonparametric tensor product smoothing spline regression and graphical models},
	author={Tan, Kean Ming and Lu, Junwei and Zhang, Tong and Liu, Han},
	journal={Journal of machine learning research},
	volume={20},
	number={119},
	pages={1--38},
	year={2019}
}

@article{williams2021bayesian,
	title={Bayesian estimation for Gaussian graphical models: Structure learning, predictability, and network comparisons},
	author={Williams, Donald R},
	journal={Multivariate Behavioral Research},
	volume={56},
	number={2},
	pages={336--352},
	year={2021},
	publisher={Taylor \& Francis}
}

@article{mardia2008multivariate,
	title={A multivariate von Mises distribution with applications to bioinformatics},
	author={Mardia, Kanti V and Hughes, Gareth and Taylor, Charles C and Singh, Harshinder},
	journal={Canadian Journal of Statistics},
	volume={36},
	number={1},
	pages={99--109},
	year={2008},
	publisher={Wiley Online Library}
}

@article{sun2015learning,
	title={Learning structured densities via infinite dimensional exponential families},
	author={Sun, Siqi and Kolar, Mladen and Xu, Jinbo},
	journal={Advances in neural information processing systems},
	volume={28},
	year={2015}
}

@article{jordan2004graphical,
	title={Graphical models},
	author={Jordan, MI},
	journal={Statistical Science},
	volume={19},
	number={1},
	pages={140--155},
	year={2004}
}

@book{lauritzen1996graphical,
	title={Graphical models},
	author={Lauritzen, Steffen L},
	volume={17},
	year={1996},
	publisher={Clarendon Press}
}

@article{wainwright2006sharp,
	title={Sharp thresholds for high-dimensional and noisy recovery of sparsity},
	author={Wainwright, Martin J},
	journal={arXiv preprint math/0605740},
	year={2006}
}

@article{ravikumar2009sparse,
	title={Sparse additive models},
	author={Ravikumar, Pradeep and Lafferty, John and Liu, Han and Wasserman, Larry},
	journal={Journal of the Royal Statistical Society Series B: Statistical Methodology},
	volume={71},
	number={5},
	pages={1009--1030},
	year={2009},
	publisher={Oxford University Press}
}

@article{yuan2007model,
  title={Model selection and estimation in the Gaussian graphical model},
  author={Yuan, Ming and Lin, Yi},
  journal={Biometrika},
  volume={94},
  number={1},
  pages={19--35},
  year={2007},
  publisher={Oxford University Press}
}

@article{sriperumbudur2017density,
	title={Density estimation in infinite dimensional exponential families},
	author={Sriperumbudur, Bharath and Fukumizu, Kenji and Gretton, Arthur and Hyv, Aapo and Kumar, Revant and others},
	journal={Journal of Machine Learning Research},
	volume={18},
	number={57},
	pages={1--59},
	year={2017}
}

@article{hyvarinen2005estimation,
	title={Estimation of non-normalized statistical models by score matching.},
	author={Hyv{\"a}rinen, Aapo},
	journal={Journal of Machine Learning Research},
	volume={6},
	pages={695--709},
	year={2005}
}

@article{liu2022estimating,
	title={Estimating density models with truncation boundaries using score matching},
	author={Liu, Song and Kanamori, Takafumi and Williams, Daniel J},
	journal={Journal of Machine Learning Research},
	volume={23},
	number={186},
	pages={1--38},
	year={2022}
}

@article{jeon2006effective,
  title={An effective method for high-dimensional log-density anova estimation, with application to nonparametric graphical model building},
  author={Jeon, Yongho and Lin, Yi},
  journal={Statistica Sinica},
  volume={16},
  number = {2},
  pages={353--374},
  year={2006},
  publisher={JSTOR}
}

@article{dawid1979conditional,
  author={Dawid, A. Philip},
  title={Conditional Independence in Statistical Theory},
  journal={Journal of the Royal Statistical Society: Series B (Methodological)},
  volume={41},
  number={1},
  pages={1--15},
  year={1979}
}

\end{document}